\documentclass
[aps,prc,twocolumn,showpacs,showkeys,amsmath,floatfix,superscriptaddress,nofootinbib]
{revtex4-1}
\usepackage{graphicx}
\usepackage{mathptmx}                
\usepackage{dcolumn}                 
\usepackage{bm}                      
\usepackage{eurosym} 			

\newcommand{\be}{\begin{equation}}
\newcommand{\ee}{\end{equation}}
\newcommand{\bea}{\begin{eqnarray}}
\newcommand{\eea}{\end{eqnarray}}

\usepackage{color}
\usepackage{soul,latexsym, amssymb, amsmath, multirow, mathrsfs, booktabs}
\usepackage{longtable}
\usepackage{rotating}

\begin{document}
\title{Unified treatment of sub-saturation stellar matter at zero and finite temperature}

\author{F. Gulminelli}
\affiliation{
CNRS/ENSICAEN/LPC/Universit\'e de Caen Basse Normandy, 
UMR6534, F-14050 Caen c\'edex, France}

\author{Ad. R. Raduta}
\affiliation{
IFIN-HH, Bucharest-Magurele, POB-MG6, Romania}

\begin{abstract}
The standard variational derivation of stellar matter structure in the Wigner-Seitz approximation is generalized 
to the finite temperature situation where a wide distribution of different nuclear species can coexist in the same
density and proton fraction condition, possibly out of $\beta$-equilibrium. 
The same theoretical formalism is shown to describe on one side the single-nucleus approximation (SNA), currently used in most core collapse 
supernova simulations,
and on the other side the nuclear statistical equilibrium (NSE) approach, routinely employed in r- and p-process explosive nucleosynthesis problems. 
In particular we show that
in-medium effects have to be accounted for in NSE to have a theoretical consistency between the zero and finite temperature modelling. The bulk part of these in-medium effects is analytically calculated in the local density approximation and shown to be different from a van der Waals excluded volume term.
This unified formalism allows controlling quantitatively the deviations from the SNA in the different thermodynamic conditions, as well as having a
NSE model which is reliable at any arbitrarily low value of the temperature, with  potential applications for neutron star cooling and accretion problems. We present different illustrative results with several mass models and effective interactions, showing the importance of accounting for the nuclear species distribution even at temperatures lower than 1 MeV.  
\end{abstract}

\today

\maketitle
 
\section{Introduction}
Since the pioneering work of G.Baym and collaborators in the early seventies~\cite{BPS,BBP}, the theoretical formalism to variationally calculate the equation of state and composition of neutron star crusts with cluster degrees of freedom is well settled and has been exploited by many authors
in the last decades~\cite{ravenhall,hashimoto,Pethick,watanabe_pasta,douchin}.
Because of the crystalline structure of low density neutron star matter, the problem of neutron star structure at a given pressure 
is indeed reduced to the composition of a simple Wigner-Seitz cell, composed of a single nucleus immersed in a gas of neutrons and electrons. The ground state of the system is then given by a set of coupled variational equations for the nucleus atomic and baryonic number (and shape, if the more exotic pasta phases are included), the volume of the cell, and the free neutron density~\cite{BPS,BBP}. 

An alternative formulation within density functional theory was developed at the same time in another seminal paper for neutron star physics by J.Negele and D.Vautherin~\cite{NV1973}. This entirely microscopic approach is in principle more appealing than a cluster model because it allows accounting for the polarisation of the neutron and electron gas. More generally, a microscopic description avoids the artificial distinction between clusters and free neutrons, and naturally accounts for the interface interaction between them. For this reason, and due to the great improvements of the predictive power of mean-field energy density functionals in the last decades~\cite{chabanat98,BSK,dutra1,dutra2}, microscopic Hartree-Fock and Hartree-Fock-Bogoliubov methods have been widely employed for the computation of the neutron star equation of state~\cite{douchin2,maruyama,baldo,pearson,pastore,constanca,stone}.
As a consequence of this important collective theoretical effort, present uncertainties on the equation of state 
of neutron star matter at zero temperature are essentially limited to the still imperfect knowledge of the density dependence of the symmetry energy~\cite{epja_book}, which is itself better and better constrained thanks to the recent improvements in ab-initio neutron matter calculations
~\cite{gandolfi,sammaruca,schwenk,hagen,gezerlis,carbone,bulgac,pederiva}.

This standard picture was recently challenged in Ref.\cite{pethick14}, where it is argued that the attractive interaction between the lattice nuclei and the surrounding free neutrons might induce lattice instabilities similar to the case of the ferroelectric phase transition in terrestrial metallic alloys. This very promising avenue is not explored in the present paper and we restrict to the standard picture where the neutrons-nucleus interaction is supposed to only lead to a modification of the surface tension. 
 
Neutron stars being born hot, a natural extension of the above mentioned works concern the consideration of finite temperature stellar matter, with applications ranging from neutron star cooling, accretion in binary systems and dynamics of supernova matter with associated nucleosynthesis problems. For these applications matter is typically out of $\beta$-equilibrium and therefore needs to be considered in a large interval of baryonic densities $\rho_B$ and proton fractions $y_p$. Finite temperature mean-field calculations in the Wigner-Seitz cells
have been largely employed~\cite{RMF_finiteT,pasta_RMF,HFBT,fortin,onsi,pasta_ETF}. However, because of the computational effort associated to these calculations, microscopic modelling of the finite temperature Wigner-Seitz cells is not adapted to the large scale calculations needed for supernova simulations, even if some large scale time-dependent Hartree-Fock
(TDHF) calculations start to be performed~\cite{pasta_tdhf1,pasta_tdhf2,pasta_tdhf3}.  
For this reason, hybrid models with cluster degrees of freedom are more appealing to address the finite temperature problem. 
The extension of the Baym et al. compressible liquid drop models to finite temperature 
was already proposed in the eighties~\cite{Pethick} and allowed the elaboration of the famous Lattimer-Swesty (LS)~\cite{LS1991} and Shen~\cite{shen} supernova equation of state models, which are still widely used in present supernova simulations.

The problem of both microscopic and liquid-drop models is that they share the so-called Single-Nucleus-Approximation (SNA), that is a unique configuration is assumed for each $(T,\rho_B,y_p)$ thermodynamic condition, against the very principle of statistical mechanics which stipulates that finite temperature corresponds to a mixing of different microstates. In particular, in the LS and Shen EoS, besides free nucleons, only one kind of light cluster ($\alpha$ particles) and one kind of heavy cluster are assumed to exist. The idea is to account in an average way for the properties of
the statistical cluster distribution. The SNA may not affect very strongly thermodynamic properties of matter in the temperature and density domains of interest~\cite{burrows}, but it  has important consequences for dynamical processes dependent on
reaction rates of specific nuclei~\cite{ec1,ec2} and for the gas-liquid phase transition. Therefore, more modern approaches rely
on an extended nuclear statistical equilibrium (NSE) concept, where the distribution of clusters over, in principle, all mass
numbers is taken into account and obtained self-consistently under conditions of statistical equilibrium~\cite{philips,botvina,souza,blinnikov}.
Originally, the NSE was introduced to describe the reaction network taking place at the end of the evolution of massive stars
in red supergiants~\cite{philips}. Being very diluted, nuclei interact weakly and are almost not modified by the surrounding medium. These
conditions naturally lead to the Saha equations. The NSE in the dense and hot matter in the core of supernovae was first applied
in the EoS of Hillebrandt and Wolff~\cite{hillebrandt}. 

In recent NSE implementations~\cite{nse_heckel,nse_hempel,nse_us,esym_paper,nse_japan},
the interactions between a cluster and the surrounding gas is treated in the so-called excluded-volume approach. The clusters and the gas of light particles do not overlap in space and the clusters’ binding energy is kept as in the free limit. It is known, however, from virial expansion at low density and quantal approaches~\cite{horowitz_schwenk,samaddar_smatrix,Typel2010,typel_hempel}, that the cluster properties are modified by the coexistence with a gas. 
In particular, the recent G.Shen et al equation of state \cite{shenG} includes 
these in-medium effects for light particles within a virial expansion insuring the proper model independent low-density limit.
 Moreover, the excluded volume treatment of cluster-nucleon interaction is not compatible with microscopic calculations in the Wigner-Seitz cell, where cluster properties are naturally modified by the surrounding gas by the density dependence of the self-consistent mean-field and the Pauli-blocking effect of occupied single-particle states. The conceptual difference between the classical excluded-volume picture and the quantal picture emerging from microscopic calculations was discussed in Ref.~\cite{Panagiota2013}. It leads to two different definitions of clusters in dense media, namely configuration-space and energy-space clusters, with different particle number and energy functionals. Including one or the other of the two definitions in a finite temperature NSE partition sum will naturally produce differences in the observables, even if the total free energy of the Wigner Seitz cell entering in the SNA approaches~\cite{LS1991,shen,shenG} does not depend by construction on the cluster definition~\cite{Panagiota2013}.
As a consequence, it is not clear if the NSE models have the correct limit towards $T=0$, where the SNA approximation becomes exact. Recent comparisons~\cite{nse_comparison}   indicate that huge differences exist among the different models even at very low temperature, suggesting that the zero temperature limit is not fully under control. Such an uncontrolled model dependence might be an important hindrance to pin down the EoS dependence of supernova dynamics~\cite{Sumiyoshi_05,Fischer_14}.

In this paper we develop an analytical unified theoretical formalism  to describe on one side the single-nucleus approximation (SNA),  
and on the other side the nuclear statistical equilibrium (NSE) approach. To this aim, we map the energetics and composition of a microscopic Wigner-Seitz cell into a model of the same cell with cluster degrees of freedom. If a density and isospin dependent modification of the cluster surface energy is included, this cluster model can thus exactly span the full spectroscopy (ground state and excited states) of the extended Thomas-Fermi (ETF) approximation~\cite{onsi} with the only uncertainty given by the employed energy density functional. 

A variational minimization of  the total free energy of the Wigner-Seitz cell with respect to the cell composition leads to the standard SNA equilibrium approach, at zero as well as finite temperature. A complete finite temperature treatment is obtained by calculating the partition sum of a system of independent cells, leading to a statistical distribution of cells with different compositions. NSE equations naturally emerge from this treatment, but energy-space clusters are demonstrated to be the correct degrees of freedom in order to get a consistent treatment towards the zero temperature limit.
We also show that a cut-off in the cluster density of states has to be applied in order to avoid double counting of scattering states.

The first part of the paper is devoted to zero temperature.
Section \ref{section:WS} defines the degrees of freedom and associated energy functionals used in this work. Section \ref{section:t0equations} gives the variational equations to be solved at zero temperature to get the ground state of stellar matter. The non-standard case where $\beta$ equilibrium is not imposed is also considered. This case is not physically realistic, but gives the reference zero temperature limit of supernova matter, thus guaranteeing the consistency of the finite temperature formalism. To maximize the predictive power of the formalism, an experimental nuclear mass table is used in section \ref{section:crust} to predict the composition of the neutron star crust, and results are compared to the rich literature available on this subject.
The equation of state is briefly addressed in Section \ref{section:EOS}, and
to conclude the zero temperature discussion, the issue of phase transitions is analyzed in 
section \ref{section:phase_transition}. 
We  confirm that the constraint of charge neutrality quenches  
the first order nuclear matter liquid-gas phase transition. 
A residual very narrow  transition region exists at densities of the order of
$\rho_0/5-\rho_0/3$, depending on the interaction, which physically corresponds to the emergence of 
pasta phases.  

In the second part of the  paper, we switch to finite temperature.  
Section \ref{section:SNA} gives the derivation of the coupled variational equations in the 
SNA approximation, as well as some applications in $\beta$-equilibrium. Sections \ref{section:NSE_can}, 
\ref{section:NSE}  build the partition sum of the model in the canonical 
and in the grandcanonical ensemble, leading to the derivation of the generalized NSE equations, 
which are compared to the SNA approximation in section \ref{section:NSE_SNA}. 
The way in which the phenomenology of dilute nuclear matter is modified, in stellar matter, 
by electrons is discussed in section \ref{sec:electrons}.
Finally section \ref{section:conclusions} gives a summary and conclusions.
  
\section{Zero temperature stellar matter}

\subsection{Energy in the Wigner-Seitz cell}\label{section:WS}

Let us define a zero temperature thermodynamic condition for compact star 
matter as a given value for the baryon density $\rho_B$ and proton fraction,
$y_p=\rho_p/\rho_B$, where $\rho_p$ is the proton density.
Since by definition there is no interaction among Wigner-Seitz cells, 
the total energy density of the system is given by:
\begin{equation}
\epsilon_{WS}^{tot}(\rho_B,y_p)=\lim_{N\to\infty}\frac{\sum_{i=1}^N E_{WS}^{tot}(i)}{\sum_{i=1}^N V_{WS}(i)} ,
\label{dense}
\end{equation}
where $ E_{WS}^{tot}(i)$ is the total energy (including rest mass contribution) of the $i-th$ WS cell.
Each cell consists of  $N_{WS}$ neutrons and $Z_{WS}$ protons in a volume $V_{WS}$.
We make the standard simplifying approximation that the cell 
consists of a single cluster with the possible addition of an homogeneous nuclear gas, 
as well as a homogeneous electron gas.
This approximation is inspired by the numerical results of microscopic 
calculations~\cite{douchin2,maruyama,baldo,pearson,pastore,constanca,stone}.
The polarisation of the nuclear gas by the cluster is shown to be small by these works. 
For this reason, this effect is generally accounted in cluster models as an in-medium modification 
of the surface tension \cite{BBP,Pethick,douchin}, see below.
Concerning the electron gas, self-consistent calculations have shown that, because of the 
high electron incompressibility, the homogeneous approximation is excellent for all densities \cite{maruyama}.
This Wigner-Seitz picture is however not fully realistic since
it is also well known that at finite temperature light clusters can 
coexist with the single heavy nucleus\cite{pasta_light,typel_light}. 
For this reason in the Lattimer-Swesty equation of state 
$\alpha$ particles are added to the nucleon gas inside the Wigner-Seitz cell\cite{LS1991}, but interactions
between the $\alpha$'s and the cluster (or the gas) are neglected in that model. 
This coexistence effect of heavy and light clusters will be automatically accounted 
by our formalism, because in the NSE model presented in section \ref{section:NSE} 
the equilibrium configuration
will consist in a mixture of different WS cells containing clusters of all species. However,  
two-body Coulomb and possibly nuclear effects due to multiple clusters inside a same cell are out of 
the scope of the present treatment, and each (light or heavy) cluster will be associated to its proper WS cell.

As we shall explicitly work out, for a given set 
$(A_{WS},I_{WS},V_{WS})$ ($A_{WS}=N_{WS}+Z_{WS}, I_{WS}=N_{WS}-Z_{WS}$)
equilibrium imposes a unique mass and composition of the cluster and of 
the gas. Five variables define this mass and composition, namely:
the cell volume $V_{WS}$, the gas density and composition
$\rho_{g}=\rho_{ng}+\rho_{pg}$, $ y_{g}=\rho_{ng}-\rho_{pg}$,  
where $\rho_{ng}$ ($\rho_{pg}$) is the density of neutrons (protons) in the gas, and   
 the neutron $N$ and proton $Z$ numbers associated to the cluster. 
The total energy in the Wigner-Seitz cell is  written as 
 $E_{WS}^{tot}=Z_{WS} m_pc^2+N_{WS}m_nc^2+E_{WS}$ 
 with:
\begin{equation}
 E_{WS}(A,Z,\rho_g,y_g,\rho_p)=E^{vac} +V_{WS} \left ( \epsilon_{HM}
 + \epsilon_{el}^{tot}\right )+\delta E. \label{energy_start}
\end{equation}
Here, $\epsilon_{HM}(\rho_g,y_g)$ is the energy density of homogeneous asymmetric nuclear matter,  
$\epsilon_{el}^{tot}(\rho_{el})$ is the total energy density (including rest mass contribution) 
of a uniform electron gas at the density 
$\rho_{el}=y_p\rho_B$ imposed by charge neutrality,
$E^{vac}(A,Z)$ is the energy of a cluster with $Z$ protons and $N=A-Z$ neutrons in the vacuum, 
and $\delta E$ is the in-medium modification due to the interaction between 
the cluster and the gas.
A part of the in-medium correction is given by the Coulomb screening 
by the electron gas, and by the Pauli-blocking effect of high energy cluster 
single particle states due to the gas ~\cite{Typel2010}. This latter effect can be 
approximately accounted for in the local density approximation by simply subtracting from the local energy density the contribution of the unbound gas states.
This local density approach is certainly insufficient for the in-medium effect on light clusters for which more sophisticated approaches have been proposed \cite{Typel2010,horowitz_schwenk,shenG}, but we expect it to represent 
the most important correction for medium-heavy nuclei. Indeed residual surface terms in that case appear to have only a perturbative effect~\cite{esym_paper}.

As in Ref. ~\cite{Panagiota2013} we introduce the left-over bound part of the cluster, 
\begin{eqnarray}
A_e&=&A\left ( 1-\frac{\rho_{g}}{\rho_{0}} \right ), \label{are} \\
Z_e&=&Z\left ( 1-\frac{\rho_{pg}}{\rho_{0p}} \right ). \label{zre}
\end{eqnarray}  
that we call "e-cluster".
$\rho_0(\delta)$ and $\rho_{0p}(\delta)$ stand for the total and, respectively, proton
densities of saturated nuclear matter of isospin asymmetry $\delta = 1 - 2 \rho_{0,p}/\rho_{0}$.
$\rho_0(\delta)$ may be calculated as\cite{Panagiota2013}
\begin{equation}
\rho_{0}(\delta) = \rho_{0}(0) \left( 1 - \frac{3 L_{sym}\delta^2}{K_{sat}+K_{sym}\delta^2} \right).
\label{eq_asym_rho0}
\end{equation}
where
$K_{sat}$ is the symmetric nuclear matter incompressibility,
and $L_{sym}$, $K_{sym}$ denote the slope and curvature of the symmetry energy at 
(symmetric) saturation $\rho_0(0)$.

In the above expressions the quantity $\delta$ represents the asymmetry in the nuclear bulk. 
It differs from the global asymmetry of the nucleus
$I/A = 1 - 2Z/A$ because of the competing effect of the Coulomb interaction and symmetry energy, 
which act in opposite directions in determining the difference between the proton and neutron radii.  
For a nucleus in the vacuum we take the estimation from the droplet model~\cite{ldm}:
\begin{equation}
\delta=\delta_0 = \frac{I + \frac{3a_c}{8Q} \frac{Z^2}{A^{5/3}}}{1+ \frac{9 J_{sym}}{4Q} \frac{1}{A^{1/3}}}.
\label{eq_asym_deltar}
\end{equation}
In this equation, $J_{sym}$ is the symmetry energy per nucleon at the saturation
density of symmetric matter,  $Q$ is the surface stiffness coefficient, and $a_c$ is the Coulomb parameter.
In the presence of an external gas of density $\rho_g$ and asymmetry 
$\delta_g=(\rho_{ng}-\rho_{pg})/\rho_g=y_g/\rho_g$,  the bulk asymmetry defined by eq.(\ref{eq_asym_deltar}) 
is generalized such as to account for the contribution of the gas as~\cite{Panagiota2013}: 
 
\begin{eqnarray}
\delta(\rho_g,y_g)&=&\left( 1-\frac{\rho_g}{\rho_0(\delta)}\right) \delta_0(Z_e,A_e)
+\frac{\rho_g}{\rho_0(\delta)}\delta_g,
\label{paperpana1:eq:deltacl}
\end{eqnarray}
where $\delta_0(Z_e,A_e)$ is the asymmetry value given by eq.(\ref{eq_asym_deltar}) 
if we consider only the bound part of the cluster, $A=A_e$, $Z=Z_e$, $I=I_e$.

For simplicity, in the following variational derivation of the equilibrium equations 
we shall initially assume $\delta=I/A$ in eq.(\ref{paperpana1:eq:deltacl}) which implies 
neglecting isospin inhomogeneities. However, we will include this effect in Section~\ref{section:NSE}.

It is interesting to observe that eq. (\ref{energy_start}) can now be conveniently written as: 
\begin{equation}
E_{WS}(A,Z,\rho_g,y_g,\rho_p)=E^e  +V_{WS} \left( \epsilon_{HM} +\epsilon_{el}^{tot}\right ) +\delta E_{surf}.
\label{energy}
\end{equation}
with $E^e$ standing for the in-medium modified cluster energy in the e-cluster 
representation:
\begin{equation}
E^e(A,Z,\rho_g,y_{g},\rho_p)=E^{vac} - \epsilon_{HM}
\frac{A}{\rho_0}+\delta E_{Coulomb}, \label{e-energy}
\end{equation}
or, alternatively,
\bea  
E_{WS}(A,Z,\rho_g,y_g,\rho_p)&=&\left ( V_{WS}-V_0\right ) \epsilon_{HM} +V_{WS}\epsilon_{el}^{tot} 
\nonumber \\
&+& E^{vac} +\delta E_{surf} +\delta E_{Coulomb}\label{energy-r},
\eea
where $V_0=A/\rho_0(\delta)$ is the equivalent cluster volume corresponding to $\delta$ isospin asymmetry.
In this representation, that we call "r-cluster" representation~\cite{Panagiota2013},
the in-medium effects only affect the surface properties of the cluster. 
The in-medium bulk term apparent in eq.(\ref{e-energy}) is here interpreted as an 
excluded volume. At variance with the classical Van der Waals model, this "excluded volume" 
is not a simple limitation of the r-space integral of the partition sum, 
but it directly affects the energetics of the Wigner-Seitz cell.

Once the dominant bulk and Coulomb in-medium effects are accounted for by the definition of the 
e-cluster representation eq.(\ref{e-energy}),  the residual in-medium energy shift 
$\delta E_{surf}$ can be shown to behave as a surface term~\cite{Panagiota2013,Aymard2014}.

The different contributions to the energy are defined as follows.
The presence of electrons in the cell modifies the cluster energy with respect to its 
vacuum value by the so called Coulomb shift,
\begin{equation}
E^{nuc}=E^{vac}+  \delta E_{Coulomb},
\label{eq:enuc}
\end{equation}
with
\begin{equation}
\delta E_{Coulomb}= a_c  f_{WS} A^{-1/3} Z^2,
\end{equation}
and the Coulomb screening function in the Wigner-Seitz approximation given by,
\begin{equation}
  f_{WS}(\delta,\rho_{el})=\frac 32\left ( \frac{2\rho_{el}}{(1-\delta)\rho_{0}}\right )^{1/3} -
\frac 12\left ( \frac{2\rho_{el}}{(1-\delta)\rho_{0}}\right ), \label{eq:screening}
\end{equation}
where we used for the average proton density inside the nucleus 
$\rho_{0,p}\left( \delta \right)=\rho_0\left(\delta\right)(1-\delta)/2$.
For the electron total energy density (containing the rest mass contribution)
we use the expression proposed in Ref. \cite{BPS} and valid above 
$10^4$ g$\cdot$cm$^{-3}$ where electrons may be considered free,
\begin{equation}
\epsilon_{el}^{tot}=\frac{m_{el}^4 c^5}{8\pi^2\hbar^3} \left [  
  \left ( 2t^2+1\right ) t \left ( t^2+1\right ) ^{1/2}
  -\ln \left (t + \left ( t^2+1\right ) ^{1/2}\right ) 
  \right ], 
\end{equation}
where $t=\hbar (3\pi^2\rho_{el})^{1/3}/m_{el} c$.
The total electron chemical potential (including the rest mass contribution)
is defined as a function of the total proton density $\rho_p=y_p\rho_B$ as  
\bea
\mu_{el}^{tot}&=&\frac{d\epsilon_{el}^{tot}}{d\rho_{el}}(\rho_{el}=\rho_p) 
= \frac{m_{el}^3 c^4}{8 (3\rho_{el}\pi^{2})^{2/3} \hbar^2}  \cdot \\
&\cdot& 
\left[(t^2+1)^{1/2}(1+6 t^2)+\frac{t^2 (2 t^2+1)}{(t^2+1)^{1/2}}
-\frac{1}{(1+t^2)^{1/2}}
\right]. \nonumber
\eea 

Unless otherwise explicitly mentioned,
we will use for the energy functional of the cluster in vacuum, $E_{vac}(A,Z)$,
the table of experimental masses of Audi {\em et al.} \cite{nudat}, 
publicly available in electronic format. 
When these latter are not known, which is typically the case close and above the drip-lines, 
we will use a liquid-drop parameterization 
\cite{Danielewicz2009} with coefficients fitted out of HF calculations using the same 
Skyrme effective interaction which is employed for the homogeneous gas. 
This parameterization, hereafter called Skyrme-LDM model, reads:
\begin{equation}
\frac{E^{vac}_{LDM}}{A}=a_v-a_s A^{-1/3}-a_a(A)\left ( 1-\frac{2Z}{A}\right )^2-a_c  \frac{Z^2}{A^{4/3}}, \label{dan}
\end{equation}
with the asymmetry energy coefficient:
\begin{equation}
 a_a(A) =\frac{a_v^a}{1+\frac{a_v^a}{a_s^a A^{1/3}}}.
\end{equation}
For the numerical applications concerning the NSE model in section \ref{section:NSE},
this parameterization will be supplemented in the case
of even mass nuclei with a simple phenomenological pairing term, $\Delta(A)=\pm 12/\sqrt{A}$ where 
+(-) corresponds to even-even (odd-odd) nuclei.
The in-medium surface correction 
$\delta E_{surf}(A,\delta,\rho_g,\delta_{g})$  due to the interaction with the surrounding gas 
can in principle be  accounted for by a density dependent modification of the surface and 
symmetry-surface coefficients. A determination of these coefficients within the extended Thomas-Fermi model~\cite{Aymard2014}
will be published elsewhere~\cite{Aymard2015}. For the numerical applications of this paper,
 we will ignore this correction, and consider that the main in-medium effect
is given by the bulk nuclear and Coulomb binding energy shift.

Below saturation, the Coulomb screening effect of the electrons is never total. 
This implies that only a finite number $N_{species}$ of nuclear species 
$(A,Z)$ can exist at zero temperature, and consequently a finite number of WS cells 
$N_{WS}(\rho_B,y_p)=N_{species}$. 
Eq. (\ref{dense}) then becomes:
\bea
\epsilon_{WS}(\rho_B,y_p)&=&\sum_{k=1}^{N_{WS}(\rho_B,y_p)} 
\frac{ E_{WS}(k)}{ V_{WS}(k)} p(k), \\ 
p(k)&=&\lim_{N_k,V\to\infty}\frac{N_k V_{WS}(k)}{V}, \label{dense1}  
\eea
where $V$ is the total volume and $p(k)$ is the number of realizations of the $k$-cell. 
The Single Nucleus Approximation (SNA) \cite{LS1991} consists in considering 
$N_{WS}(\rho_B,y_p)=1, p(1)=1$. This approximation is exact at zero temperature in the absence of 
phase transitions, and in principle should fail at finite temperature, 
even in the absence of phase transitions. In the following we explore if phase transitions 
are there or not, and the degree of violation of SNA at finite temperature.

\subsection{Zero temperature solution in the SNA}\label{section:t0equations}

The variational formalism to obtain the composition of stellar matter at zero temperature 
has been proposed long ago~\cite{BPS,BBP} and regularly employed since then, using 
more sophisticated models for the nuclear energetics
\cite{croute_haensel,croute_haensel2,croute_hempel,RocaMaza}.
We will follow the very same strategy, but at variance with the seminal papers~\cite{BPS,BBP} , 
we will determine the optimal configuration for each given $(\rho_B,y_p)$ point without 
implementing $\beta$-equilibrium in the variational constraints.
This choice will allow us keeping the same formalism for neutron star crust and 
the finite temperature supernova matter, which is not in $\beta$-equilibrium. 
Concerning the specific application to the NS crust, we will determine in a second step 
the $y_p(\rho_B)$ relation imposed by $\beta$-equilibrium.

The variables to be variationally found are $(V_{WS},A,\delta,\rho_g,y_g)$. 
The two constraints, which will lead to the introduction of two chemical potentials,  
can be written as:
\begin{eqnarray}
\rho_g&=&\frac{1}{V_{WS}}\left ( A_{WS}-A_e \right ), \label{consA} \\
y_g&=&\frac{1}{V_{WS}}\left ( I_{WS}-I_e \right ). 
\label{consy}
\end{eqnarray}  

Using the relations (\ref{are}),(\ref{zre}) between r-clusters and e-clusters
we can write the auxiliary function to be minimized:
\begin{eqnarray}
{\cal{D}}(A,\delta,\rho_g,y_g,V_{WS})&=& 
\epsilon_{HM}+E^e/ V_{WS}   \nonumber \\
&-&\alpha \rho_g \left ( \rho_0 -A/ V_{WS}    \right ) + \alpha \rho_0 \left ( \rho_{B}-A/ V_{WS}
\right ) \nonumber \\
&-&\beta y_g \left ( \rho_0 -A/ V_{WS}    \right ) + \beta \rho_0 
\rho_{B}(1-2y_p) \nonumber \\
&-& \beta \rho_0\frac{A\delta} {V_{WS}} ,
\label{lagrange-D}
\end{eqnarray}
where $\alpha$ and $\beta$ are Lagrange multipliers.
An additional complication comes from the fact that at zero temperature the gas 
can only be a pure gas. 

Indeed within the neutron and proton drip-lines a pure nucleus solution is by definition 
more bound than a solution where one particle would be in the gas. 
The drip-lines are defined by the lowest $N$ ($Z$) solution of the equations:
\bea
E^{nuc}(N+1,Z,\rho_{el})&-&E^{nuc}(N,Z,\rho_{el}) \geq 0; \nonumber \\
E^{nuc}(N,Z+1,\rho_{el})&-&E^{nuc}(N,Z,\rho_{el}) \geq 0 .
\label{drip}
\eea
Notice that because of the electron screening the drip-lines in the neutron star crust 
are displaced with respect to nuclei in the vacuum, and in particular the fission 
instability line does not exist. However  eqs.(\ref{drip}) admit a solution for any 
$N$, $Z$, meaning that when the equilibrium solution is below that line the nucleus 
will be in equilibrium with the vacuum. Above the neutron (proton) drip-line, 
we will have an equilibrium with a neutron (proton) vacuum gas.
This $T=0$ anomaly is very well known in nuclear matter. An equilibrium with the vacuum 
does not impose an equality between two chemical potentials, because the vacuum has $\mu=0$. 
If a system with $A$ particles and energy $E(A)$ is in equilibrium with the vacuum, 
its chemical potential is defined by a one-sided Maxwell construction between 
$E(A=0)=0$ and $E(A)$ with slope $E(A)/A$. 
The chemical potential of this particular equilibrium is given by:
\begin{equation}
\mu\equiv\frac{d E}{dA}=\frac{E}{A},
 \end{equation}
which implies $d(E/A)/dA=0$, that is the equilibrium solution minimizes the energy per particle 
(and not the total energy, as it is the case in a finite well defined volume $V$).

Coming back to the minimization of the auxiliary function eq.(\ref{lagrange-D}),  
the minimization with respect to the gas densities gives the definition 
of the neutron (proton) chemical potential $\mu_n$ ($\mu_p$) as:
\bea
\alpha+\beta&=&\frac{\mu_n}{\rho_0}=\frac{\mu_{ng}}{\rho_0}\;\;\; if\;\;\; \rho_{ng}>0 ; \\
\alpha-\beta&=&\frac{\mu_p}{\rho_0}=\frac{\mu_{pg}}{\rho_0}\;\;\; if\;\;\; \rho_{pg}>0 ,
\eea
with $\mu_{n(p)g}=\partial\epsilon_{HM}/\partial\rho_{n(p)g}$.
If one of the two densities is zero, that is below the corresponding drip-line, 
we lose one equation but also one unknown variable, and one can use one of the 
conservation equations to determine the missing variables.
The result is a system of four coupled equations:
\begin{eqnarray}
\rho_{Bp(n)}&=&\frac{A (1\mp \delta )}{2V_{WS}}, \label{eq1}\\
\rho_{Bn(p)}&=&\rho_g   \left (1-\frac{A}{\rho_0 V_{WS}} \right )+   \frac{A (1\pm \delta )}{2V_{WS}} ,
\label{eq2}\\
\frac{\partial (E^{nuc}/A)}{\partial A}|_{\delta}&=&0, \label{eq3}\\
\frac{1}{A}\frac{\partial E^e}{\partial \delta}|_{A} &\pm& \frac{1}{1\mp\delta}
\frac{\partial E^e}{\partial A}|_{\delta}=
\pm\mu_g \frac{1}{1\mp\delta} \left ( 1-\frac{\rho_g}{\rho_0}\right )
+ \nonumber \\
&+&\mu_g\frac{\rho_g}{\rho_0^2}\frac{d\rho_0}{d\delta}.
\label{eq4}
\end{eqnarray}
The upper (lower) sign refers to a neutron (proton) gas,
$\rho_{Bp(n)}$  indicates the proton (neutron) baryon density for a neutron (proton) gas, $\rho_g=\rho_{n(p)g}$,  $\mu_g=\mu_{n(p)g}$.
The two last equations suppose that $E^{nuc}$ is a differentiable function of $A$
and $\delta$, which is obviously not the case if we take experimental masses. 
In this case the derivatives have to be interpreted as finite differences.
We can see that the Coulomb screening effect of the electrons enters in the 
equilibrium equations, while the kinetic energy of the electrons does not play any 
role in the equilibrium sharing. This is the reason why this term is usually 
disregarded out of $\beta$ equilibrium. 
However we will see that it does play a role, determining 
the possible existence of phase transitions.

Different observations are in order. 

First, from eq.(\ref{eq3}) we can see that, both below and above the  drip, 
the minimization conditions correspond to the minimization of the energy per nucleon 
with respect to the nucleus size, at the isospin value imposed by the constraint and the 
chemical equilibrium with the gas eq.(\ref{eq4}).
Concerning eq.(\ref{eq4}), the coupling of the isoscalar to the isovector sector is trivially 
due to the fact that we are using $(A,\delta=I/A)$ as isoscalar and isovector variables 
instead of $(A,I)$ which would be the more natural choice if we did not have 
$\rho_0=\rho_0(\delta)$. With this choice of variables, if we consider the 
textbook example of two ideal gases composed of two different species of molecules 1, 2, 
$E=E_1(A_1,I_1)+E_2(A_2,I_2)$ fulfilling the conservation equations
\begin{equation}
A=A_1+A_2 \; ; \; I=I_1+I_2,
\end{equation}
using the same Lagrange multiplier method as before and defining $\mu,\mu_3$ 
as the conjugated chemical potentials of component 2,
we find the classical equality of chemical potentials if we work with the variables $(A, I)$:
\begin{equation}
\frac{\partial E_1}{\partial A_1}|_{I_1}=\frac{\partial E_2}{\partial A_2}|_{I_2}=\mu  \; ; \;
\frac{\partial E_1}{\partial I_1}|_{A_1}=\frac{\partial E_2}{\partial I_2}|_{A_2} = {\mu_3},
\end{equation}
while we have a coupling to the isovector sector for the mass sharing  equation if we work 
with $(A,\delta)$:
\begin{equation}
\frac{\partial E_1}{\partial A_1}|_{\delta_1}=\mu +  \delta_1  {\mu_3} \; ; \;
\frac{\partial E_1}{\partial \delta_1}|_{A_1}=A_1   {\mu_3}.
\end{equation}
This is a very natural result, because with this choice of variables the two constraints 
are not independent any more, that is the constraint associated to the $\beta$ multiplier 
contains the variable $A_1$.

Second, the factor $(1-\rho_g/\rho_0)$  introduces a coupling between the two subsystems cluster 
and gas, that is an interaction. This comes from that fact that our two systems are 
in fact coupled: the energy of the cluster depends on the composition of the gas as it can be 
seen from eq.(\ref{e-energy}). 
This coupling is due to the fact that a part of the high density part of the Wigner-Seitz 
cell is constituted by the gas. 
From eq.(\ref{energy-r}) ,  we can see that this is an effect of the excluded volume 
which enters the mass conservation constraint. 
In the e-cluster language (see eq.(\ref{energy})), we can equivalently say that it is 
an effect of the self-energy shift of the e-cluster inside the gas. 
This shows that the excluded volume indeed acts as an interaction. 
This effect goes in the direction of reducing the effective chemical potential with respect 
to the non-interaction case, that is reducing the cluster size. 
If we account for the cluster compressibility, that is the $\delta$ dependence of $\rho_0$, 
an extra effective coupling emerges (last term in the r.h.s. of eq.(\ref{eq4})).

In the case of moderate asymmetries below the neutron drip, the set of coupled equations 
eqs.(\ref{eq1}),(\ref{eq2}),(\ref{eq3}),(\ref{eq4}) reduces to the single 
equation eq.(\ref{eq3}) giving the most stable isotope for a given asymmetry. 
If  we assume a functional form as eq.(\ref{dan}) for the cluster energy functional, 
this equation admits an analytical solution:
\begin{equation}
a_s-\frac{a_a^2(A)}{a_s^a}\frac{I^2}{A^2}=2 a_c (1-f_{WS}) 
\frac{Z^2}{A}.
\end{equation}
The solution is particularly simple in the case of symmetric nuclei $I=0$:
\begin{equation}
A^{eq}(I=0)=\frac{2a_s}{a_c (1-f_{WS})}.
\end{equation}
In the vacuum $\rho_{el}=0, f_{WS}=0$ and we get a nucleus around $A\approx 55$, $A^{eq}=2a_s/a_c$,
while at saturation density $\rho_{el}=\rho_{0p}=\rho_0/2$ $A^{eq}\to\infty$, showing that we do 
obtain the homogeneous matter limit at saturation.

\subsection{The structure of the neutron star crust} \label{section:crust}
The different solutions of eqs.(\ref{eq1})-(\ref{eq4}) lead to a unique composition 
$(A,\delta,\rho_g, V_{WS})$
for each couple of external constraints $(\rho_B,y_p)$. 
Let us consider the energy density, $\epsilon_{WS}^{tot}(\rho_B,y_p)=E_{WS}^{tot}/V_{WS}$. 
It contains a baryonic $\epsilon_B$ and a leptonic $\epsilon_{el}^{tot}$ part, 
$\epsilon_{WS}^{tot}(\rho_B,y_p)=(\rho_p m_p+\rho_n m_n)c^2+\epsilon_B  +
\epsilon_{el}^{tot}$, with $\epsilon_B(\rho_B,y_p)=E^e/V_{WS}+\epsilon_{HM}$
(see eq.(\ref{energy})).  
\begin{figure}
\begin{center}
\includegraphics[angle=0, width=0.7\columnwidth]{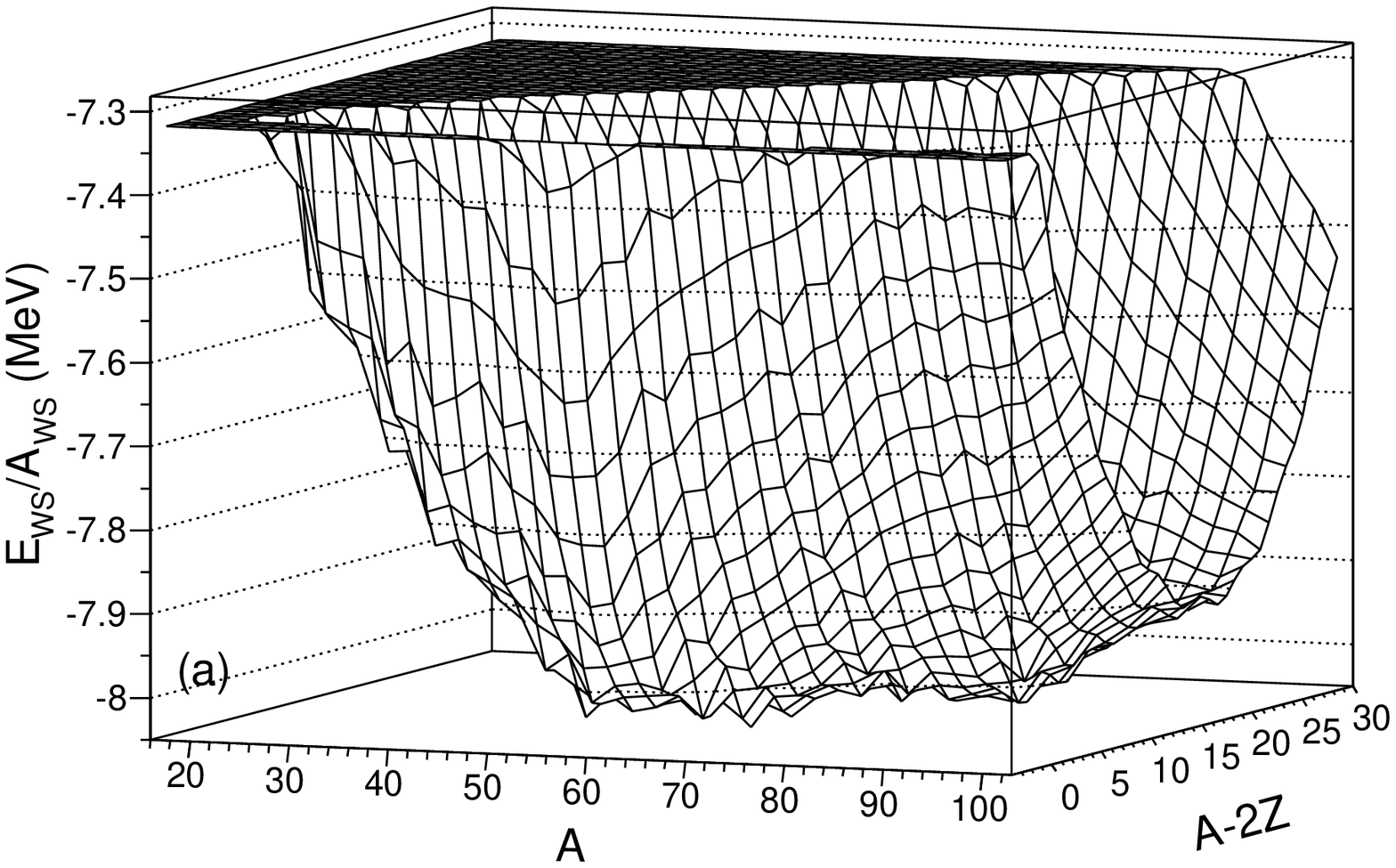}
\includegraphics[angle=0, width=0.7\columnwidth]{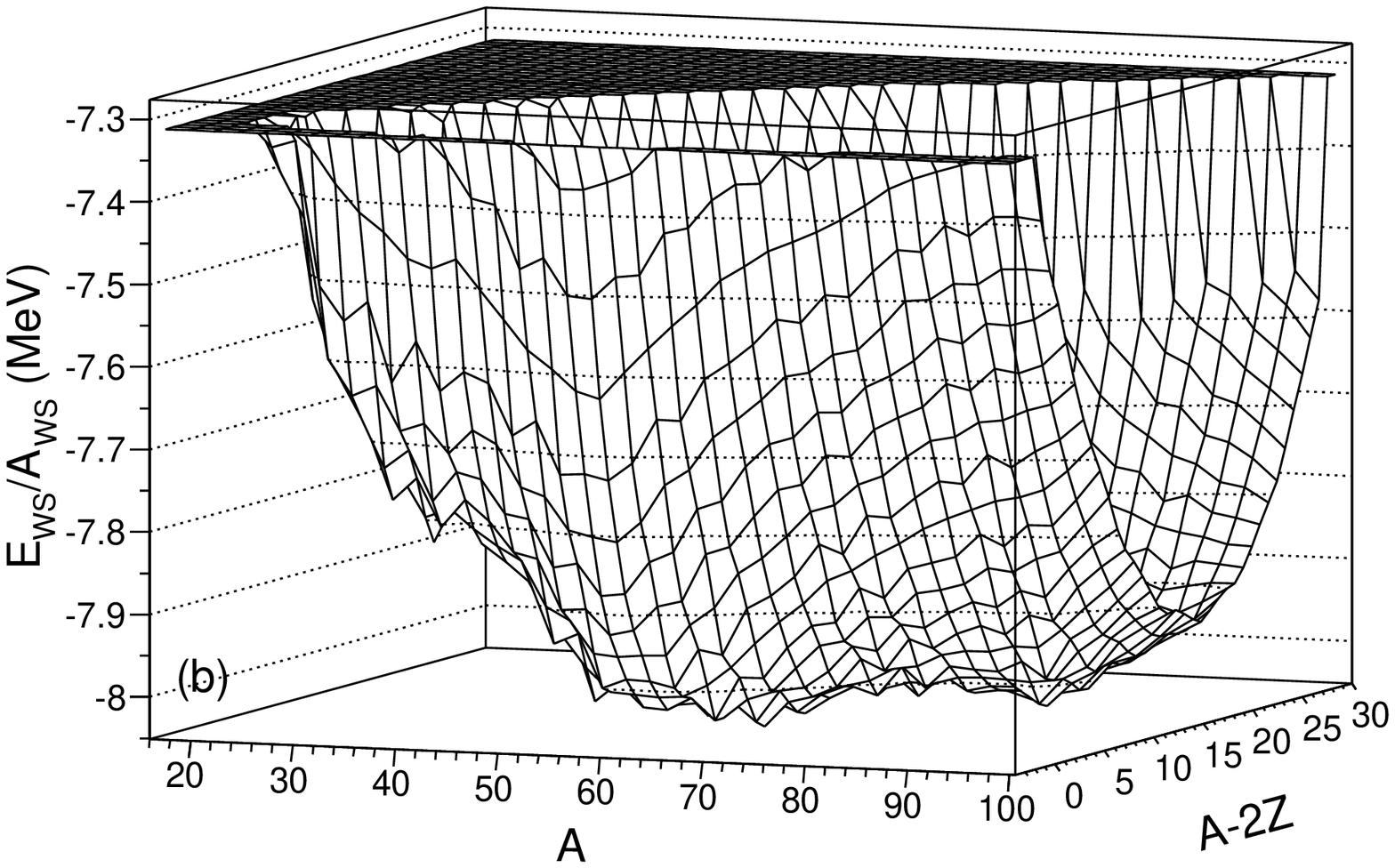}
\end{center}
\caption{Surface of the energy per baryon $E_{WS}/A_{WS}$ at 
$\rho_B = 10^{{-}7}$  fm$^{{−}3}$ and different values of proton fractions. 
The cluster energy is calculated according to FRDM \cite{FRDM} (a) and 
using the experimental database \cite{nudat} (b).}
\label{fig0}
\end{figure}  
In the long-lived neutron star, the proton and neutron densities do not correspond to 
separate conservation laws because weak processes transforming protons into neutrons 
are in complete equilibrium.
The structure of the neutron star crust is then obtained by choosing, among all the different 
$Z_{WS}$ values corresponding to  different values of the proton fraction $y_p$, 
the one leading to an absolute minimum of the energy density.
This minimization condition reads:
\begin{equation}
\epsilon_{WS}^{\beta eq}(\rho_B)=\min_{y_p} \left ( \epsilon_{WS}^{tot}(\rho_B,y_p) \right ). \label{min_tot}
\end{equation}
In the inner crust above the neutron drip the densities are continuous variables and 
the energy density is a differentiable function. 
The minimization then trivially gives the usual chemical $\beta$-equilibrium condition
\begin{eqnarray}
\frac{\partial \epsilon_{WS}^{tot}}{\partial \rho_n} - 
\frac{\partial \epsilon_{WS}^{tot}}{\partial \rho_p}
&=& m_n c^2+ \frac{\partial \epsilon_{B}}{\partial \rho_n} - m_p c^2 - 
\frac{\partial \epsilon_{B}}{\partial \rho_p}
 - \frac{\partial \epsilon_{el}^{tot}}{\partial \rho_p} \nonumber \\
&=&\mu_n^{tot}-\mu_p^{tot}-\mu_{el}^{tot}=0,
\end{eqnarray}
where $\mu^{tot}_i=\partial \epsilon_{B}/\partial \rho_i+m_i c^2$
with $i=n,p$ 
is the chemical potential including the rest mass contribution.
 
Below the drip-line (outer crust), the baryonic energy density 
is simply given by $\epsilon_B=E^{nuc}/V_{WS}=E^e(\rho_g=0)/V_{WS}$, and 
the minimization condition (\ref{min_tot}) reduces to:
\begin{equation}
\epsilon_{WS}^{\beta eq}(\rho_B)=\min_{Z}
\left (\frac{E^{nuc}}{V_{WS}}+\rho_n m_n c^2+
\rho_p m_p c^2+\epsilon_{el}^{tot} \right ).
\end{equation}
 The value of $Z$ leading to the minimal energy $Z=Z_{\beta eq}(A)$ corresponds
to the equilibrium nucleus.
This is still a $\beta$-equilibrium condition, but it has to be
interpreted as the ensemble of two inequalities: 
\begin{eqnarray}
\mu_n^{tot}(N-1,Z+1)-\mu_p^{tot}(N,Z)-\mu_{el}^{tot}(Z)&<&0, \\
\mu_n^{tot}(N,Z)-\mu_p^{tot}(N+1,Z-1)-\mu_{el}^{tot}(Z-1)&>&0.
\end{eqnarray}
In this set of inequalities,  
$\mu_n^{tot}(N,Z)d\rho_n={\cal E}_B(N+1,Z)-{\cal E}_B(N,Z)$
with   ${\cal E}_B(N,Z)=(E^{nuc}(N,Z)+N m_nc^2+Z m_pc^2)/V_{WS}(N,Z)$.
An equivalent relation holds for $\mu_p^{tot}$.

Most of the published studies on the composition of the neutron star crust employ empirical 
mass formulas~\cite{BPS,BBP} or microscopic functionals from self-consistent 
HFB calculations~\cite{croute_haensel,croute_haensel2,croute_hempel,RocaMaza,pearson}  
to describe the cluster energy functional $E^{vac}$. 
A functional approach is unavoidable in the inner crust, because no experimental measurement 
exists above the drip-line. Conversely, in the outer crust the predictive power of the 
approach entirely depends on the quality of the mass formula to describe experimental data. 
Now, it comes out that the energy surface in the presence of the electron 
gas has a huge number of quasi-degenerate minima.

This is shown for an arbitrary chosen density within the outer crust 
$\rho_B=10^{-7}$ fm$^{-3}$ in Figure \ref{fig0}. This figure shows the energy surface of the 
equilibrium Wigner-Seitz cells corresponding to different $y_p$ values
obtained using the FRDM parameterization  by Moller and Nix from ref.~\cite{FRDM} as well as the measured 
nuclear masses from ref. \cite{nudat}. 
We can see that, though the FRDM predictions are very close to the measured mass,
the obvious tiny differences can affect the determination of the absolute minimum.
This means that even modern highly predictive mass formulas describing nuclear masses 
within 0.5 MeV or even less can lead to inexact results when applied to the outer crust.

\begin{figure}
\begin{center}
\includegraphics[angle=0, width=0.9\columnwidth]{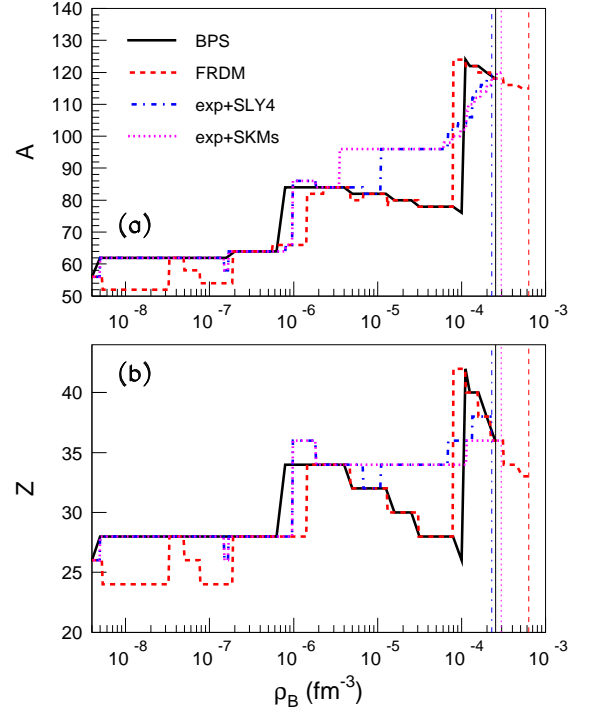}
\end{center}
\caption{(Color online). Outer crust composition at T = 0: Baryonic (top panel) and atomic (lower panel) 
numbers of the $\beta$-equilibrium nucleus as a function of the baryonic density. 
BPS corresponds to predictions by BPS\cite{BPS}; FRDM and exp+SLY4(SKMs) stand
for present model predictions when nuclear masses are calculated according to 
Finite-Range Droplet Model of Ref.\cite{FRDM} and, respectively, 
atomic mass data of Ref.\cite{nudat} + LDM-SLY4 (SKMs) model of 
Ref.\cite{Danielewicz2009}. The vertical lines mark the drip-line in the stellar medium.
} 
\label{fig1}
\end{figure}

The importance of this model dependence  is shown in Figure~\ref{fig1} and Table I.

\begin{table}\label{table:crust}
\caption{Composition of the outer layer of the outer crust of a cold neutron star 
as a function of baryonic density. FRDM and exp+SLY4  stand
for  model predictions when nuclear masses are calculated according to 
Finite-Range Droplet Model of Ref.\cite{FRDM} and, respectively, 
atomic mass data of Ref.\cite{nudat} + SLY4  model of 
Ref.\cite{Danielewicz2009}. The atomic and mass numbers in italics for the set exp+SLY4 
correspond to nuclides for which experimental mass
evaluations (or extrapolations) do not exist.
}
\begin{tabular*}{\linewidth}{ @{\extracolsep{\fill}} ll *{9}c @{}}
\hline
 \multicolumn{3}{c}{FRDM} & \multicolumn{3}{c}{exp+SLY4}  \\
\hline
$\rho_B$ (fm$^{-3}$)& $A$ & $Z$ &$\rho_B$ (fm$^{-3}$)& $A$ & $Z$  \\
\hline
 $1.000 \cdot 10^{-10}$ & 56        & 26       & $1.000 \cdot 10^{-10}$  &  56  &  26 \\
 $4.467 \cdot 10^{-9}$  & 52        & 24       & $5.012 \cdot 10^{-9}$   &  62  &  28 \\
 $3.388 \cdot 10^{-8}$  & 62        & 28       & $1.513 \cdot 10^{-7}$    &  58  &  26 \\
 $4.786 \cdot 10^{-8}$  & 58        & 26       & $1.698 \cdot 10^{-7}$   &  64  &  28 \\
 $7.585 \cdot 10^{-8}$  & 54        & 24       & $8.128 \cdot 10^{-7}$   &  66  &  28 \\
 $1.950 \cdot 10^{-7}$  & 64        & 28       & $9.772 \cdot 10^{-7}$   &  86  &  36 \\
 $5.623 \cdot 10^{-7}$  & 66        & 28       & $1.862 \cdot 10^{-6}$   &  84  &  34 \\
 $1.479 \cdot 10^{-6}$  & 82        & 34       & $6.761 \cdot 10^{-6}$   &  82  &  32 \\ 
 $2.291 \cdot 10^{-6}$  & 84        & 34       & $1.096 \cdot 10^{-5}$   & {\em 96}  & {\em 34} \\  
 $4.786 \cdot 10^{-6}$  & 80        & 32       & $6.918 \cdot 10^{-5}$   & {\em 102} & {\em 36} \\ 
 $6.761 \cdot 10^{-6}$  & 82        & 32       & $9.120 \cdot 10^{-5}$   & {\em 104} & {\em 36} \\
 $1.349 \cdot 10^{-5}$  & 80        & 30       & $1.148 \cdot 10^{-4}$   & {\em 106} & {\em 36} \\
 $3.090 \cdot 10^{-5}$  & 78        & 28       &\\
 $7.943 \cdot 10^{-5}$  & {124} & {42} & \\
 $ 1.097 \cdot 10^{-4}$ & {122} & {40} & \\ 
 $ 1.585 \cdot 10^{-4}$ & {120} & {38} & \\ 
\hline
\end{tabular*}
\end{table}

Let us first discuss the outer crust, on the left of the vertical lines in Fig.\ref{fig1}.
We can see that the use of an experimental mass table leads to sizable differences even with 
sophisticated mass formulas like the FRDM model~\cite{FRDM}. 
The solution of the variational equations for densities up to about  $\rho_B=10^{-5}$ fm$^{-3}$, 
leads to an equilibrium nucleus whose mass has been experimentally measured. 
This means that, up to that density, fully model independent results can be obtained 
using the experimental mass table, as it is done for the values noted as "exp+Sly4" in the table. 
If the FRDM model is used instead (results labelled "FRDM" in the table), differences appear 
even at very low density. Not only the density at which the transition from a nuclear species 
to another is observed is not correctly reproduced (column 1 and 4, respectively),  
but the isotope (column 2 and 5) and element (column 3 and 6) sequence is not correct either. 
These differences are due to the imperfect reproduction of nuclear mass measurements by the model, 
and stress the importance of using experimental values for the nuclear mass when studying 
the crust composition.

In the inner depths of the outer crust, the equilibrium nucleus is so neutron rich that 
no mass measurement exists.  Then
the crust composition depends on the theoretical cluster functional employed, and more 
specifically on its properties in the isovector channel, which are still largely unknown. 
As it is well known, this induces a strong model dependence on the composition.
As shown in Table I, the lowest density at which this model 
dependence appears is of the order of $\rho_B=10^{-5}$ fm$^{-3}$.
At that density the solution of the variational equations solved using the SLY4 functional when experimental masses are not available, 
produces as preferred isotope  $^{96}Se$ $(A=96, Z=34)$. Now, the smallest Z for which an experimental mass exists for A=96 is 
Z=35 showing that this solution is due to a mismatch between the prediction of the SLY4 functional and the experimental data. 
The results in italic in Table I are therefore not reliable.
The observed deviation in Fig.\ref{fig1} between exp+SLY4 and exp+SKMs 
is similarly due to the fact that the mismatch is bigger 
with the less performing SKMs functional. 
The full model independence of the outer crust composition is confirmed by the fact that 
our results for the outer crust are in agreement with refs.\cite{croute_haensel2,croute_hempel}. 
This essentially shows that our variational equations are correctly solved. In ref.\cite{croute_haensel2} the model-independent region is slightly
larger than in our work, because they have used the FRDM model to complement the experimental information when unavailable, and this latter, as we have already stressed, has a smaller mismatch with experimental data. 

\begin{figure}
\begin{center}
\includegraphics[angle=0, width=0.9\columnwidth]{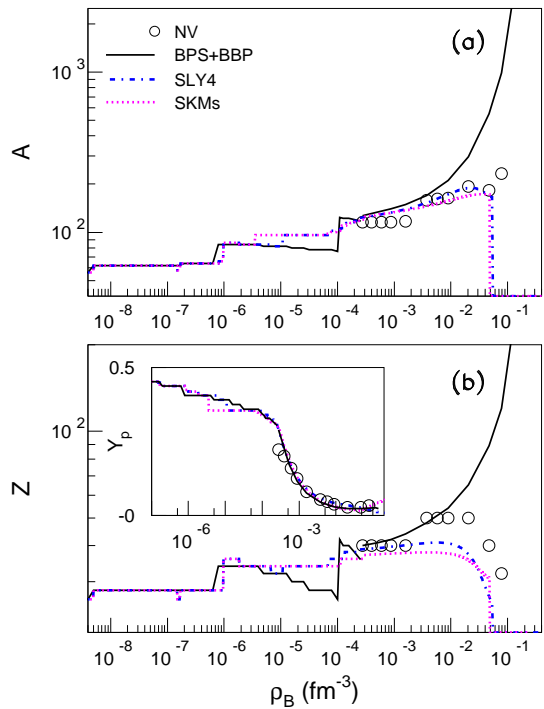}
\end{center}
\caption{  (Color online).
Crust composition at T = 0: Baryonic (top panel) and atomic (lower panel) numbers 
of the equilibrium nucleus as a function of the baryonic density. 
NV stands for predictions by Negele and Vautherin \cite{NV1973}; 
BPS+BBP corresponds to predictions by BPS\cite{BPS} and BBP\cite{BBP} ; 
exp+SLY4(SKMs) stand for present model predictions when nuclear masses are calculated 
according to atomic mass data of Ref.\cite{nudat} + LDM-SLY4 (SKMs) model of 
Ref.\cite{Danielewicz2009}.
The inset in the bottom panel depicts the evolution with baryonic density
of the total proton fraction.
}
\label{fig2}
\end{figure}  
Whatever the predictive power of the mass model, a model dependence is unavoidable
in the inner crust, where the equation of state 
of the pure neutron gas directly enters in the minimization equations \cite{pearson}.
To illustrate this point, we show in Fig.\ref{fig2} the total composition of the 
neutron star crust obtained with different models. 
Whatever the equation of state, the predictions of eqs.(\ref{eq1})-(\ref{eq4})
show that, in the inner crust, the mass and charge of the unique nucleus
of the WS cell continously increase with baryonic density and then suddenly fall to zero.
The abrupt cluster disappearence occurs because, depending of the employed interaction, 
at a density of the order of $\rho_0/5-\rho_0/3$
homogeneous matter becomes energetically more favorable than clusterized matter. 

The precise value of the density corresponding to cluster dissolution 
depends on the effective interaction mainly through the $L_{sym}$ parameter of the equation of state~\cite{constanca} but also on the exact prescription
for the cluster surface tension, particularly its isospin dependence which cannot be unambigously extracted from mean field calculations~\cite{dan03,reinhard,pei,douchin_npa}. A more sophisticated expression for the surface symmetry energy, different from the one of ref.\cite{Danielewicz2009} was variationally calculated in  ref.\cite{douchin} for some selected Skyrme models, and slightly higher transition densities are consequently reported. 

Because of the abrupt behavior shown by Fig.\ref{fig2}, the crust-core transition was typically considered as (weakly) first order
in the literature~\cite{BBP}. For this reason the density of cluster melting is still known in the literature as the "transition density".
It is however nowadays well established that at the density of nuclei dissolution non-spherical pasta can be energetically favored, making the transition continuous from the thermodynamic point of view. We will come back to this
point in section \ref{section:phase_transition}.

The effect of the nuclear matter equation of state in the prediction
of the composition of the inner crust has been studied in detail in the recent years~\cite{pearson,croute_hempel}. 
It leads to the difference in Fig.\ref{fig2} between the dotted and dash-dotted line, 
which represent two characteristic equations of state. 
A more extensive study of the different Skyrme interactions is beyond the scope of this paper, 
however some extra results on this subject can be found in ref.~\cite{esym_paper}.
It is interesting to notice that, at variance with $A$ and $Z$, $y_p(\rho_B)$ plotted in the insert of 
Fig. \ref{fig2}(b) shows no sensitivity to the equation of state. This means that the energetics
of electrons dominates over the details of the nucleon-nucleon interaction.
The significant difference in atomic number between our model (irrespective of the effective interaction) 
and the original Negele-Vautherin Hartree-Fock calculation~\cite{NV1973} is due to the fact that 
our cluster model with the energy density functional~\cite{Danielewicz2009} eq.(\ref{dan}) 
contains only the smooth part of the cluster energy. The neglected shell effects are responsible of the 
emergence of the magic number $Z=40$ in the Hartree-Fock calculation. 
We can see that the knowledge of shell closures for extremely neutron rich nuclei 
is much more important for the description of the inner crust than the isovector equation of state, and 
it is clear that to be predictive, the model at zero temperature should be augmented of 
realistic proton shell effects, as it is done in the Strutinsky approximation by S.Goriely and 
collaborators~\cite{pearson}, at the obvious price of a greatly increased numerical effort. 
This  limitation of eq.(\ref{dan}) will however not be a serious problem  for the finite temperature 
applications for which the model has been conceived, and which will be studied in the second part of this paper.

\begin{figure}
\begin{center}
\includegraphics[angle=0, width=0.9\columnwidth]{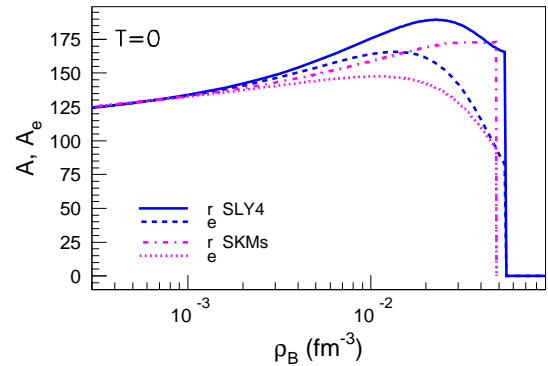}
\end{center}
\caption{(Color online).
Behavior of the cluster mass number as a function of the baryonic density in the inner crust 
using two different Skyrme functionals (SKMs \cite{bartel82} and SLY4 \cite{chabanat98})
for both the free neutron energy density and the nuclear masses  
according to the LDM-Skyrme model of  Ref.~\cite{Danielewicz2009}. 
The total mass of the cluster is compared to the bound part of the cluster, 
obtained by simply subtracting the number of free neutrons according to eq.(\ref{are}). 
} 
\label{fig2_bis}
\end{figure}

The most striking feature of Fig.\ref{fig2} is the huge qualitative discrepancy at high density with 
the original inner crust BBP model~\cite{BBP}.
To understand the origin of this difference Fig.\ref{fig2_bis} displays the behavior as a function 
of the baryonic density of the mass of the energy-cluster
\cite{Panagiota2013} from eq.(\ref{are}). 
SKMs \cite{bartel82} and SLY4 \cite{chabanat98} effective interactions have been considered.
We can see that the difference between BBP and our 
approach starts when the e-cluster size starts to depart from the r-cluster size, that is when the 
contribution of the neutron gas becomes important. In this situation one can expect a modification of 
the surface energy of the cluster according to eq.(\ref{e-energy}). In BBP, the in-medium 
modified surface energy is assumed to be a monotonically decreasing function of the gas density, 
independent of the isospin, exactly vanishing when the density of the gas reaches the density of the 
cluster~\cite{BBP}. 
In the language of the present paper, this happens when $A_e=0$ (see eq.(\ref{are})). 
In such a condition BBP clusters are liquid drops with bulk only, and their size naturally diverges. 
However, this approach  neglects the energy cost of the isospin jump at the cluster-gas interface.
It is shown in refs.~\cite{esym_paper,Aymard2014}, in the framework of the extended Thomas-Fermi theory, 
that the in-medium correction to the surface energy shows a complex dependence on the isospin, and 
specifically behave very differently in symmetric nuclear matter and in the equilibrium with a 
pure neutron gas. Only in the case of symmetric nuclei immersed in a symmetric gas
the transition to the homogeneous core can be seen as the simple vanishing of surface energy 
with diverging size of the nuclei; conversely, in the case of $\beta$-equilibrium matter, 
the inclusion of in-medium surface effects leads to a weak decrease of the average cluster size 
and a slightly advanced dissolution of clusters in the dense matter.

Finally, it is important to stress that results at densities higher than about one fifth of normal 
nuclear matter density are not reliable in any of the presented models because of the lack of 
deformation degrees of freedom which could allow the appearance of pasta phases~\cite{ravenhall,hashimoto}.

\subsection{Equation of state} \label{section:EOS}

A quantity of primary importance when discussing the sensitivity of stellar matter
energetics to the details of the nucleon-nucleon interaction or linking
nuclear parameters with astronomical observables is the equation of state and,
in particular, the total energy density - total pressure dependence. 

\begin{figure}
\begin{center}
\includegraphics[angle=0, width=0.8\columnwidth]{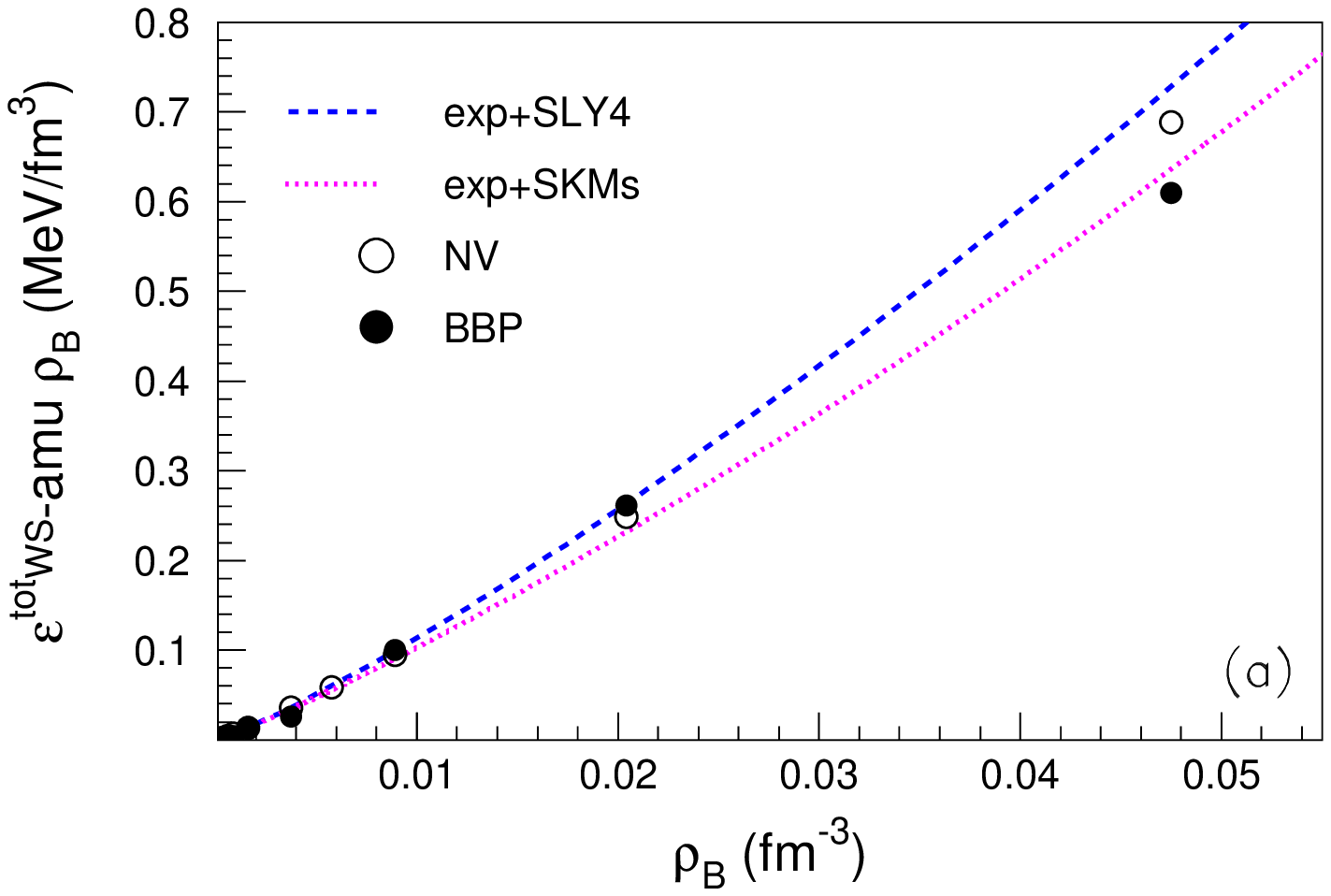}
\includegraphics[angle=0, width=0.8\columnwidth]{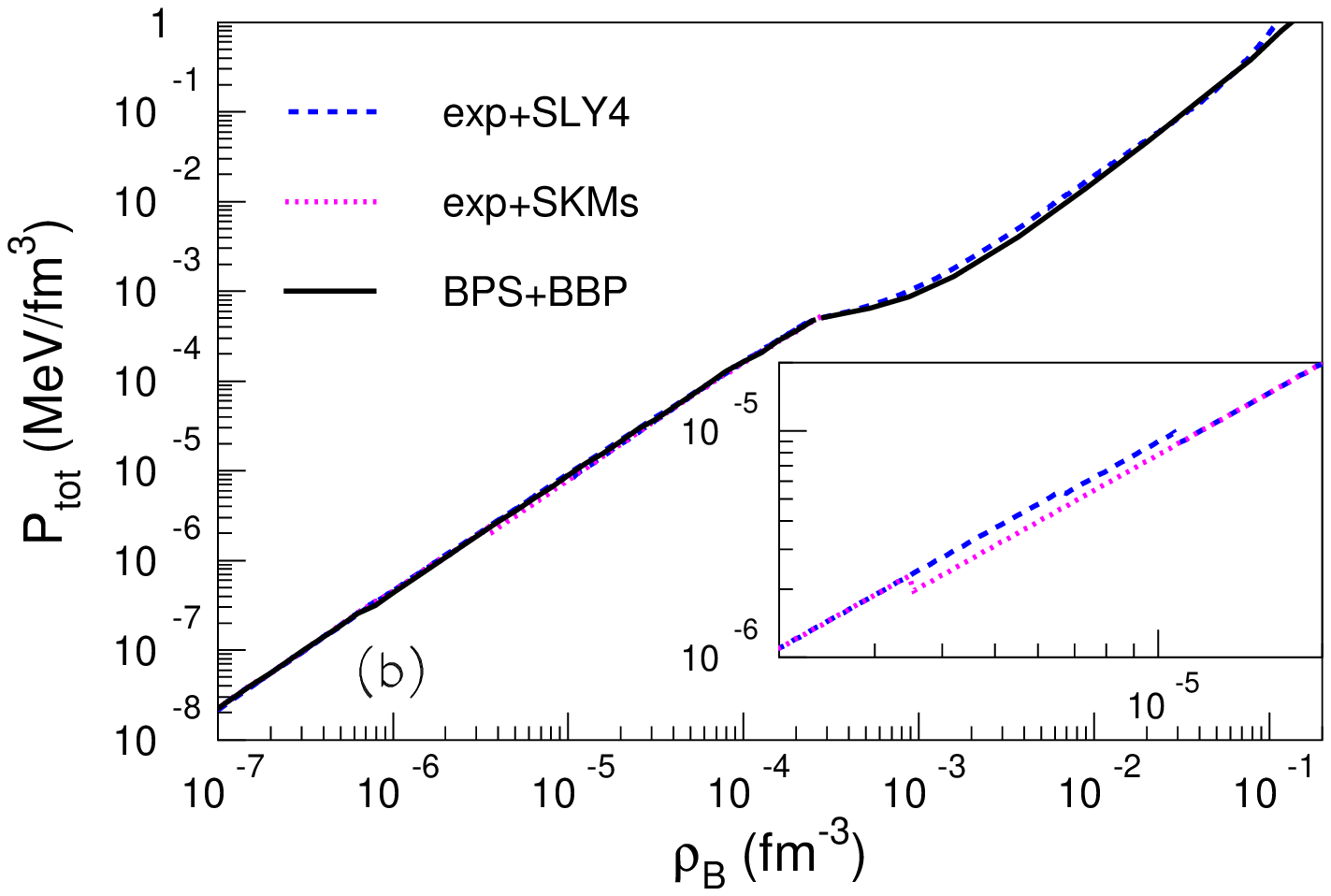}
\end{center}
\caption{ (Color online).
Total (baryonic+leptonic) energy density and pressure as a function of baryonic density
corresponding to the neutron star crust (T=0, $\beta$-equilibrium).
Experimental \cite{nudat} and LDM binding energies \cite{Danielewicz2009}
have been used in the outer and, respectively, inner crust. 
The employed nuclear effective interaction are SLY4 \cite{chabanat98} and SKMs \cite{bartel82}.
Present results are confronted with those of NV \cite{NV1973} and BBP \cite{BBP}. 
}
\label{fig:etot_crust}
\end{figure}

The total energy density and pressure of the WS cell are plotted in 
Fig. \ref{fig:etot_crust} as a function of baryonic density, in comparison with the result 
from the macroscopic BBP~\cite{BBP} and the microscopic Negele-Vautherin~\cite{NV1973} model. 
We can see that the quantitative value of the energy density obviously depends on the model,
and more specifically on the effective interaction, but in all cases over the 
considered density range the energy density surface is convex.
This means that there is no way to minimize the system energy by state mixing,
such that the system is thermodynamically stable. 
The discontinuous change of the crust composition due to the shell effects only leads to very 
tiny backbendings in the baryonic pressure as shown in the insert of Fig.\ref{fig:etot_crust},
and already observed by different authors~\cite{croute_haensel,croute_haensel2,croute_hempel,RocaMaza,pearson}.
These structures can be formally interpreted as phase transitions, but are so small that 
are not expected to have any thermodynamic consequence and can simply be understood 
as an interface effect.

The absence of a phase coexistence region covering a broad density domain, 
well known in the astrophysical context, is surprising from the nuclear physics viewpoint 
because it is  in clear contrast with the phenomenology of pure baryonic matter, which is 
dominated at sub-saturation densities by the nuclear liquid-gas phase transition\cite{camille1}.
One may wonder if this difference is due to the fact that we are limiting our analysis to a limited part of the 
two-dimensional baryon density space that is explored in $\beta$-equilibrium. Indeed the $\beta$-equilibrium 
trajectory corresponds to very neutron rich matter, and it is well known that the coexistence zone in the 
nuclear matter phase diagram shrinks with increasing asymmetry. We therefore turn to demonstrate that the 
difference between stellar matter and nuclear matter thermodynamics
is not restrained to $\beta$-equilibrium. 

\subsection{Phase transitions in the inner crust?}\label{section:phase_transition}

In the previous section we have assumed that a one-to-one correspondence exists between
baryonic density and chemical potential, that is a unique Wigner-Seitz configuration can be 
systematically associated to each pressure and chemical potential field inside the star. 

This is only correct in the absence of phase transitions, and it is in principle possible 
that a mixture of different Wigner-Seitz configurations might lead to a lower energy density 
than a periodic repetition of the same cell. The highly degenerate energy minima showed by the experimental 
energy surface even without the (more model dependent) inclusion of unbound neutrons beyond the drip-lines
(see figure~\ref{fig0}) evoke the possibility  that first order phase transitions could even appear 
at finite temperature in the outer crust.

More generally, it is well known that such a first-order phase transition covers almost the whole phase diagram
of sub-saturation neutral nuclear matter \cite{camille1} and has baryonic density as 
order parameter. It is therefore natural to ask whether such a phase transition persists in the stellar context.
As a matter of fact, the existence of such first order phase transition is systematically assumed
in most seminal papers on the stellar matter equation of state\cite{BBP,Pethick}, 
and in particular it is implemented in the publicly available and popularly used LS tables \cite{LS1991}. 
Even more modern equations of state of supernova matter\cite{nse_hempel,stone_2014} 
invoke the persistence of the nuclear liquid-gas  phase transition in the stellar context, 
based on the fact that the baryonic energy density of star matter is unstable with respect both to 
thermodynamic\cite{camille1,providencia_inst} 
and to finite size density fluctuations\cite{pethick_finitek,douchin,camille3}.

On the other side, it was shown in different works that the liquid gas phase transition in stellar
matter is quenched by the very strong incompressibility of the electron background
\cite{providencia_inst,camille3,ising_star,inequivalence,maruyama}, and microscopic 
modelling of the Wigner-Seitz cell has confirmed a continuous transition from the solid crust to the liquid 
core through a sequence of inhomogeneous pasta phases
\cite{douchin2,maruyama,constanca,stone,pasta_RMF,pasta_ETF,pasta_tdhf1,pasta_tdhf2,pasta_tdhf3,pasta_md1,pasta_md2}.

It is therefore important to examine this question in further detail.

We have already seen in section \ref{section:crust} that the solution of eqs.(\ref{eq1})-(\ref{eq4})
is always unique, even if many different solutions can be very close in energy per nucleon 
(see Figure \ref{fig0}).

This means that at zero temperature a unique cluster-gas configuration can be associated to a given value of 
$\rho_B^{WS}=A_{WS}/V_{WS}$,  $y_p^{WS}=Z_{WS}/A_{WS}$. 
This statement can of course be model dependent, as we have seen that very small variations of the mass functional
can lead to very different results.  However, even if multiple solutions of the cluster configuration would occur
(which will indeed be the case at finite temperature), this would not lead to a first order phase transition.
For a first-order phase transition to occur, solutions corresponding to different densities should be degenerate
in (constrained) energy. Then, the absolute energy density minimum  would be 
obtained by mixing these degenerate configurations with different $\rho_B^{WS}, y_p^{WS}$. 
If this was the physical result, the single-nucleus approximation would fail, and
even at zero temperature one should account for a distribution of different Wigner-Seitz cells.

Thermodynamic instabilities and eventual phase transitions in systems with 
more than one component have in principle to be studied in the full N-dimensional density space
\cite{glendenning}. In our case this means that the energy density has to be studied
in the full two dimensional $(\rho_n,\rho_p)$ plane, and the $\beta$-equilibrium condition has to be 
applied only after the Gibbs construction is performed (indeed $\beta$-equilibrium has to be 
imposed only at the macroscopic level, and can very well be violated at the microscopic level of a single cell). 
However the problem simplifies if the order parameter is known. In that case 
it is useful to introduce a  Legendre transformation of the thermodynamic potential with
respect to the chemical potentials of all the densities except the order parameter \cite{camille1}.
Then the multi-dimensional Gibbs construction exactly 
reduces to a one-dimensional Maxwell construction on the residual density.

In the case of stellar matter the neutrality condition $\rho_p=\rho_{el}$ allows a variable 
change $(\rho_n,\rho_p) \to (\rho_B=\rho_n+\rho_p,\rho_L=\rho_{el})$.
Due to the very huge electron incompressibility it is reasonable to expect that the
two coexisting phases, if any, would not present any jump in electron density \cite{camille3}.
Microscopic calculations \cite{maruyama} have convincingly shown that the electron polarization
by the proton distribution is negligible, as long as the clusters have linear dimensions of the order
of the femtometer. Then, we can safely perform a Legendre transform with respect to $\rho_L$
and introduce the constrained energy density
\begin{equation}
\bar \epsilon_{WS}(\rho_B, \mu_L)=\epsilon_{WS} -\mu_L \rho_L,
\end{equation}
where $\epsilon_{WS}(\rho_B, \rho_L)=E_{WS}/V_{WS}$, $\mu_L$ stands for lepton chemical potential  and $\rho_L$ is the 
value taken by the lepton density at chemical potential $\mu_L$, $\rho_L=\rho_L(\mu_L)$.
Note that $\mu_L=0$ corresponds to $\beta$-equilibrium in lack of neutrinos, 
$\mu_n^{tot}=\mu_p^{tot}+\mu_{el}^{tot}$. We do not therefore need to examine the whole $\mu_L$
plane, but can limit ourselves to the single point $\mu_L=0$.

We can then conclude that we can identify the possible presence of phase transitions in
the neutron star crust by simply considering the $\rho_B$ density behavior of the energy 
density in $\beta$-equilibrium, $\epsilon_{WS}(\rho_B, \rho_L(\rho_B,\mu_L=0))$.
As in section \ref{section:t0equations}, $\epsilon_B=\epsilon_{WS}-\epsilon_{el}^{tot}$ 
is obtained  solving, for each condition $(\rho_B,y_p)$
the coupled equations (\ref{eq1})-(\ref{eq4}).

\begin{figure}
\begin{center}
\includegraphics[angle=0, width=0.8\columnwidth]{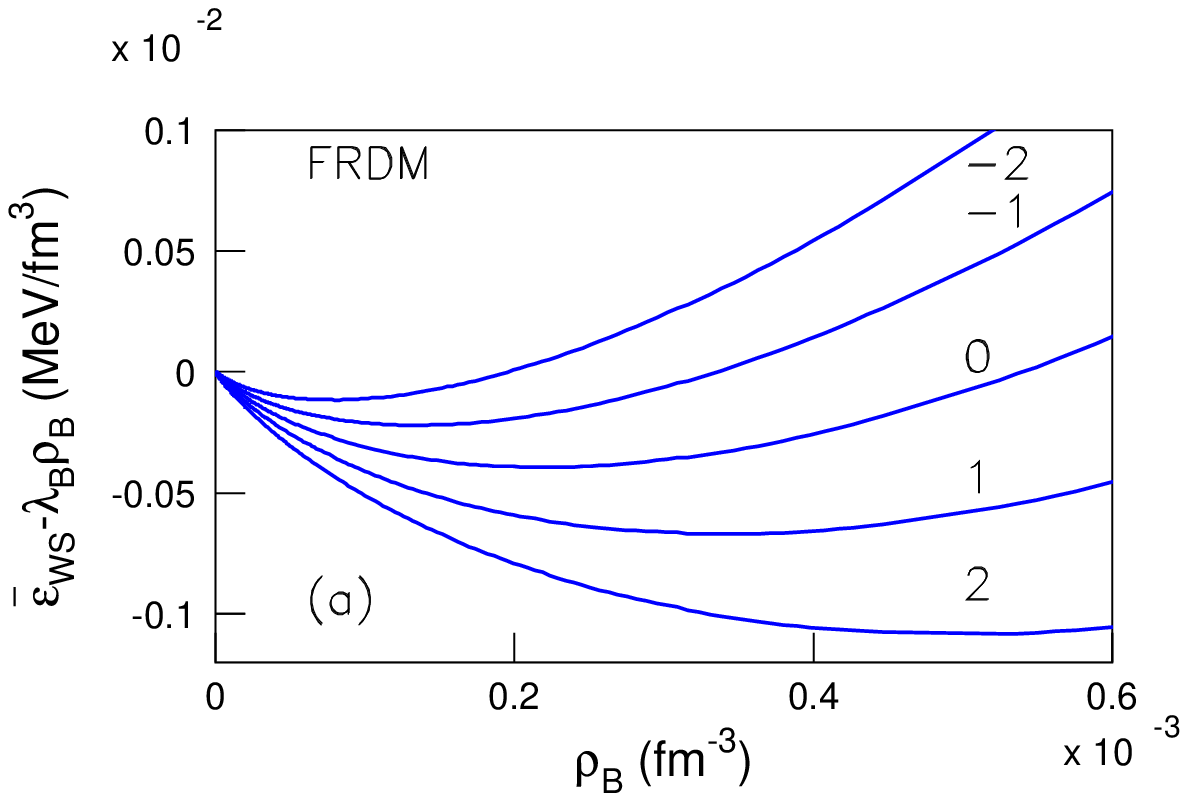}
\includegraphics[angle=0, width=0.8\columnwidth]{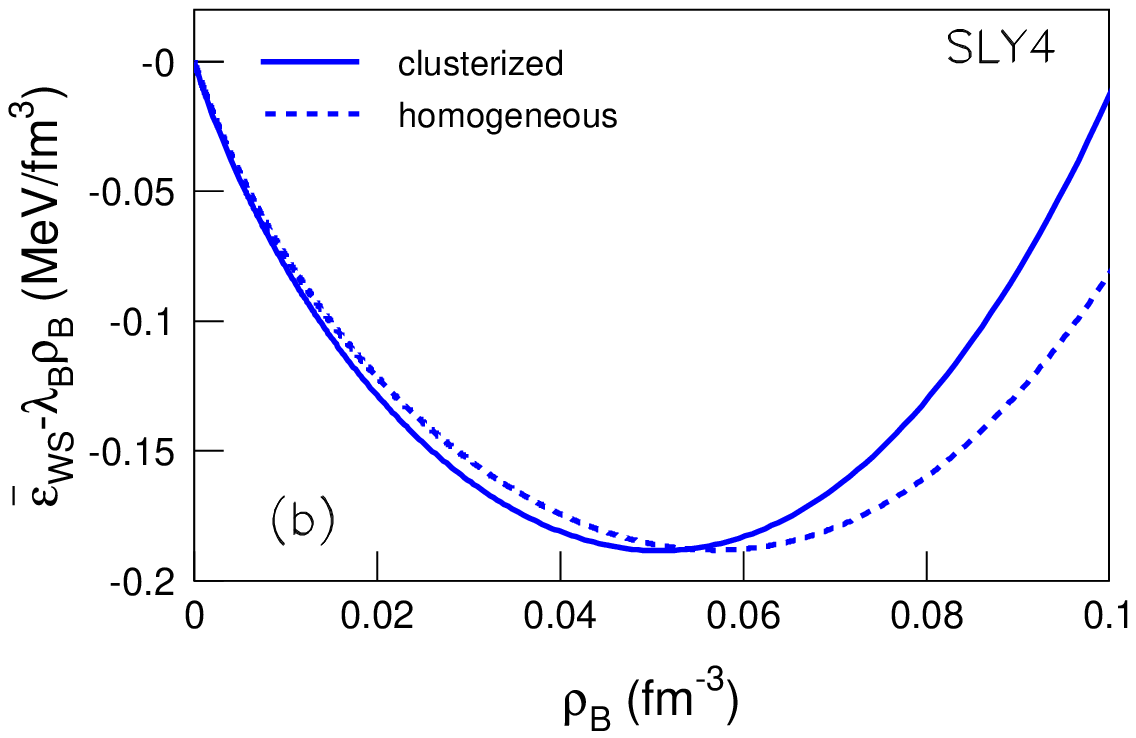}
\end{center}
\caption{ (Color online).
Constrained energy densities as a function of density.
(a) Outer crust as obtained using FRDM. The numbers accompanying the curves are in 
MeV and stand for $\lambda_B$.
(b) Crust (thick line) and  $\beta$-equilibrated homogeneous matter
(thin line) at T=0 corresponding to SLY4 and $\lambda_B=11.25$ MeV.
} 
\label{fig2_ter}
\end{figure}  

To better evidence possible convexities, it is useful to introduce a linear bias
with slope $\lambda_B$:
\begin{equation}
\bar \epsilon_{WS,\lambda_B}(\rho_B,\mu_L=0)=\bar \epsilon_{WS}(\rho_B, \mu_L=0)
-\lambda_B\rho_B .\label{bias}
\end{equation}
For each $\lambda_B$ value, which plays the role of an external chemical potential field, 
the equilibrium density of star matter is given by the minimum of this function.
If the function $\bar \epsilon_{WS}$ is convex, it will be characterized by a single mimimum value giving 
the usual relation between intensive and extensive variables
\begin{equation}
\lambda_B=\frac{\partial \bar \epsilon_{WS}}{\partial\rho_B} =\mu'_B . 
\label{chemical}
\end{equation}
In this equation, we have introduced a prime symbol on the chemical potential to indicate that 
the electron contribution is included in $\epsilon_{WS}$.
However, if $\bar \epsilon_{WS}$ has concave region(s) on  the baryonic density axis $\rho_B$,
it will be possible to find one or more values  of $\lambda_B$ such that 
two (or more) different configurations correspond to the same value of the constrained energy density. 
This will indicate a first order phase transition, and the associated $\lambda_B$
value will correspond to the transition chemical potential.
The constrained energy density eq.(\ref{bias}) for clusterized matter in the crust 
is displayed in Fig. \ref{fig2_ter} for some chosen values 
of $\lambda_B$ corresponding to minima in the outer (a) and inner (b), respectively.
We can see that both in the outer and inner crust the constrained energy surface is smooth
and that the
equilibrium configuration is given by a single Wigner-Seitz cell, 
thus justifying our variational procedure.
The FRDM mass model is limited to nuclei below the dripline and cannot be used for calculations 
in the inner crust. For this reason we have switched to the Skyrme-LDM mass model~\cite{Danielewicz2009} 
to produce the right panel. 
Again, a unique clusterized solution characterizes the equilibrium up to 
a chemical potential of the order of 10 MeV ($\lambda_B=11.25$ MeV for Sly4). 
At that point, the corresponding equilibrium density is of the order of $\rho_0/3$.
As we have already discussed commenting Fig.\ref{fig2}, the precise value depends on the EoS 
and on the surface tension.  
We can see from Fig. \ref{fig2_ter} that at this transition value of the chemical potential 
the constrained energy density of the clusterized system is equal to the one of homogeneous matter, meaning that  it is 
possible to put in equilibrium the two phases. 

This defines a tiny region of first-order phase transition, much less extended than  
the liquid-gas phase transition of normal nuclear matter. Indeed this latter covers the whole 
sub-saturation density region. Moreover, we believe that this residual phase transition might 
be an artifact of the present model which does not account for deformation degrees of freedom. 
It is well known that in this density domain deformed pasta structures have to be accounted 
for~\cite{watanabe_pasta}. For this reason, we do not perform any Gibbs construction and simply 
put to zero the cluster mass at the transition point, assuming that pasta would take over. 
The results of Fig.\ref{fig2_ter} show that the thermodynamics of $\beta-$ equilibrated matter 
is completely different from the one of nuclear matter.

As previously discussed in refs.\cite{camille3} within mean-field arguments, this difference
is due to the huge electron gas incompressibility which quenches the phase transition in stellar matter.
 
To demonstrate this point in the framework of the present model, we turn to consider the 
behavior of the baryonic part of the energy density $\epsilon_B=\epsilon_{WS}-\epsilon_{el}^{tot}$  in the full 
$(\rho_B,\rho_p)$ plane. 

To better spot convexities in the two-dimensional space,
we introduce again a constrained energy density  
\begin{equation}
 \bar \epsilon_{\lambda_B,\lambda_3}(\rho_B,\rho_p)=\epsilon_B(\rho_B,\rho_p)
-\lambda_B\rho_B-\lambda_3(\rho_B-2\rho_p), \label{e_constr}
\end{equation}
where  $\lambda_B$ and $\lambda_3$ represent an isoscalar and isovector external chemical potential field. 

Again, phase transitions will be signalled by the existence of
%
%
one or more values  $(\lambda_B,\lambda_3)$ such that 
two (or more) different configurations correspond to the same value of the constrained energy density. 

\begin{figure}
\begin{center}
\includegraphics[angle=0, width=0.9\columnwidth]{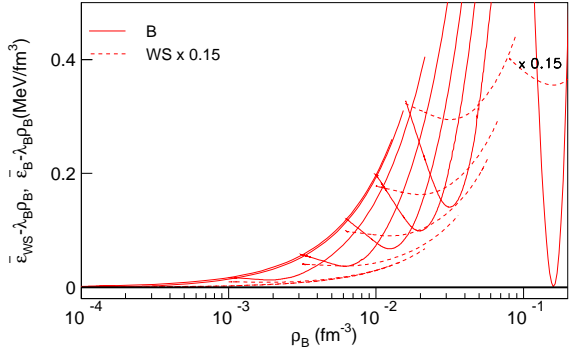}
\end{center}
\caption{(Color online). Constrained energy density (eq. (\ref{e_constr})) in the 
thermodynamic condition corresponding to the 
liquid-gas phase transition of symmetric matter given by $\lambda_B=-15.97$ MeV
and  $\lambda_3=0$. 
The different curves represent different values of the total proton density $\rho_p$.
Solid (dashed) curves correspond to constrained baryonic (baryonic+leptonic) 
energy density.
The considered effective interaction is SLY4.
}
\label{fig3}
\end{figure}

In the case of uncharged nuclear matter, we know that the dominant part of the 
$(\lambda_B,\lambda_3)$ plane is characterized by concavities. 
It is therefore not surprising to see that this is clearly the case for the $\epsilon_B$ 
function plotted in Fig. \ref{fig3} corresponding to $\lambda_B=-15.97$ MeV
and $\lambda_3=0$. 

This figure displays the energy density obtained solving the variational equations 
eqs.(\ref{eq1}-\ref{eq4}), biased by an external chemical potential field according to eq.(\ref{e_constr}).   
The value of the total proton density $\rho_p$ is the same in each point of the different curves plotted in the figure.
Because of the neutrality constraint $\rho_p=\rho_{el}$, each curve  represents a given screening factor 
to the cluster Coulomb energy according to eq.(\ref{eq:screening}). 
The minimum of each curve then gives the ensemble of optimal Wigner-Seitz cells corresponding 
to the chosen $\lambda_3$ value, and to different baryon density. The absolute minimum corresponds 
to  the equilibrium Wigner-Seitz cell  associated to the couple $(\lambda_B,\lambda_3)$. 
The $(N,Z)$ sequence of the corresponding cluster gives the composition, as a function of baryonic density, 
of matter at that $\lambda_3$ chemical potential.

The choice $\lambda_3=0$  selects the equilibrium solutions for symmetric matter.  
A unique point is the absolute minimum for all choices of $\lambda_B$ except $\lambda_B=-15.97$ MeV
which is shown in the figure. 
At that chemical potential, if the electron part of the energy density is not taken into account 
(solid curves), two different points correspond to the same constrained energy.
This corresponds to  the well-known nuclear matter phase transition which at zero temperature 
takes place at a chemical potential equal to the saturation energy,  $\mu_B=-15.97$ MeV 
for the SLY4 functional chosen in Fig. \ref{fig2}. 
We can notice that the low density phase, which is predicted to be the vacuum phase in mean-field
calculations, is obtained here at a low but finite density, corresponding to the most stable $N=Z$ 
isotope $^{56}Ni$. This is due to the limitation
of mean-field calculations that do not account for clusterization at low density.

In the stellar matter case, because of the Coulomb coupling between protons and electrons, 
the lepton part of the energy density is not independent of the baryon part. 
This means that the energy in the Wigner-Seitz cell has to include  the electron zero point 
energy as written in  eq.(\ref{energy}) .
The total energy densities are given by dashed curves in Fig.\ref{fig2}.
This contribution is a simple constant shift of each curve because of the condition 
$\rho_p=\rho_{el}$, and therefore does not change the sequence of optimal compositions as a function of the density.
From a thermodynamic point of view, we can say \cite{inequivalence} that the canonical solution is the same 
as without the electron contribution. 
However, the electron energy density is a monotonically increasing function of $\rho_{el}=\rho_p$,
and the optimal $\rho_p$ monotonically increases with $\rho_B$ in this symmetric matter situation we are considering.
As a consequence, no value of $\lambda_B$ can be found such that two different Wigner-Seitz cells 
can be put in equilibrium, and the phase transition disappears. 
This can be easily understood mathematically considering that the optimal energy density  gains an extra term as
\begin{equation}
\epsilon_B \to \epsilon_{WS}=\epsilon_B +\epsilon_{el}^{tot}(\rho_p) .  
\end{equation}
The relations (\ref{chemical}) between density and chemical potential are shifted because of the electron contribution
\begin{equation}
\lambda_B \to \mu'_B=\mu_B +\frac{1}{2} \mu_{el}^{tot} \;\; ;\;\; \lambda_3\to \mu'_3=\mu_3-\frac{1}{2}  \mu_{el}^{tot},
\end{equation}
and the curvature of the constrained energy density
 becomes:
\begin{equation}
\frac{\partial^2 \bar \epsilon_{WS} }{\partial \rho_B^2} = \frac{\partial \mu_{B}}{\partial \rho_B}+ \frac{1}{2}
\frac{\partial \mu_{el}^{tot}}{\partial \rho_{el}}.
\end{equation}
Because of the very high electron incompressibility, the convexity observed in the baryonic part of the 
energy density is not present any more in the total thermodynamic potential.
This is known in the literature as the quenching of the phase transition due to 
Coulomb frustration \cite{camille3,ising_star}, and shows \cite{inequivalence} that 
convexities in the (free) energy density do not necessarily correspond to instabilities
in the physical system.

This shows that if one wants to formulate the equilibrium problem in the grandcanonical ensemble, one has 
to account for the electron zero point motion. This is a triviality for the zero-temperature problem, 
since the Wigner Seitz cell is naturally defined in the canonical ensemble. 
However, in the finite temperature NSE problem, which is typically treated grandcanonically, 
this kinetic contribution is usually disregarded with the argument that, the electrons being an ideal gas, 
the corresponding partition sum is factorized \cite{nse_hempel,nse_us}. 
It is then important to stress that  a negative eigenvalue
in the baryonic energy curvature matrix should not be taken as a sign of a first order phase transition: 
stellar matter inside the convex region is clusterized but perfectly stable, and no Maxwell or 
Gibbs constructions should be performed to get the equation of state.

We will come back on this point in the second part of this paper. 
 
To summarize the results of this section, first order phase transitions could be possible in zero 
temperature stellar matter only if the function
$\epsilon_{WS}=E_{WS}/V_{WS}$, with $E_{WS}$ defined in eq.(\ref{energy-r}) and the values of $A,I,\rho_g, V_{WS}$ obtained from
eqs.(\ref{eq1})-(\ref{eq4}), presents a convexity anomaly as a function of $\rho_B$, for fixed values
of $\rho_p$. 
As it can be seen in Figs.\ref{fig2_ter}, \ref{fig3}, 
except the narrow density domain
where deformation degrees of freedom have to be accounted for,
this is not the case even using energy functionals for the 
nuclear masses which include shell effects.
In particular, the minimum of the constrained energy density for a given set of chemical potential is 
systematically associated to a single Wigner-Seitz cell, characterized by a unique composition 
in terms of cluster and gas mass and composition.
This means that the SNA is perfectly adequate to deal with the zero temperature problem.

\section{Finite temperature stellar matter}

In the second part of this paper we extend to the finite temperature regime the modelling of the 
Wigner-Seitz cell with cluster degrees of freedom, presented in the first part. 
We will start by deriving the classical equations corresponding to the single nucleus approximation. 
This approach is at the origin of most extensively used equations of state for 
supernova matter\cite{LS1991,shen,shenG}.
Then the main part of the paper is devoted to the derivation of an extended nuclear statistical 
equilibrium model, which by construction reproduces the results of SNA if only the most probable 
cluster is considered, and the same chemical potentials are considered. Since the SNA naturally 
converges at T=0 to the standard  modelling of the neutron star crust, the consistency between 
the theoretical treatment of neutron star crust and finite temperature supernova matter will thus be guaranteed.  

\subsection{Single nucleus approximation}\label{section:SNA}
 
The natural extension at finite temperature of the model presented in the first part of this paper 
consists in keeping the SNA approach, 
and replace the variational problem of the energy density minimization with the variational problem 
of the free energy density minimization. In the e-cluster representation eq.(\ref{energy})
the energy is additive and we can write for a given configuration 
$k=\{V_{WS}^{(k)},A^{(k)},\delta^{(k)},\rho_g^{(k)},y_g^{(k)}\}$
\bea
F_{WS}(A,Z,\rho_g,y_g,\rho_p)&=&F_\beta^{e} -T V_{WS} \ln z_\beta^{HM}    -T V_{WS} \ln z_\beta^{el}  \nonumber \\
&+&\delta F_{surf},
\label{fenergy}
\eea
where $\beta=T^{-1}$ is the inverse temperature,
$z_\beta^{HM}(\rho_g,y_g)$ is the mean-field partition sum for homogeneous matter and
$F_\beta^{e}$ represents the cluster free energy in the Wigner Seitz cell (defined below). 
 
The electron contribution  is independent of the different configurations and the associated 
partition sum $z_\beta^{el}(\rho_p)$ will be factorized out. Similar to the previous section, 
we will neglect the surface in-medium corrections to the free energy, though they might 
turn out to be important in the situations where the gas contribution is not negligible.
We note that in this section, at variance with the notations used in Section II, thermodynamical
quantities corresponding to homogeneous matter component and electron gas bare the $HM$ and $el$
labels as superscripts.
 
If we consider temperatures higher than the solid-gas phase transition temperature,
the free energy of a cluster defined by the variable couple $(A,Z)$ or equivalently $(A,\delta)$ is 
different from its ground state energy because at 
finite temperature the cluster can be found in different translational and internal states. 

To calculate this term, one has to consider that within the $A_{WS}$ total number of particles, 
a number $A_g=\rho_g V_{WS}$ belongs to the gas
part. The entropy associated to these particles is already contained in the term $\ln z_\beta^{HM}$. 
To avoid a double-counting of the number of states, 
the canonical partition sum of the cluster 
must thus be defined summing up the statistical weight of the different energy states associated 
to this reduced particle number (see eqs.(\ref{are}),(\ref{zre})): 
\bea 
Z^{cl}_{WS}(A,\delta,\rho_g,y_g)&\equiv& \exp \left ( -\beta F_\beta^e \right )
 \\ &=&
\sum_{\vec{p}}\sum_{E^*} \exp \left [ -\beta 
 \left ( \frac{p^2}{2mA_e} + E^e + E^* \right )\right ] . \nonumber
\eea
The cluster center-of-mass motion is a plane wave. The first sum is thus given by the plane 
wave density of states, with periodic boundary conditions at the cell borders. This is simply
\begin{equation}
\sum_{\vec{p}} =\frac{V_{WS}}{(2\pi\hbar)^3} \int d^3p.
\end{equation}
Notice that the available volume for the center of mass is the whole Wigner-Seitz volume, 
and there is no excluded volume effect.
The center-of-mass momentum integral is a Gaussian integral
\begin{equation}
 \int d^3p \exp \left(-\beta  \frac{p^2}{2mA_e}\right )=\left ( \frac{2\pi m A_e}{\beta}\right )^{3/2} .
\end{equation}
The sum over the cluster excited states has to be cut at the average particle separation energy, 
to avoid double counting with the gas states.
This leads to a temperature dependent degeneracy factor defined by:
\bea
 \sum_{E^*} \exp \left (-\beta  E^* \right )&=&\int _0^{<S>}dE^* \rho_{A,\delta}(E^*)
\exp \left ( -\beta  E^*\right ) \nonumber \\
&=&g_\beta(A,\delta,\rho_g,y_g),  \label{deg_bucurescu}
\eea
where $<S>=\min(<S_n>,<S_p>)$ is the average particle separation energy. 
For the numerical applications of this paper, we will use for simplicity a different 
higher energy cut-off for each cluster species 
$<S>\approx <S>(A,\delta,\rho_{el})=\min(<S_n>(A,Z,\rho_{el}),<S_p>(A,Z,\rho_{el}))$, 
with separation energies calculated from the smooth part of the cluster energy functional, 
given by $E_{LDM}^{vac}+\delta E_{Coulomb}$.
In the zero temperature limit $g_\beta \to g_{GS}=2J_{GS}+1$ gives the spin degeneracy of the 
cluster ground state. 

For the numerical applications of this section, in order to  be able to study the whole 
subsaturation density domain in $\beta$ equilibrium without any mismatch in the 
cluster energy functional, we have systematically used 
the Skyrme-LDM model for nuclear mass~\cite{Danielewicz2009} and considered $g_{GS}=1$. 
The level density $\rho_{A,\delta}(E)$ is here taken for simplicity with 
a simple Fermi gas  formula~\cite{nse_us}. A more realistic choice
will be presented in section \ref{section:NSE}.

The cluster free energy in the Wigner Seitz cell than reads, 
\begin{eqnarray}
F_\beta^{e}&=& F_\beta^{nuc}+T\ln z_\beta^{HM}(\rho_g,y_g) \frac{A}{\rho_0}\label{fenergy_cl_ws}  \\
&=&E^e(A,\delta,\rho_g,y_{g},\rho_p)-T \ln  V_{WS} 
-T \ln c_\beta - \frac{3}{2} T  \ln A_e, \nonumber
\end{eqnarray}
with $F_\beta^{nuc}=F_\beta^{vac}+\delta F_{coul}$ (the equivalent of eq. (\ref{eq:enuc})), 
$c_\beta=g_\beta(mT/(2\pi\hbar^2))^{3/2}$, and $m$ the nucleon mass. 

\begin{figure}
\begin{center}
\includegraphics[angle=0, width=0.8\columnwidth]{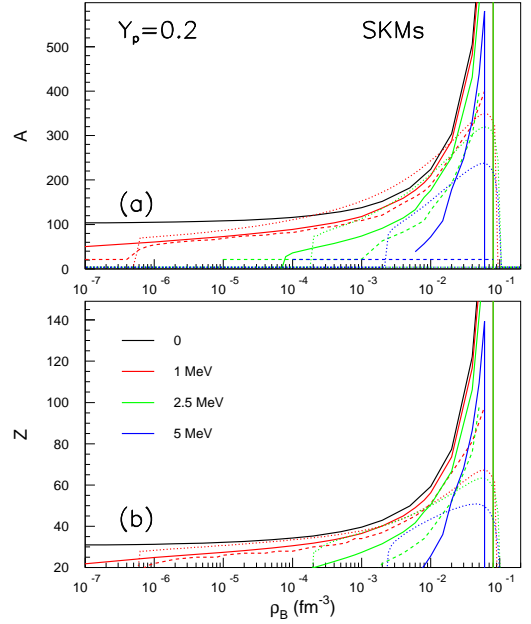}
\end{center}
\caption{(Color online).
Mass (top panel) and atomic (bottom panel) numbers of the unique nucleus of the WS cell 
as a function of baryonic density
for $Y_p$=0.2 and different temperatures $T=0, 1, 2.5, 5$ MeV.
Predictions of present SNA model (solid lines) are compared with the LS results 
\cite{LS1991} corresponding to LS220  
as calculated in Ref. \cite{Compose_LS220} (dotted lines) as well as 
with the NSE prediction for the most probable cluster (dashed lines).
The LDM-SKMs model of  ref.\cite{Danielewicz2009} is used for the  cluster energy functional. 
}
\label{fig:compo_Yp=0.2}
\end{figure}  

\begin{figure}
\begin{center}
\includegraphics[angle=0, width=0.8\columnwidth]{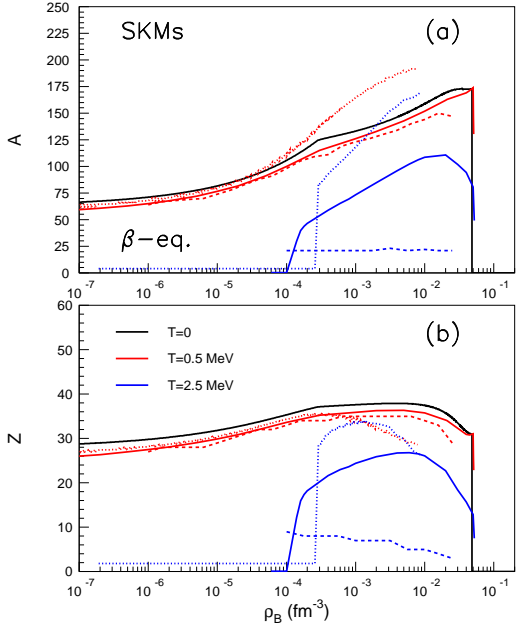}
\end{center}
\caption{(Color online).
The same as in Fig. \ref{fig:compo_Yp=0.2}
for $\beta$-equilibrium and different temperatures $T=0, 0.5, 2.5$ MeV.
Predictions of the present SNA model (solid lines) are compared with the LS
results \cite{LS1991} corresponding to LS220 as calculated in Ref. \cite{Compose_LS220} 
(dotted lines) as well as with the NSE predictions for the most probable cluster (dashed lines).
The LDM-SKMs model of  ref.\cite{Danielewicz2009} is used for the  cluster energy functional. 
}
\label{fig:compo_betaeq}
\end{figure}  
 The auxiliary function to be minimized is the extension of eq.(\ref{lagrange-D}) including the entropy terms:
\begin{eqnarray}
{\cal{D_\beta}}(A,\delta,\rho_g,y_g,V_{WS})&=& 
\frac{F_{WS}(A,\delta,\rho_g,y_g,\rho_p)} {V_{WS}}  \label{lagrange-T} \\
&-&\alpha \rho_g \left ( \rho_0 -A/ V_{WS}    \right ) + \alpha \rho_0 \left( \rho_{B}-A/ V_{WS} \right) \nonumber \\
&-&\beta y_g \left ( \rho_0 -A/ V_{WS}    \right ) + \beta \rho_0   \rho_{B}(1-2y_p) \nonumber \\
&-&\beta \rho_0 A\delta/ V_{WS}
.
\nonumber
\end{eqnarray}
The variational equations result:
\begin{eqnarray}
  &&\frac{\partial E^e}{\partial A}|_{\delta,\rho_g,y_g} = \mu_B \frac{\rho_0-\rho_g}{\rho_0} + 
{\mu_3} \frac{\rho_0\delta-y_g}{\rho_0} \label{teq1} \\
&&+\frac{3T}{2A} \frac{\rho_0 V_{WS}}{\rho_0 V_{WS}-A} +
T  \frac{\partial \ln c_\beta}{\partial A}|_{\delta,\rho_g,y_g}, \nonumber \\
&&\frac{\partial E^e}{\partial \delta}|_{A,\rho_g,y_g} =
 \mu_3 A+
\frac{d\rho_0}{d\delta}\frac{A}{\rho_0} (\mu_B \frac{\rho_g}{\rho_0}+ \mu_3 \frac{y_g}{\rho_0}) \label{teq2} \\
 &&+\frac{3}{2}T  \frac{d\rho_0}{d\delta}\frac{ \rho_g V_{WS}}{(\rho_0-\rho_g)(\rho_0 V_{WS}-A)}
+T  \frac{\partial \ln c_\beta}{\partial \delta}|_{A,\rho_g,y_g}, 
\nonumber \\
&&\frac{\partial (F_\beta^{0}/A)}{\partial A}|_{\delta, V_{WS}}= 0, \label{teq3}\\
&&\mu_B \equiv -T\frac{\partial \ln z_\beta^{HM}}{\partial \rho_g}\label{teq4}\\
&&{\mu_3} \equiv -T\frac{\partial \ln z_\beta^{HM}}{\partial y_g}. \label{teq5}
\end{eqnarray}

The finite temperature predictions of SNA are plotted in Figs. \ref{fig:compo_Yp=0.2} and
\ref{fig:compo_betaeq} along predictions of zero-temperature SNA of Section II,
results of Lattimer-Swesty model \cite{LS1991} as available in Ref. \cite{Compose_LS220}
and NSE results (see Section III.C).
Different density, temperature and proton fraction conditions are considered.
The considered effective interactions is SKMs \cite{bartel82}.

Fig.\ref{fig:compo_Yp=0.2} corresponds to the case where the proton fraction is kept constant and equal to 0.2.
It shows, as expected, a monotonic decrease of the cluster size as a function of temperature. 
More interesting, the results converge for $T\to 0$ to our zero temperature results in the 
Wigner-Seitz cell, which we know to be exact at the thermodynamic limit, 
and model independent below neutron drip.
Fig. \ref{fig:compo_betaeq} illustrates cluster mass and charge numbers as a function of baryonic density
for different temperatures at $\beta$-equilibrium.
The observed non-monotonic behavior is due to the strong decrease of proton fraction.
Indeed, with increasing density, the proton fraction becomes 
so low that loosely bound hydrogen and helium resonances dominate over heavy clusters which 
dissolve into homogeneous matter.
At constant proton fraction this effect is not apparent, meaning that the in-medium bulk energy shift is 
not enough to suppress the cluster binding. In that case, the preferential cluster size monotonically increases with 
density up to the point where homogeneous matter is energetically prefered. As in the previous section, 
we have indicated that point by putting to zero
the $A(\rho_B)$ and $Z(\rho_B)$ curves.
We cannot exclude that the inclusion of surface in-medium effects, neglected in this paper, could change this behavior. 
However preliminary results\cite{esym_paper} indicate that this effect is small.
 
The qualitative behavior of the cluster size and charge with density is similar to the one of the 
LS220 Lattimer-Swesty equation of state, plotted with dotted lines in Figs.\ref{fig:compo_Yp=0.2}
and \ref{fig:compo_betaeq}.
Quantitative differences exist nevertheless.
On one hand they could be due to the (slightly) different equation of state parameters and cluster surface tension model.
Effects of employed effective interaction on clusters have been already seen in Fig. \ref{fig2_bis}
where SKMs and SLY4 have been considered. 
Probably more important, the SNA model of Lattimer-Swesty additionally accounts for alpha-particles 
that can be present in the WS cell together  with heavier clusters. 
The presence of an isospin-symmetric bound component in the gas obviously modifies the cluster size and composition.  
Finally, to obtain the emergence (at low density) and dissolution (at high density) of clusters, 
first order phase transitions to an $\alpha$ particle gas and to homogeneous matter respectively,  
are implemented in the LS model\cite{LS1991}. 
We also note that at the highest temperatures, our SNA clusters tend to be smaller than in LS. 
This is probably due to the high energy cut in the density of state integral eq.(\ref{deg_bucurescu}) 
implemented in order to avoid double counting of the continuum states, which reduces the 
statistical weight of heavy clusters. 
We will discuss in section \ref{section:NSE} that the inclusion of the proper statistical weight of clusters 
of all sizes naturally leads to the emergence of an important fraction of light particles, 
and to the disappearance of heavy nuclei in the dense medium, without invoking any phase transition. 

\begin{figure}
\begin{center}
\includegraphics[angle=0, width=0.8\columnwidth]{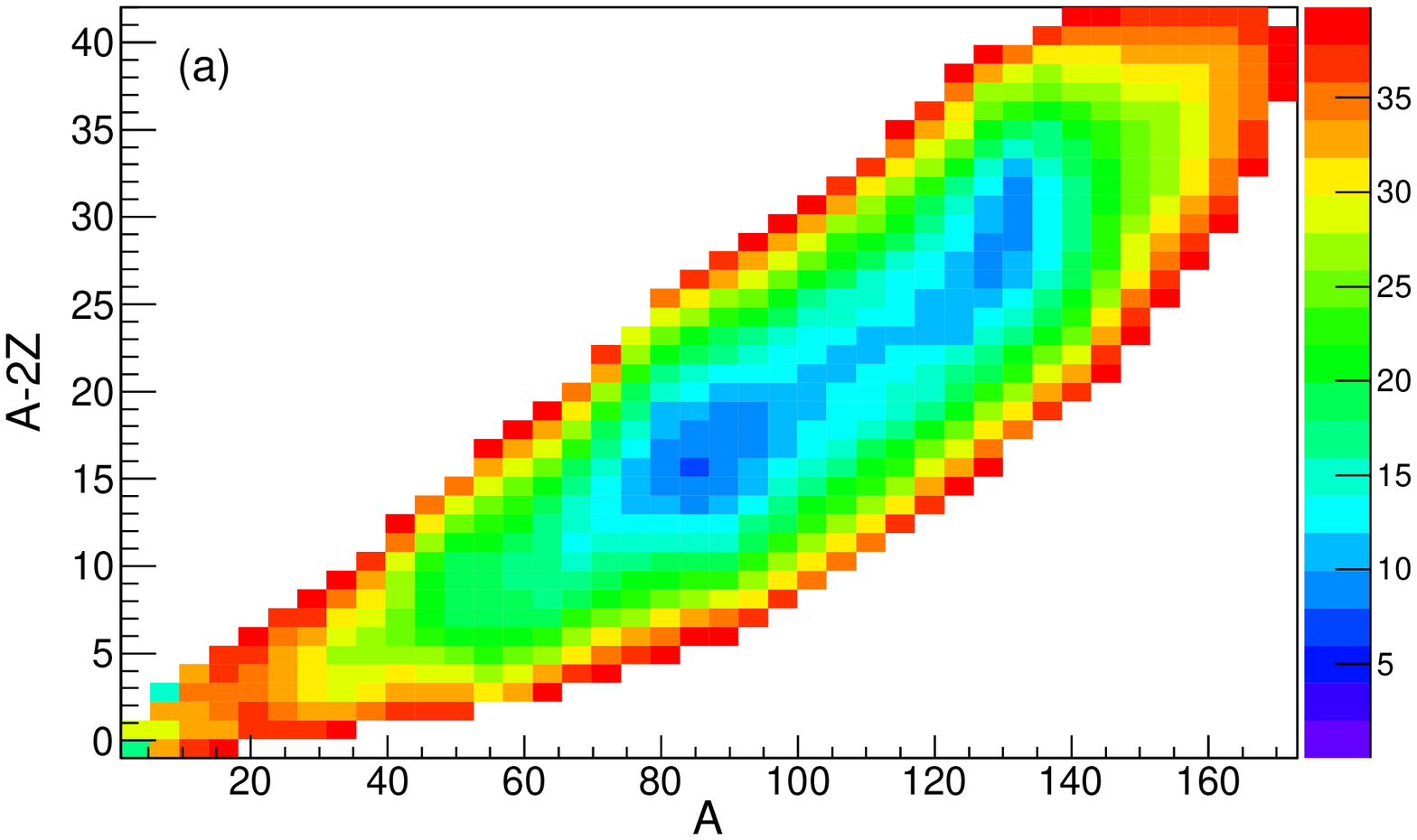}
\includegraphics[angle=0, width=0.8\columnwidth]{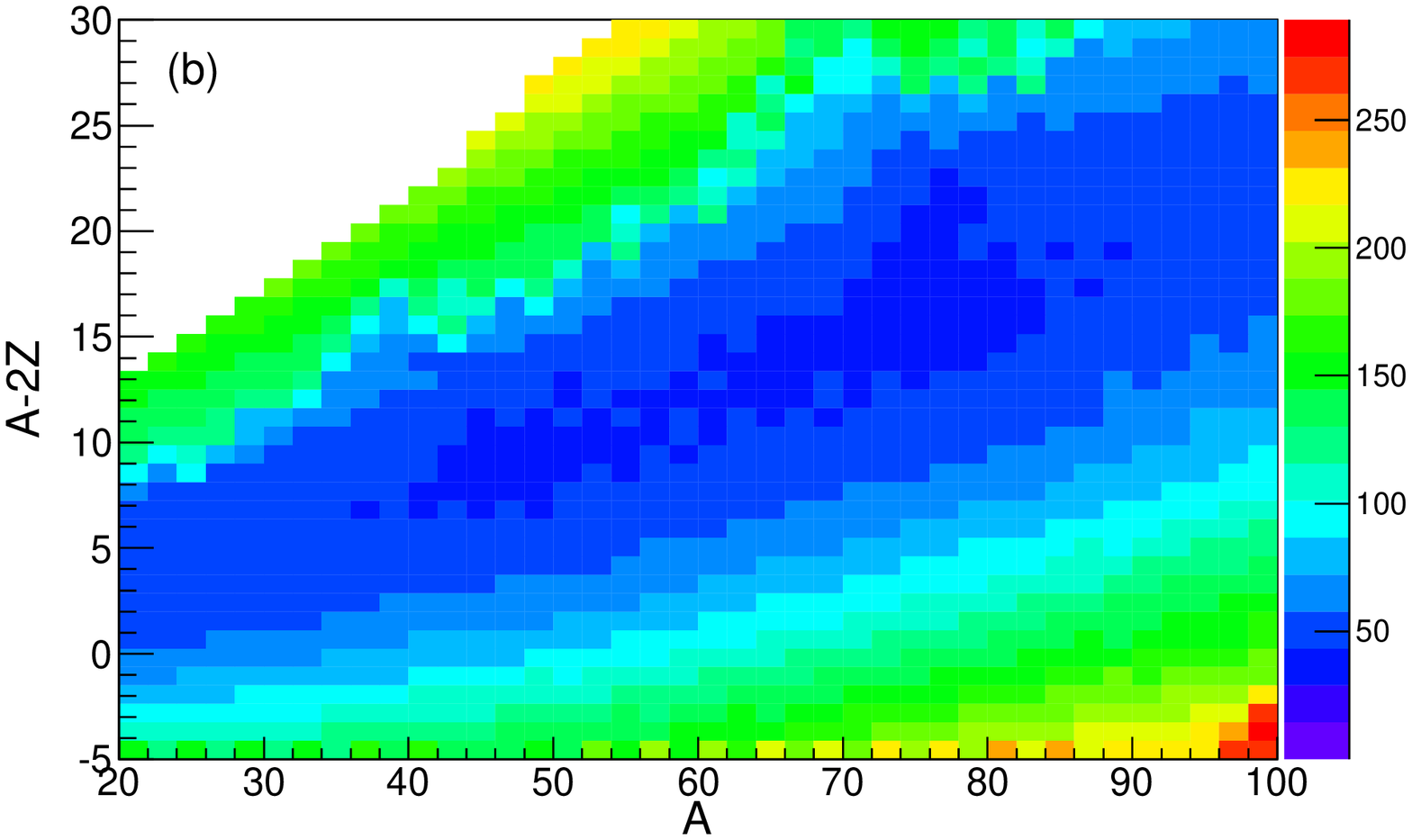}
\end{center}
\caption{(Color online). Contours of cluster constrained free energy
$F_{\beta}^e-\lambda_B A -\lambda_3 (A-2Z)$ (in MeV).
The considered thermodynamical conditions are:
$\rho_B=10^{-3}$ fm$^{-3}$, $Y_p$=0.39 and T=0.5 MeV (a) and, respectively, 2.4 MeV (b).
The values of the external isovector and isoscalar chemical potentials
($\lambda$, $\lambda_3$) are (-10.06 MeV, 3.98 MeV) and, respectively,
(-11.03 MeV, 5.86 MeV).
Experimental values \cite{nudat} and predictions of the 10-parameter mass model 
of Duflo-Zuker \cite{DZ} have been used for the binding energies. 
}
\label{fig:surfFe}
\end{figure}

In the treatment of finite temperature we have presented in this section we have assumed that, 
similar to the zero temperature case, stellar matter in a given thermodynamic condition 
$(\rho_B,y_p,T)$ is characterized by a single well defined Wigner-Seitz cell. 
This is of course an approximation, since what has to be minimized at equilibrium 
is the total free energy density, and not the single-cell free energy density. 
The two variational approaches will give approximately the same result if the free-energy landscape 
has a single deep minimum.
If, by contray, different Wigner-Seitz cells correspond to comparable free energies, 
they will all be present in the equilibrium configuration, even if with different probabilities. 
As we could have anticipated from inspection of Fig.\ref{fig0}, the free energy landscape is highly 
degenerate for stellar matter.
This is confirmed by Fig.\ref{fig:surfFe} where, for two arbitrary representative 
thermodynamic conditions,  
$\rho_B=10^{-3}$ fm$^{-3}$, $Y_p=0.39$, $T$=0.5 MeV and $T$=2.4 MeV,
the constrained cluster free energies $F_{\beta}^e-\lambda_B A -\lambda_3 (A-2Z)$
with conveniently choosen values for $(\lambda_B, \lambda_3)$ are depicted. 
In each of these plots the different nuclei are  
immersed in the same neutron, proton and electron gas. 
At the lowest considered temperature we can see that, though a single minimum exists, 
which corresponds to the solution of the SNA variational coupled equations (\ref{teq1})-(\ref{teq5}) 
for this given set of constraints, different (heavy as well as light) nuclei might lead
to comparable free energy densities, and will therefore be present at equilibrium.
The higher considered temperature shows a different pattern. Indeed, a plethora of nuclei with masses
and isospin symmetries spanning important ranges have close values of the constrained free energies.
It is easy to anticipate that if pairing and shell effects were ignored, the minimum of the contrained
free energy would have been even flatter.

 \subsection{Thermodynamic limit in the canonical ensemble}\label{section:NSE_can}

The very principle of statistical mechanics tells us that at non-zero temperature different realizations
of the Wigner-Seitz cell will be possible
within the same constraint of total density and proton fraction.

If we consider a very large volume $V$ which contains a number $n\to \infty$ of different Wigner-Seitz cells 
for a total number of particles $A_{tot}$ and a total isospin asymmetry $I_{tot}$, a possible realization of the 
system is now characterized by $k=\{  n_i^{(k)}, i=1,\dots,\infty \}$ 
where $n_i^{(k)}$ is the number of realizations, within the volume $V$, of an arbitrary Wigner-Seitz cell constituted
of a cluster with particle numbers  $A_e^{(i)},I_e^{(i)}=A_e^{(i)}-2Z_e^{(i)}$ and a gas with particle numbers 
$A_g^{(i,k)}=V_{WS}^{(i)}\rho_g^{(k)}, I_g^{(i,k)}=V_{WS}^{(i)}y_g^{(k)}$.
Notice that the gas density and isospin can in principle depend on the realization $(k)$ but do not depend 
on the cell $(i)$. Indeed the nucleon gas density is uniform over the volume because we have divided 
it in cells only for convenience, and the variation of gas particle numbers is just due to the variation of 
Wigner-Seitz volumes.

The total number of particles 
in the  cell $A_{WS}^{(i)}, I_{WS}^{(i)}$ varies from one cell to the other, but the total number of particles
in the volume $V$ is the same for each realization $(k)$:
\begin{eqnarray}
A_{tot}&=&\sum_i  n_i^{(k)}(A_e^{(i)}+V_{WS}^{(i)}\rho_g^{(k)} ), \\
I_{tot}&=&\sum_i  n_i^{(k)}(I_e^{(i)}+V_{WS}^{(i)}y_g^{(k)} ), \\
V&=&\sum_i  n_i^{(k)} V_{WS}^{(i)}.
\end{eqnarray}
Since we are at the thermodynamic limit, these three conditions are in reality only two
\begin{eqnarray}
\rho_B&=&\frac{A_{tot}}{V}=\frac{\sum_i  n_i^{(k)}(A_e^{(i)}+V_{WS}^{(i)}\rho_g^{(k)} )}
{\sum_i  n_i^{(k)} V_{WS}^{(i)}}, \\
y_B&=&\frac{I_{tot}}{V}=\frac{\sum_i  n_i^{(k)}(I_e^{(i)}+V_{WS}^{(i)}y_g^{(k)} )} {\sum_i  n_i^{(k)} V_{WS}^{(i)}}.
\end{eqnarray}

We can then characterize a realization $(k)$ by the fragment distribution and 
the gas isoscalar and isovector densities  
$k=\{  n_i^{(k)}, i=1,\dots,\infty,\rho_g^{(k)},y_g^{(k)} \}$ where now  $ n_i^{(k)}$ is the number of occurrences 
of the Wigner-Seitz cell $(i)$ constituted of a gas $\rho_g^{(k)},y_g^{(k)}$, a cluster  $A_e^{(i)},I_e^{(i)}$
and a volume $V_{WS}^{(i)}$ uniquely defined by the neutrality condition which has to be fulfilled in each cell
\begin{equation}
\frac{Z_e^{(i)}+V_{WS}^{(i)}\rho_{pg}^{(k)} }{V_{WS}^{(i)}}= \rho_p=y_p\rho_B.
\label{volume}
\end{equation}
We can further simplify the problem considering that, for a given macroscopic set of constraints
$(T,\rho_B,y_p)$ we will have a unique partitioning in the macroscopic volume between the cluster fraction
and the gas fraction, that is the one that minimizes the total free energy.
It is very easy to improve on this approximation if necessary, by considering the canonical probability 
associated to each partitioning. We do not do it because it comes out that there are very 
few combinations of $\rho_{cl}=\sum_i  n_i A_e^{(i)}/V$ and $\rho_g$ which lead to the same $\rho_B$. 
This means that we consider that $\rho_g$ and $y_g$ 
do not depend on $(k)$ but only on the  macroscopic constraints. Then the conservation law simplifies to:
\begin{eqnarray}
\rho_B&=&\frac{A_{tot}}{V}=\frac{1}{V} \sum_i  n_i^{(k)}A_e^{(i)}+\rho_g =\rho_{cl}+\rho_g,  \label{cons1}
 \\
y_B&=&\frac{I_{tot}}{V}=\frac{1}{V} \sum_i  n_i^{(k)}I_e^{(i)}+y_g =y_{cl}+y_g;  \label{cons2}
\end{eqnarray}
and the different realizations of the set of constraints $(T,\rho_B,y_p,\rho_g,y_g)$  are defined by 
$k=\{  n_i^{(k)}, i=A_e^{(i)}, I_e^{(i)} \}$ .

The  probability $p_{\beta}(k)$ of this realization  is determined by the usual maximization of the information 
entropy under the constraint of the average energy  and a sharp constraint on the mass 
$A_{tot}$ and isospin $I_{tot}$ eqs.(\ref{cons1}),(\ref{cons2}).
%
%
%
%
We define the total~\footnote{note that here ``total'' has another meaning than in Eq. (\ref{dense})}  
free energy of each realization $(k)$ as: 
\begin{eqnarray}
{F_{tot}(k)}=F_{cl}(k)- T V \ln \left ( z_\beta^{HM}(k) z_\beta^{el} \right ) , \label{ftot}
\end{eqnarray}
with
\begin{eqnarray}
F_{cl}(k)&=& \sum_i  n_i^{(k)}  F_\beta^e(i); \\
F_{\beta}^{e}(i)&=&
E^e -T \ln  V 
-T \ln c_\beta - \frac{3}{2} T  \ln A_e.
\label{fenergy_cl_vtot}
\end{eqnarray}

It is interesting to remark that the cluster free energy at the thermodynamic limit eq.(\ref{fenergy_cl_vtot})
differs from the cluster free energy in the Wigner-Seitz cell eq.(\ref{fenergy_cl_ws}). 
Indeed the number of states for the center-of-mass motion has to be calculated over the whole volume:
\begin{equation}
\sum_{\vec{p}} \exp \left[-\beta  \frac{p^2}{2mA_e}\right]=
\frac{V}{(2\pi\hbar)^3}\left ( \frac{2\pi m A_e}{\beta}\right )^{3/2} .
\end{equation}
This is well known from solid state physics and leads to the Bloch theorem: even if the ions are 
localized at fixed positions in the Coulomb lattice, their center of mass motion is a plane wave over 
the whole volume\cite{bloch_chamel,bloch_us}.

Thanks to the thermodynamic limit, the partition sums are now factorized
\begin{equation}
Z_\beta(\rho_B,y_p) = \sum_k \exp \left[- \beta  F_{cl}(k)\right] \left ( z_\beta^{HM}   z_\beta^{el}\right )^V.
\label{zcano}
\end{equation}
The probability of realization $(k)$ is then simply given by:
\begin{equation}
p_{\beta}(k)=\frac{1}
{Z_\beta^{cl} } \exp \left[- \beta   F_{cl}(k) \right] ,
\end{equation}
with 
\begin{equation}
Z_\beta^{cl}(\rho_B,y_p,\rho_g,y_g) = \sum_k \exp \left[- \beta  F_{cl}(k)\right].
\end{equation}

$ Z_\beta^{cl}$ is the canonical partition sum of an ensemble of fully independent 
clusters, for a total mass number $A_{cl}=V\rho_{cl}$ and isospin $I_{cl}=Vy_{cl}$.   
We note by passing  that we can easily extend this result to the case where we allow mixing of different $\rho_g,y_g$
by considering eq.(\ref{zcano}) as a constrained partition sum.
In that case, the  total partition sum has to be defined as
\begin{equation}
Z_\beta(\rho_B,y_p) = \sum_I Z_\beta^{cl}(\rho_B,y_p,\rho_g^I,y_g^I)  
\left (z_\beta^{HM} (\rho_g^I,y_g^I) \right )^V  \left ( z_\beta^{el}\right )^V.
\end{equation}

The explicit calculation of $ Z_\beta^{cl}$  is a classical problem\cite{dasgupta,nse_us}, and its solution is given by:
\begin{equation}
Z_{\beta}^{cl} =\sum_{(k)} \prod_{A,Z}
\frac{\left (\omega_\beta^e (A,Z)\right )^{n_{A,Z}^{(k)}}}{n_{A,Z}^{(k)}!},
\end{equation}
where 
\begin{equation}
\omega_\beta^e(A,Z) = \exp \left[-\beta F_\beta^e (A,\delta,\rho_g,y_g)\right],
\end{equation}
the sum runs over all possible realizations of the system such that 
the total number of particles is $A_{cl}$, $n_{A,Z}^{(k)}$ is the number of occurrences 
of cluster $A,Z$ in the realization $k$,
and the product runs over r-clusters $A, Z$ or e-cluster $A_e, Z_e$ equivalently, 
since the two are scaled by a factor which is constant if $\rho_g$ and $y_g$ are constant. 
This partition sum can be calculated with a Monte-Carlo technique\cite{nse_us} 
or also analytically via a recursive relation\cite{dasgupta,inequivalence}. 

Notice that for a finite system the total volume $V_{tot}=\sum_i n_i^{(k)} V_i$ is a fluctuating quantity, and only 
$A_{cl}$ is the same event by event. However this is a not problem, because  the conservation 
law is applied to the total density.

It is instructive to consider the SNA limit of a representative cluster. 
Let us suppose that the average multiplicity density
$<n_{AZ}>/V\approx1/<V_{WS}(A,Z,\rho_B,y_B,\rho_g,y_g)>$ for a given $A=\bar{A}$, $Z=\bar{Z}$ and 
$<n_{AZ}>\approx0$,  $\forall A\neq \bar{A}$, $Z\neq \bar{Z}$, or equivalently let us suppose that we 
consider only the most probable cluster in the partition sum.  
Since $A_{cl}=n\bar{A}$, $I_{cl}=n(\bar{A}-2\bar{Z})$ we immediately get
\begin{equation}
Z_\beta^{cl}(n\bar{A},n\bar{Z})=\frac{ \left (\omega_\beta^e (\bar{A},\bar{Z})\right )^n}{n!},
\end{equation}
and 
\begin{equation}
\ln z_\beta^{cl}(n\bar{A}/V,n\bar{Z}/V)=\frac{1}{V_{WS}}\ln \left[\frac{\omega_\beta^e (\bar{A},\bar{Z})}{n}\right]^{n},
\end{equation}
where we have used the Stirling approximation neglecting the $-n$ term:
$\ln (n!)\approx n\ln n-n\approx n\ln n$,
and we have introduced the free energy densities as $-T\ln z_{\beta}=-T\ln Z_\beta/V$.
Using eq.(\ref{zcano}) and eq.(\ref{fenergy}) 
the partition sum becomes 
\begin{equation}
-T \ln z_\beta(\rho_B,y_p) =\frac{1}{V_{WS}} F_{WS}(\bar{A},\bar{Z},\rho_g,y_g,\rho_p)   .
\end{equation}
We can see that we recover a SNA expression which we have already shown converges towards the exact result 
at zero temperature.

The value of $\bar{A},\bar{Z}$ can  be deduced from the equations of state 
\begin{eqnarray}
\mu_B &=& -{T}\frac{\partial \ln z_{\beta}}{\partial \rho_B}|_{y_B}; \\
{\mu}_ 3 &=& -{T}\frac{\partial \ln z_{\beta}}{\partial y_B}|_{\rho_B}, 
\end{eqnarray}
which can also be written as
\begin{eqnarray}
0 &=& \frac{\partial  (-T\ln z_{\beta}-\mu_ B \rho_B)}{\partial \rho_B}|_{y_B}; \\
0 &=& \frac{\partial (-T\ln z_{\beta}-{\mu}_ 3 y_B)}{\partial y_B}|_{\rho_B} .
\end{eqnarray}
Integrating these equations leads to:
\begin{eqnarray}
d\left (-T\ln z_{\beta}-\mu_ B \rho_B- k(y_B)\right ) &=& 0; \\
d\left (-T\ln z_{\beta}-{\mu}_ 3 y_B- h(\rho_B)\right ) &=& 0,
\end{eqnarray}
or also
\begin{equation}
d\left (-T\ln z_{\beta}-\mu_B \rho_B- {\mu}_3 y_B\right ) = 0. 
\end{equation}
This is exactly the same minimization problem as in section \ref{section:SNA}, with the difference that now the variables are
$A,\delta,V_{WS}$ because $\rho_g, y_g$ are fixed.
This physically means that the fact of considering a large number of Wigner Seitz cell has 
eliminated the conservation constraint between $A, \delta$ and $\rho_g, y_g$: 
density and isospin fluctuations are allowed
in each Wigner-Seitz cell because the conservation law applies only to the macroscopic system.
 
As a consequence, the equilibrium sharing equations are slightly modified:
\begin{eqnarray}
  \frac{\partial E^e}{\partial A}|_{\delta,\rho_g,y_g} &=& \mu_B \frac{\rho_0-\rho_g}{\rho_0} + 
{\mu}_3 \frac{\rho_0\delta-y_g}{\rho_0} +\frac{3T}{2A}  \nonumber \\
&+&T  \frac{\partial \ln c_\beta}{\partial A}|_{\delta,\rho_g,y_g} ; \label{thermo}\\
\frac{\partial E^e}{\partial \delta}|_{A,\rho_g,y_g} &=&
 \mu_3 A+
\frac{d\rho_0}{d\delta}\frac{A}{\rho_0} (\mu_B \frac{\rho_g}{\rho_0}+ \mu_3 \frac{y_g}{\rho_0})\nonumber \\
 &+& \frac{3}{2}T \frac{\rho_g}{\rho_0(\rho_0-\rho_g)}\frac{d\rho}{d\delta}  
+T  \frac{\partial \ln c_\beta}{\partial \delta}|_{A,\rho_g,y_g} ,
\end{eqnarray}
with  
\begin{eqnarray}
\mu_B &\equiv& -T\frac{\partial \ln z_\beta^{HM}}{\partial \rho_g}; \\
{\mu}_3 &\equiv& -T\frac{\partial \ln z_\beta^{HM}}{\partial y_g}.
\end{eqnarray}
 We can also notice that in the limit $T\to 0$ the sharing equations at $T=0$ that we have obtained, 
by imposing exact conservation laws within the cells, are recovered as they should. 
Indeed in this limit the system is periodic and the global conservation law is equivalent to a local 
(within the cell) conservation law.

\subsection{the grandcanonical NSE}\label{section:NSE}

The canonical treatment of the previous section is formally correct, but has the disadvantage of being 
extremely expensive from the computational point of view.
For this reason, a grandcanonical formulation appears more appealing and has been preferentially invoked in the
star matter literature\cite{botvina,souza,blinnikov,nse_heckel,nse_hempel,esym_paper,nse_japan}.

To formulate this problem, we consider a very large volume $V\to \infty$ which contains a number 
$n\to \infty$ of a-priori different Wigner-Seitz cells, 
and introduce two external Lagrange multipliers to impose the average isoscalar and isovector densities over the whole
volume. 
As in the previous section, a possible realization of the system is noted by an index
 $k=\{  n_i^{(k)}, i=1,\dots,\infty \}$ 
where $n_i$ is the number of occurrences, within the volume $V$, of an arbitrary Wigner-Seitz cell constituted
of a cluster with particle numbers  $A^{(i)},I^{(i)}$ and a gas with particle numbers 
$A_g^{(i)}=V_{WS}^{(i,k)}\rho_g^{(k)}, I_g^{(i)}=V_{WS}^{(i,k)}y_g^{(k)}$. 
The total number of particles 
in the  cell $A_{WS}^{(i)}, I_{WS}^{(i)}$ varies from one cell to the other, but the total number of particles
in the volume $V$ (or more precisely, the total density and proton fraction, since we are at the thermodynamic
limit $V\to \infty$), are fixed by the externally imposed chemical potentials $\mu$ and ${\mu_3}$.
These densities, as well as the average cluster multiplicities $<n_i>_{\beta\mu{\mu_3}}$ and 
the gas densities $<\rho_g>,<y_g>$, is what we want to calculate.

As we have already discussed, the gas density and isospin could in principle depend on the realization $(k)$ 
 but do not depend on the cell $(i)$. 
The Wigner-Seitz volume is uniquely defined by the neutrality condition eq.(\ref{volume}) in the cell
\begin{equation}
V_{WS}^{(i,k)}=\frac{Z^{(i)}}{\rho_p-\rho_{pg}^{(k)}}  .
\end{equation}
The total Helmholtz free energy  of each realization $(k)$ is given by eq.(\ref{ftot}):
\begin{equation}
{F_{tot}(k)}=  \sum_i  n_i^{(k)} \left ( F_\beta^e(i)  -TV_{WS}^{(i,k)} 
 \ln   \left ( z_\beta^{HM}(k)  z_\beta^{el} \right )\right ) .
\end{equation}
We can see that, because of the dependence on $\rho_p$ of the electron free energy, this equation defines 
a self-consistency problem. Indeed we have:
\begin{equation}
\rho_p=\sum_i n_i^{(k)} \frac{Z_e^{(i)} +\rho_{pg}^{(k)} V_{WS}^{(i,k)} }{V},
\end{equation}
showing that our variational variables $\{n_i^{(k)} \}$ are not independent variables. 
As it is well known in the framework
of the self-consistent mean-field theory~\cite{balian}, an equivalent one-body problem can be formulated 
corresponding to the same information entropy, therefore to the same set of occupations 
as in the self-consistent problem, but with a different  free energy 
corresponding to independent particles, which contains rearrangement terms. 
These rearrangement terms will explicitly appear in the one-body occupations of the 
self-consistent problem.
The free energy of the equivalent one-body problem is given by:
\begin{equation}
{F^{1b}_{tot}(k)}= -TV \ln z_\beta^{el}(\rho_p) +\sum_i  n_i^{(k)}  F_\beta^{1b}(i) ,
\end{equation}
with
\begin{eqnarray}
F_{\beta}^{1b}(i)&=& \frac{\partial  F_{tot} }{\partial n_i^{(k)}} \nonumber \\
&=& F_\beta^e(i)  -TV_{WS}^{(i,k)} 
 \ln  z_\beta^{HM}(k) + \mu_{el} 
\left( Z_e^{(i)} +\rho_{pg}^{(k)} V_{WS}^{(i,k)} \right),\nonumber \\
\end{eqnarray}
where the derivation is taken at constant $\rho_g, y_g, n_j^{(k)}, j\neq i$.
The grand-canonical occupations $n_i^{(k)}$ are determined by the free energy of the equivalent one-body problem, 
meaning that they directly depend on the electron chemical potential.
Notice that in principle also $ F_\beta^e(i)$ depends on the total proton density through the Coulomb 
screening term, therefore it should also give rise to extra rearrangement terms. 
However this extra term
$\partial F_\beta^e/\partial\rho_p\cdot \partial \rho_p/\partial n_i \propto V^{-1}$, 
is negligible in the thermodynamic limit.  
The total Gibbs one-body free energy of each realization $(k)$ is obtained by 
Legendre transformation with respect to the total baryon number and isospin. 
This amounts to introducing as usual two chemical potentials $\mu'_B,\mu'_3$ according to: 
\begin{equation}
G_{tot}^{1b}(k)=  {F^{1b}_{tot}(k)} - 
 \sum_i  n_i^{(k)} \left( \mu'_B A_{WS}^{(i)}+\mu'_3 I_{WS}^{(i)}
\right ) 
\end{equation} 
We can see that we can define auxiliary chemical potentials as
\begin{equation}
\mu_B=\mu'_B-\mu_{el}/2 \;\; ;\;\; \mu_3=\mu'_3+\mu_{el}/2
\end{equation}
such as to make formally disappear the electron contribution in the cluster free energy:
\begin{eqnarray}
G_{tot}^{1b}(k)&=&    - T V \ln z_\beta^{el}  +
 \sum_i  n_i^{(k)} \left( F_\beta^e(i)  -TV_{WS}^{(i,k)} \ln z_\beta^{HM}(k) \right) \nonumber \\
&-& \sum_i  n_i^{(k)} \left[ \mu_B \left(   A_e^{(i)}+V_{WS}^{(i,k)}\rho_g^{(k)} \right)
+ \mu_3 \left(   I_e^{(i)}+V_{WS}^{(i,k)}y_g^{(k)} \right) 
\right] .\nonumber \\
\end{eqnarray}
Using the mean-field relations of uniform nuclear matter: 
\begin{equation}
\ln z_{\beta\mu_B{\mu_3}}^{HM}=\ln z_\beta^{HM} (\rho_g,y_g) +\beta\mu_B\rho_g+\beta{\mu}_3 y_g.
\label{equiv_gas}
\end{equation}
we can see that the gas densities are uniquely determined by the external chemical potentials, and independent of the realization, as we could expect:
\begin{eqnarray}
\rho_g &=& {T}\frac{\partial \ln z_{\beta\mu_B{\mu_3}}^{HM}}{\partial \mu_B}|_{\mu_3}; \label{gas1}\\
y_g &=& {T}\frac{\partial \ln z_{\beta\mu_B{\mu_3}}^{HM}}{\partial \mu_3}|_{\mu_B} . \label{gas2}
\end{eqnarray}
We can then write
$V_{WS}^{(i,k)}=V_{WS}^{(i)}$, $\rho_g^{(k)}=\rho_g$, $y_g^{(k)}=y_g$,  and:
\begin{eqnarray}
G_ {tot}^{1b}(k)&=& - T V \ln z_\beta^{el} \nonumber \\
&+&\sum_i  n_i^{(k)} \left[ G^{e}_{\beta\mu_B\mu_3}(A_e^{(i)},I_e^{(i)}) -T V_{WS}^{(i,k)} 
\ln z^{HM}_{\beta\mu_B\mu_3} \right],\nonumber \\
\end{eqnarray}
where we have defined the in-medium modified cluster Gibbs energy:
\begin{equation}
  G^{e}_{\beta\mu_B\mu_3}(A,\delta,\rho_g,y_g,\rho_p)= F^{e}_\beta- \mu_B A_e- \mu_3  I_e.
\end{equation}
The thermodynamic limit implies that all the realizations correspond to the same (infinite) volume:
\begin{equation}
V=\sum_i n_i^{(k)}V_{WS}^{(i,k)},
\end{equation}
meaning that the gas contribution becomes completely independent of the cluster contribution, 
and fully determined by the chemical potentials
\begin{equation}
  G_ {tot}^{1b}(k)= - T V   \ln \left ( z_\beta^{el} z^{HM}_{\beta\mu_B\mu_3} \right ) 
  +
  \sum_i  n_i^{(k)}   G^{e}_{\beta\mu_B\mu_3}(A_e^{(i)},I_e^{(i)}),
\end{equation}
We are ready to calculate the one-body equivalent grandcanonical partition sum
\begin{equation}
Z^{1b}=\sum_k \exp \left[- \beta G_{tot}^{1b}(k)\right]
=\left (z_\beta^{el}z^{HM}_{\beta\mu_B\mu_3}\right )^V
Z^{cl}_{\beta\mu_B\mu_3},
\end{equation}
with
\begin{eqnarray}
Z^{cl}_{\beta\mu_B\mu_3}&=&\sum_k \exp \left[- \beta \sum_i  n_i^{(k)} G^{e}_{\beta\mu_B\mu_3}(i) \right]\\
&=& \prod_i \sum_{n=0}^\infty \frac{\left[ \exp \left(-\beta  G^{e}_{\beta\mu_B\mu_3}(i) \right) \right]^n}
{n!}\\&=& \prod_i \exp \omega_{\beta\mu_B\mu_3}(i) . \label{zcl_nse}
\end{eqnarray}

With eq.(\ref{zcl_nse}) have recovered a NSE-like expression for the cluster multiplicities
\begin{eqnarray}
<n_i>_{\beta,\mu_B,\mu_3} &=&\omega_{\beta\mu_B\mu_3}(i) \label{n_i_gc} \\
&=&
\exp\left[-\beta  \left (
F_\beta^e(A,\delta,\rho_g,y_g,\rho_p) 
-\mu_B A_e - \mu_3  I_e
\right )\right], \nonumber \\
\label{mult_nse}
\end{eqnarray}
where the electron energy density and entropy density are known.

It is interesting to notice that 
the baryonic component (clusters as well as gas) only depend on the baryonic part of 
the total chemical potentials $\mu_B,\mu_3$.  These chemical potentials are not the thermodynamic potentials
conjugated to the densities $\mu'_B,\mu'_3$ which determine the thermodynamics; 
indeed they are shifted of the electron contribution. This explains why 
the phase transition is quenched in stellar matter even if the baryonic chemical potential 
$\mu_B$ has a backbending behavior as a function of the baryonic density. 
This point will be further discussed in section \ref{sec:electrons}.
The backbending behavior of $\mu_B$ was observed in refs.\cite{inequivalence,stone_2014}, 
but it was interpreted as a sign of ensemble inequivalence \cite{inequivalence} or 
of instability\cite{stone_2014}, since the rearrangement terms coming from the electron contribution 
were not discussed in those papers. 

From a practical point of view, the numerical implementation of the NSE model is simpler than the one if its approximation,
namely the SNA. Indeed the variational character of the approach is fully exhausted by the construction of the partition sum, 
and no extra  variation of the energy functional has to be performed. 
This means that we can easily use fully realistic functionals for the cluster free-energies with no extra numerical cost. 
For the applications shown in this section,  we use the tables of experimental masses of Audi {\em et al.} \cite{nudat} 
and, in order to extend the pool of nuclei for which pairing and shell effects are accounted for, 
evaluated masses of Duflo-Zuker \cite{DZ} for the vacuum energies, the full list of low-lying resonances for light nuclei
for the degeneracy factor $g_\beta$, and realistic level densities fitted from experimental data from ref.\cite{Bucurescu2005} 
in eq.(\ref{deg_bucurescu}). 
Only when this information is not available (or to make the quantitative comparisons with SNA as in 
section \ref{section:NSE_SNA}), we switch to the Skyrme-LDM mass model. 
Moreover, we explicitly consider isospin inhomogeneities in the spatial distribution 
of clusters due to Coulomb and skin effects.
This is done considering that the bulk asymmetry $\delta$ entering in the in-medium correction to the 
cluster energies does not coincide with the global asymmetry $I/A=1-2Z/A$, as proposed in ref.\cite{ldm} 
(see eq.(\ref{eq_asym_deltar})). 
This equation is consistently solved with eq.(\ref{eq_asym_rho0}) which
gives the isospin dependence of the saturation density~\cite{Panagiota2013}.
%
%
%
%
It was shown in ref.~\cite{Panagiota2013,Aymard2014} that accounting for the difference between bulk and global 
asymmetry is a crucial point to obtain, within a cluster model, energy functionals compatible with microscopic calculations. 
Again, the difference between $\delta$ and $I/A$ is neglected in the numerical applications of 
section  \ref{section:NSE_SNA}, in order to compare the NSE and SNA approaches within the same 
definitions for the physical ingredients.

Our final result, eqs.(\ref{gas1}),(\ref{gas2}),(\ref{mult_nse}) is formally very close to the different existing versions 
of grandcanonical extended NSE\cite{botvina,souza,blinnikov,nse_heckel,nse_hempel,esym_paper,nse_japan}.
This is not surprising, since these equations simply state that all the different baryonic species are quasi-ideal 
gases of independent particles. However, some specificity of the proposed approach should be stressed.

It is clear from the microscopic treatments of the Wigner Seitz cell at zero temperature 
that any realistic finite temperature model has to include in some way interactions between the clusters. 
The way of implementing these in medium effects is not unique, and the different treatments lead to a considerable 
spread in the predictions of extended NSE models\cite{nse_comparison}.

The viewpoint we have taken in this paper is that the very definition of a Wigner-Seitz cell implies that WS cells are 
the correct variables that can be treated as independent degrees of freedom. This fully fixes the in-medium effect under 
the unique hypothesis that each cell contains only one bound cluster. 
As we have discussed in the introduction, this hypothesis, which is employed by all the existing models 
in the literature, is certainly not completely correct in general and some cluster-cluster interaction should be 
taken into account\cite{typel_light} to improve the present description.

The result of building a model on independent WS cells is that a NSE-like expression can be recovered for 
the cluster abundances, but with some specific features which insure that the zero temperature limit is properly obtained.
Specifically, we can see from eq.(\ref{mult_nse}) that the variable conjugated to the chemical potentials is not 
the physical cluster size $(A,Z)$ but the reduced value $(A_e,Z_e)$ (eqs.(\ref{are}),(\ref{zre})) which 
represents its bound part. Moreover, the cluster free energy has to be modified according to eq.(\ref{fenergy_cl_ws}) 
if one wants that in-medium effects are limited to a modification of the surface tension.

\subsection{NSE versus SNA}\label{section:NSE_SNA}

To compare in greater detail the SNA to the NSE results, we can evaluate the most probable 
cluster mass and isospin $\bar{A},\bar{I}$ predicted by the NSE. This is obtained by 
maximizing the argument of the exponential in eq.(\ref{n_i_gc}): 
\begin{eqnarray}
dG^e_{\beta,\mu_B,\mu_3}= d\left (F_\beta^e -\mu_B A_e- {\mu}_3 I_e\right ) = 0.
\end{eqnarray}
Since $\rho_g,y_g$ are fixed, we can equivalently put to zero the partial derivatives with respect to 
$A_e$, $I_e$, or with respect to $A$, $\delta$. The first choice leads to:
\begin{eqnarray}
 {\mu}_3&=&\frac{\partial F_\beta^e}{\partial I_e} |_{A_e,\rho_g,y_g}; \\
\mu_B&=&\frac{\partial F_\beta^e}{\partial A_e}  |_{I_e,\rho_g,y_g}  .
\end{eqnarray}
These equations look very different from the equilibrium equations (\ref{teq1}),(\ref{teq2}) 
corresponding to the SNA. 
However they are far from being in a closed form. Indeed,  
the dependence on $A_e,I_e$ of $G^e$ is highly non trivial:
\begin{equation}
F_\beta^e(A,\delta)=F_\beta^e(A(A_e,\delta(A_e,I_e)),\delta(A_e,I_e)),
\end{equation}
where the dependence of $\delta$ on $A_e,I_e$ is obtained from the solution of the two coupled equations
\begin{eqnarray}
\frac{I_e}{A_e}&=&\frac{\rho_0(\delta)}{\rho_0(\delta)-\rho_g}\left ( \delta -\frac{y_g}{\rho_0(\delta)}\right ); \\
A_e&=&A\frac{\rho_0(\delta)-\rho_g}{\rho_0(\delta)}.
\end{eqnarray}
 This coupling will induce an effective coupling between the isoscalar and isovector chemical potential.
After some algebra we get: 
\begin{eqnarray}
  \frac{\partial E^e}{\partial A}|_{\delta} &=& \mu_B \frac {\rho_0-\rho_g}{\rho_0} 
+ {\mu}_3 \frac{\rho_0\delta-y_g}{\rho_0} +\frac{3T}{2A}
+T  \frac{\partial \ln c_\beta}{\partial A}|_{\delta,\rho_g,y_g} ;  \\
\frac{\partial E^e}{\partial \delta}|_{A}  &=&  {\mu}_3 A  \left ( 1+  \frac {y_g}{\rho_0^2}\frac{d\rho_0}{d\delta} \right )
+ \mu_B A  \frac {\rho_g }{\rho_0^2}\frac{d\rho_0}{d\delta} \nonumber \\
&+&\frac{3}{2}T\frac{\rho_g}{\rho_0}\frac{1}{\rho_0-\rho_g}\frac{d\rho_0}{d\delta}
+T  \frac{\partial \ln c_\beta}{\partial \delta}|_{A,\rho_g,y_g} .
\end{eqnarray}
These equations are similar, but not identical to the SNA equations (\ref{teq1}),(\ref{teq2}).
The difference arises from the fact that the Wigner-Seitz volume as a variational variable 
in the SNA approach induces a complex coupling between the different equations. 
In the NSE, the most probable Wigner-Seitz volume is trivially defined by the condition
\begin{equation}
\bar{V}_{WS} =\frac{\bar{Z}_e}{\rho_p-\rho_{pg}}.
\end{equation}
%


Conversely, we have seen that the same result as in this grandcanonical approach is obtained if we consider 
a canonical problem with a large number of Wigner-Seitz cells.
This is not surprising, because the neighboring cells act as a particle bath.

This result implies that we do not necessarily expect that the most probable cluster obtained
in the complete NSE model exactly coincides with the result of the SNA approximation.
This effect however turns out to be very small. 
A much more important source of difference between SNA and NSE is expected when the NSE distribution has multiple peaks
of comparable height. In that case the $(\rho_B,y_p)$ of the total distribution for a given set of chemical potentials 
is not the same as the one of the most probable cluster. 
This induces a non negligible shift between SNA and NSE even at very low temperatures.

\begin{figure}
\begin{center}
\includegraphics[angle=0, width=0.9\columnwidth]{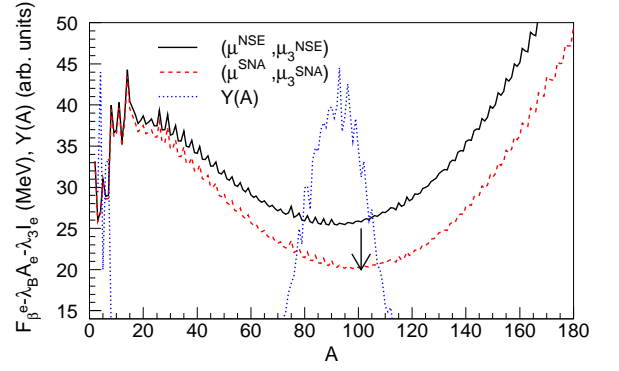}
\end{center}
\caption{(Color online). Constrained cluster free energy for $\rho_B=10^{-3}$ fm$^{-3}$, $T=1.5$ MeV in $\beta$-equilibrium 
with chemical potentials corresponding to the NSE (full line) and SNA (dashed line) model, for a fixed cluster proton 
fraction corresponding to the minimum of the constrained free energy. The arrow gives the SNA solution. 
The dotted line gives the NSE multiplicity distribution in arbitrary units.}
\label{fig:YA_NSE_T=0.5_betaeq}
\end{figure}  

This point is explained in Fig.\ref{fig:YA_NSE_T=0.5_betaeq} which shows the in-medium modified cluster 
free energy eq.(\ref{fenergy_cl_ws}) as a function of the cluster size. 
The free energy has been constrained with two Lagrange multipliers, corresponding to the chemical potentials 
obtained in the SNA and NSE model at an arbitrarily chosen thermodynamic point belonging to the 
$\beta$-equilibrium trajectory,  $\rho_B=10^{-3}$ fm$^{-3}$,  $T=1.5$ MeV, $y_p=0.08503$.
To allow a one-dimensional representation, a cut has been done  with a plane whose $(1-2 \cdot \bar{Z}/\bar{A})$-value
is as close as possible to the corresponding value of the constrained free energy minimum. 
The observed staggering stems from discrete values of $(1-2 \cdot \bar{Z}/\bar{A})$ and, for $A \leq 16$,
structure effects accounted for in the
experimental binding energy. The NSE abundancies are also represented in arbitrary units. 
We can see that the NSE abundancies correctly follow the shape of the constrained free energy, 
as implied by eq.(\ref{mult_nse}). This means that for identical values of the chemical potentials in the 
two models, the optimal SNA cluster (indicated by an arrow in the example shown in the figure) 
should exactly coincide with the most probable NSE cluster. 
However, allowing clusters of any arbitrary size and composition obviously alters the mapping between 
density and chemical potential. The deviation of chemical potentials is typically very small 
(for the example shown in the figure, we have  $\mu_B=-12.929 (-12.889)$ MeV ,
$\mu_3= 13.547 (13.507)$ MeV for NSE (SNA)), 
but it is sufficient to modify the position of the constrained energy minimum. 
As a consequence, a SNA treatment cannot correctly identify the most probable cluster.

NSE predictions were already compared to the SNA approximation, both at fixed proton fraction and 
in $\beta$-equilibrium, in Figs. \ref{fig:compo_Yp=0.2} and  \ref{fig:compo_betaeq} above.
We have seen that, except the very low densities where light cluster degrees of freedom are important, 
at low temperature the NSE model is very close to SNA. 
However the consideration of clusters of all sizes naturally leads to a reduction 
of the cluster size  at high density and high temperature, similar to the  LS equation of 
state because of the particular treatment of $\alpha$ particles in that model. 
It is however important to notice that in the complete NSE $\alpha$ particles are only abundant 
for matter close to isospin symmetry, while more neutron rich hydrogen and helium isotopes 
prevail in neutron rich matter.  This aspect, which by construction cannot be addressed in the LS model, 
will be discussed in greater detail later. 
 
At higher temperature the NSE distribution is spread over a large domain of cluster sizes and isospin 
(see the panel (b) of Fig. \ref{fig:surfFe}), 
and the deviation both with SNA and with the LS equation of state becomes very large. 
In particular, the abundances are dominated by light resonances and the heavy cluster yield becomes 
increasingly negligible with increasing temperature. 

\begin{figure}
\begin{center}
\includegraphics[angle=0, width=0.49\columnwidth]{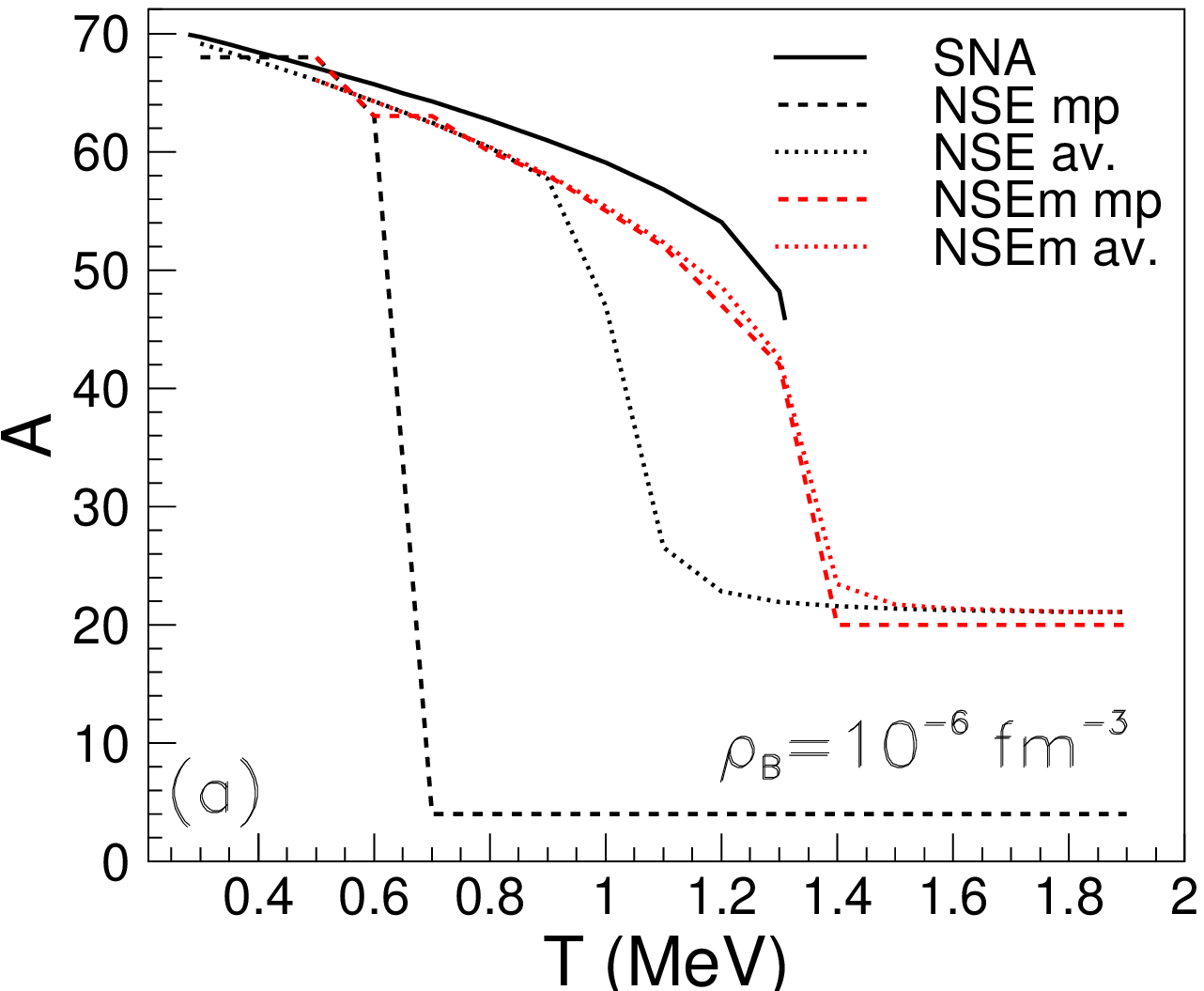}
\includegraphics[angle=0, width=0.49\columnwidth]{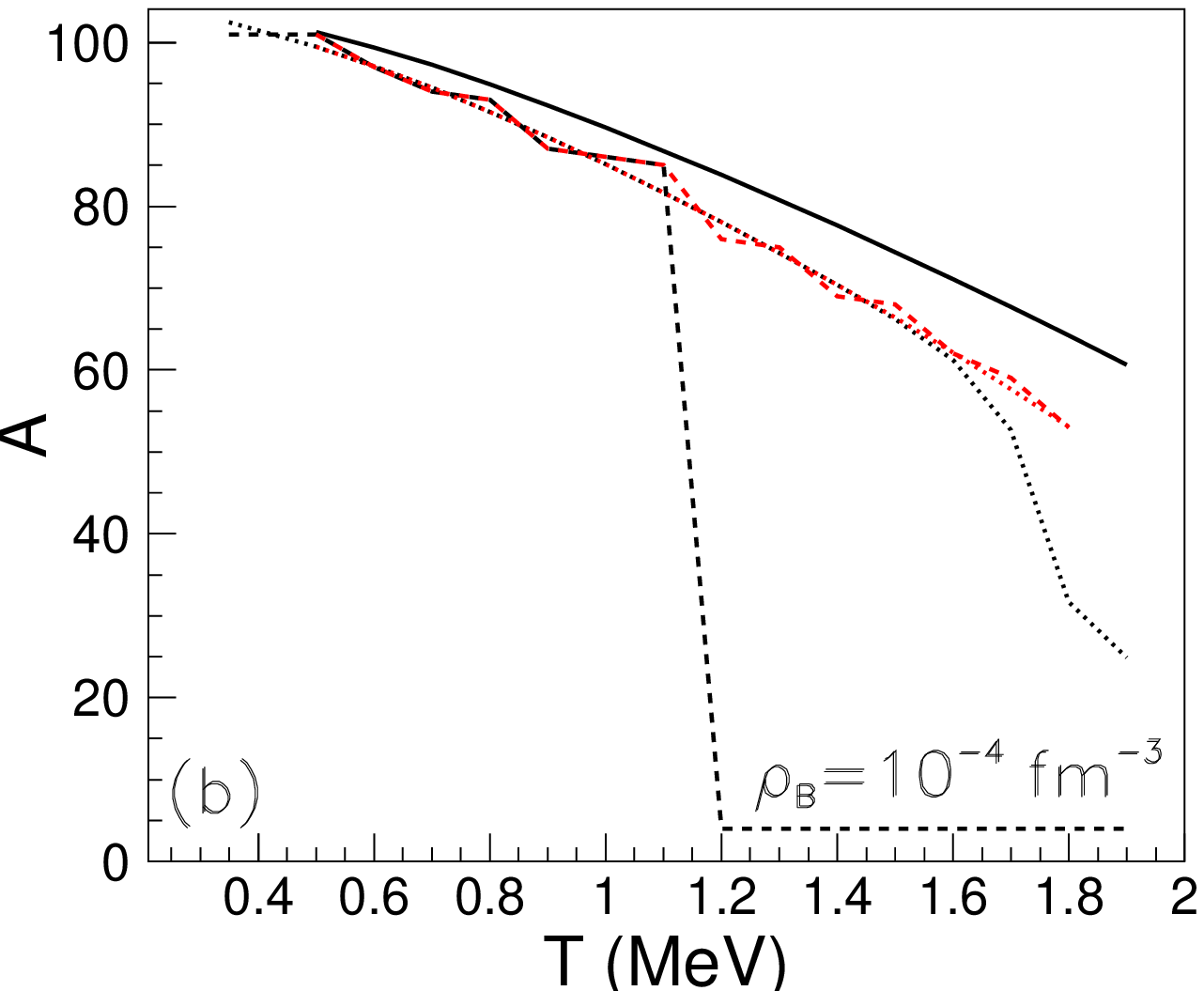}
\includegraphics[angle=0, width=0.49\columnwidth]{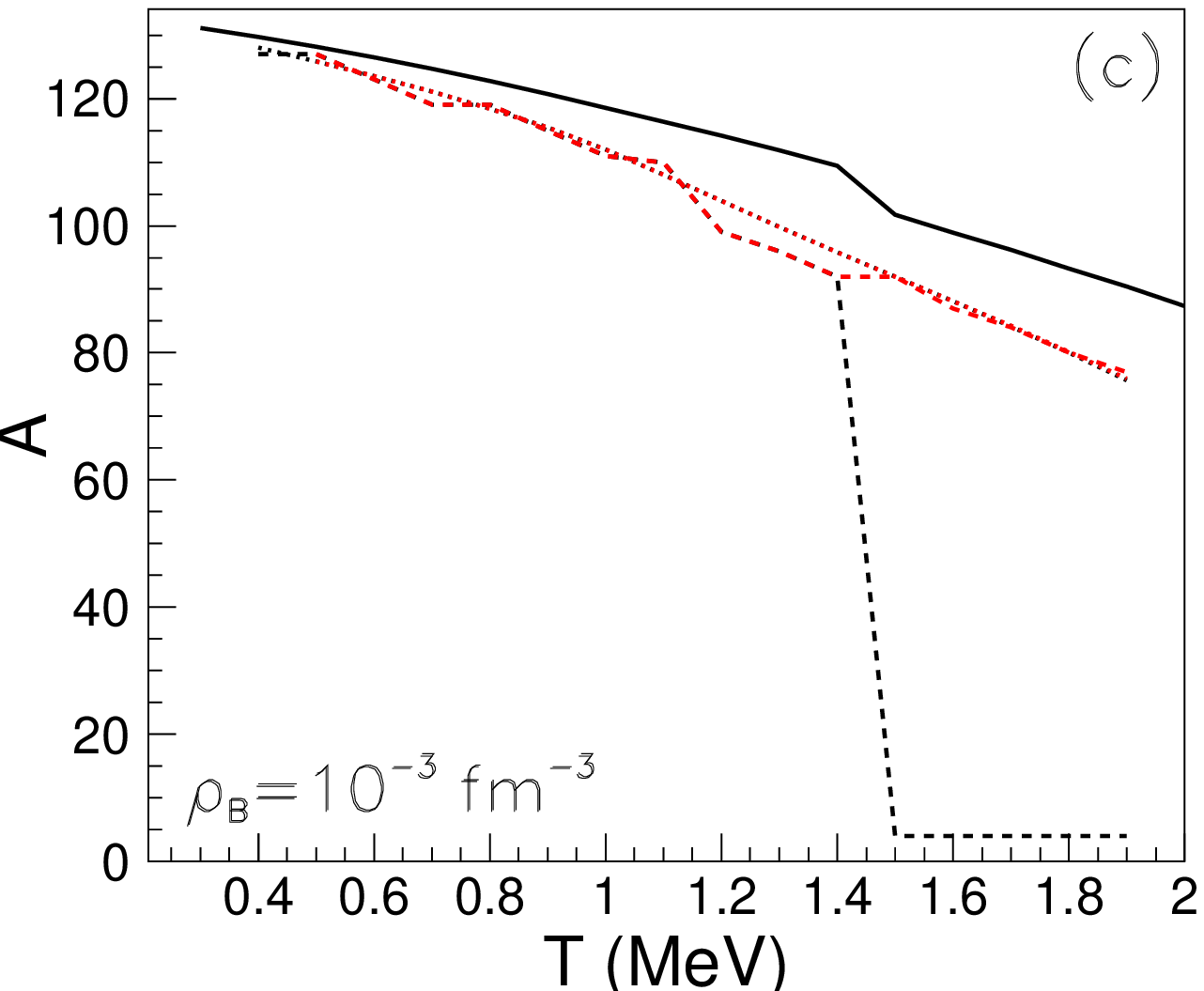}
\includegraphics[angle=0, width=0.49\columnwidth]{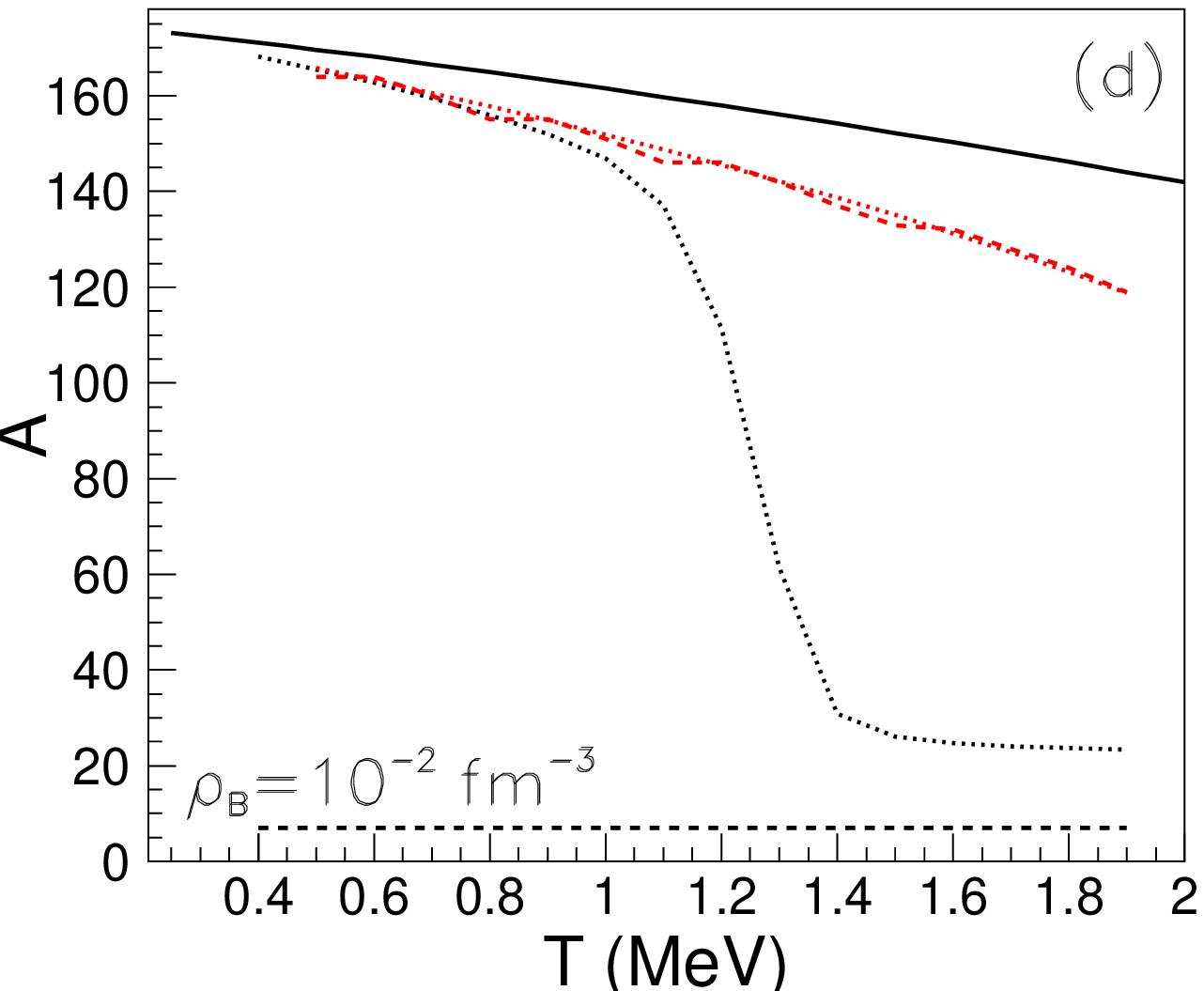}
\end{center}
\caption{(Color online). Structure of the (P)NS crust 
at $\beta-$equilibrium as a function of temperature
for different values of the baryonic density $\rho_B=10^{-6}, 10^{-4}, 10^{-3}, 10^{-2}$ fm$^{-3}$.
Solid and dashed lines correspond to predictions of SNA and, respectively, most abundant cluster in NSE.
Dotted lines correspond to the average heavy ($A \geq 20$) cluster atomic mass in NSE. 
Standard NSE predictions are plotted agaist predictions of a modified NSE (NSEm)
where no cluster lighter than $A=20$ is allowed to exist. 
In SNA Skyrme-LDM \cite{Danielewicz2009} binding energies have been used.
In NSE, experimental data \cite{nudat} have been used for the binding energies 
of nuclides with $A<16$ and Skyrme-LDM \cite{Danielewicz2009} predictions otherwise.
The considered effective interaction is SLY4.
}
\label{fig:NSEvsSNA_A_vs_T_betaeq}
\end{figure}

\begin{figure}
\begin{center}
\includegraphics[angle=0, width=0.49\columnwidth]{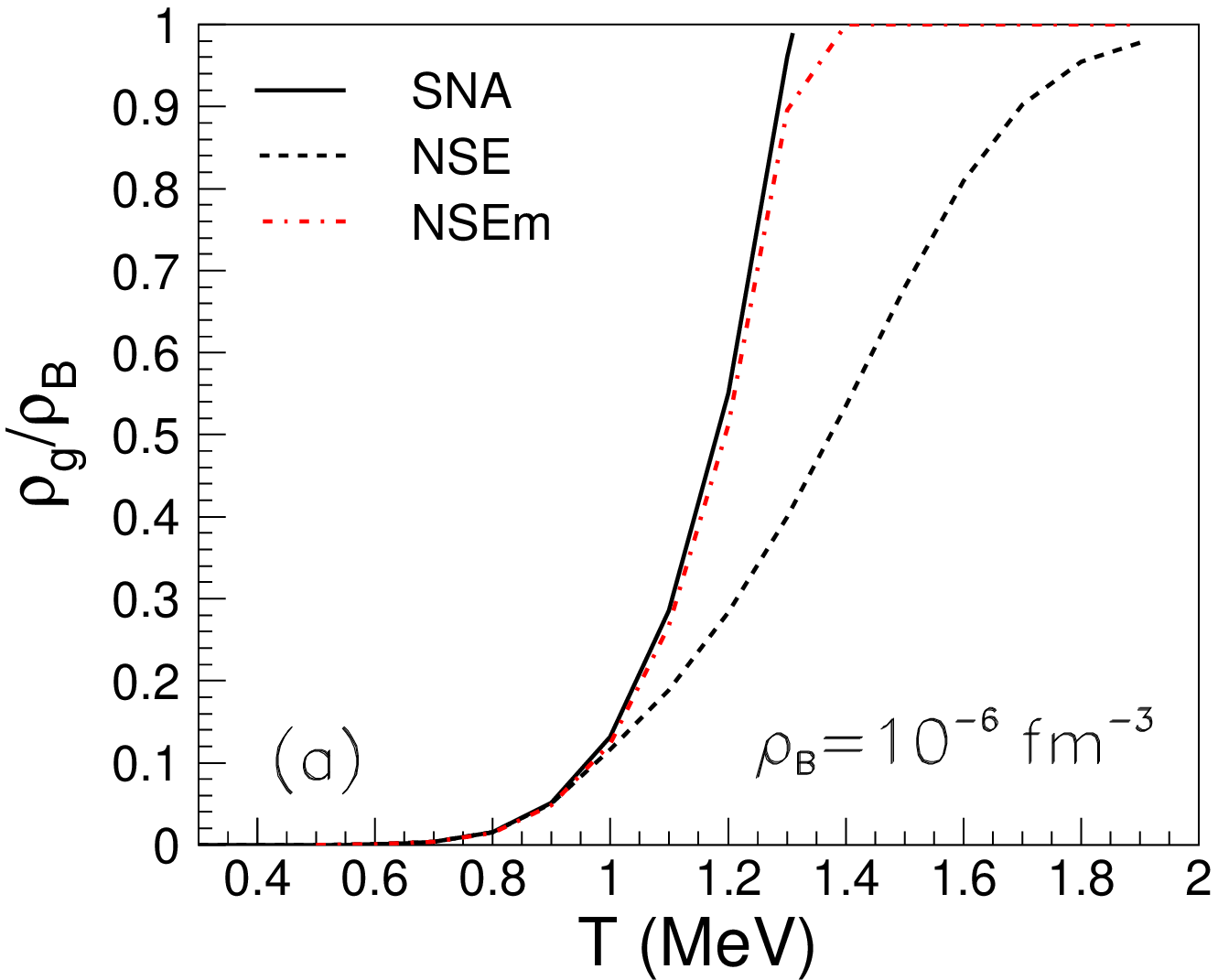}
\includegraphics[angle=0, width=0.49\columnwidth]{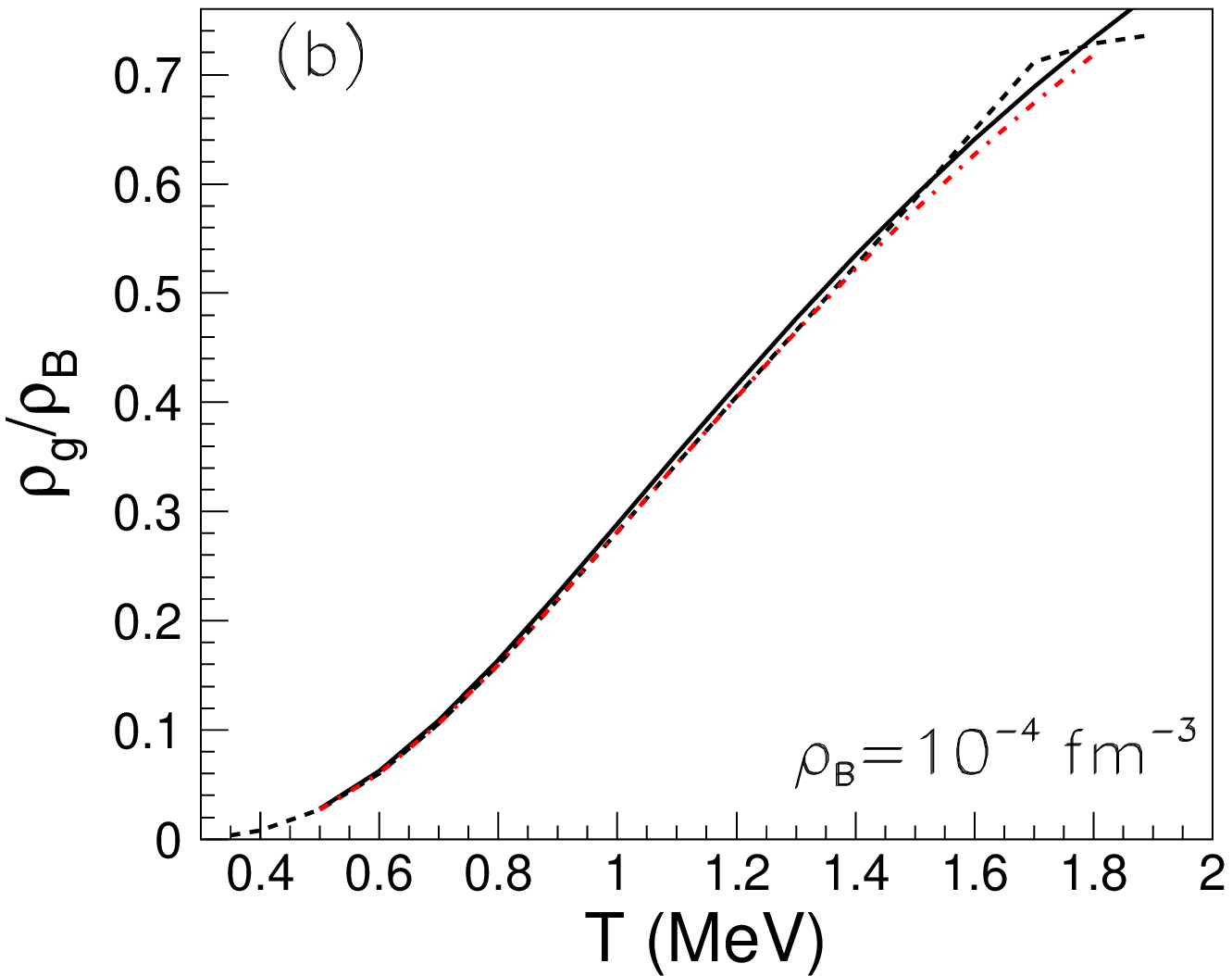}
\includegraphics[angle=0, width=0.49\columnwidth]{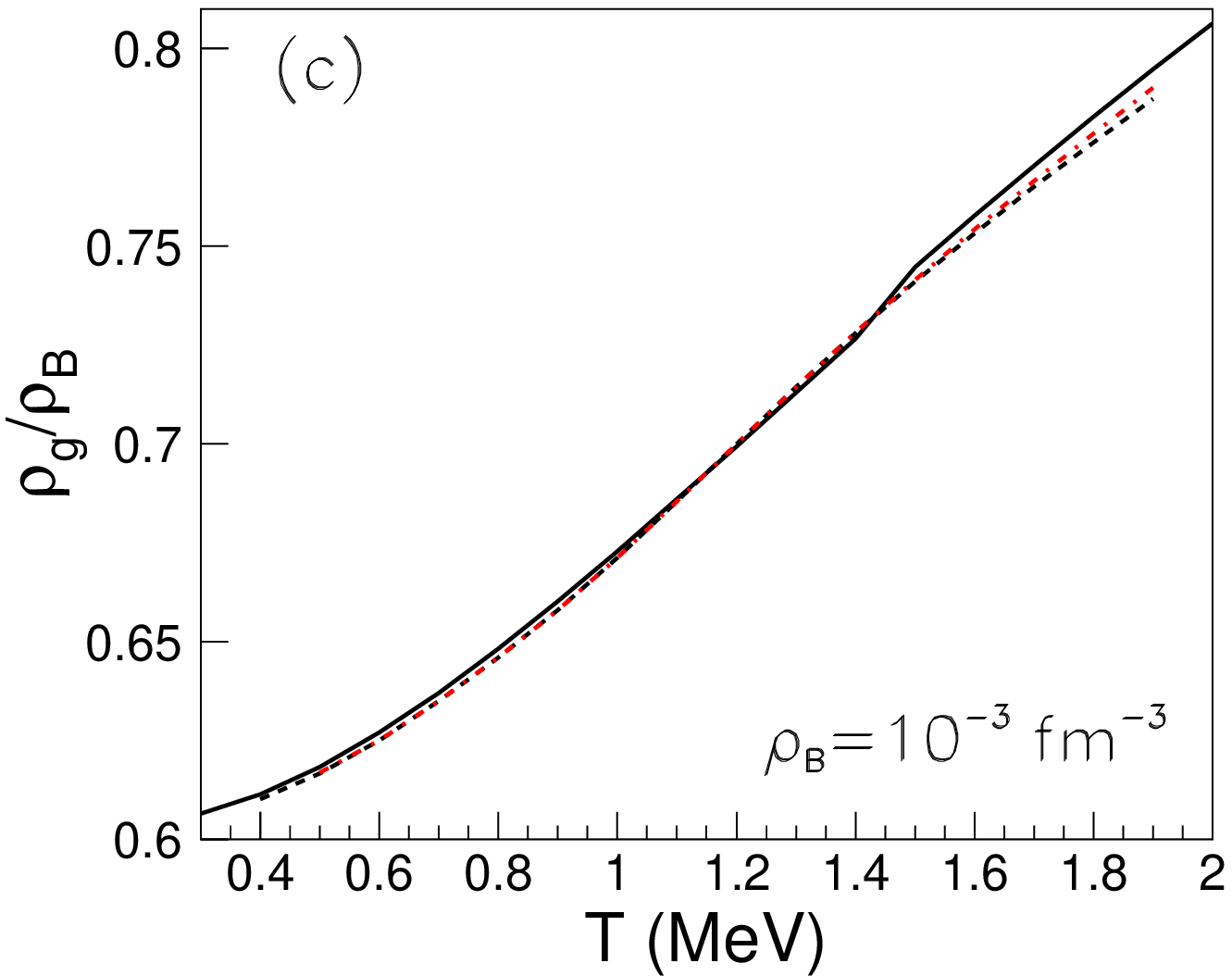}
\includegraphics[angle=0, width=0.49\columnwidth]{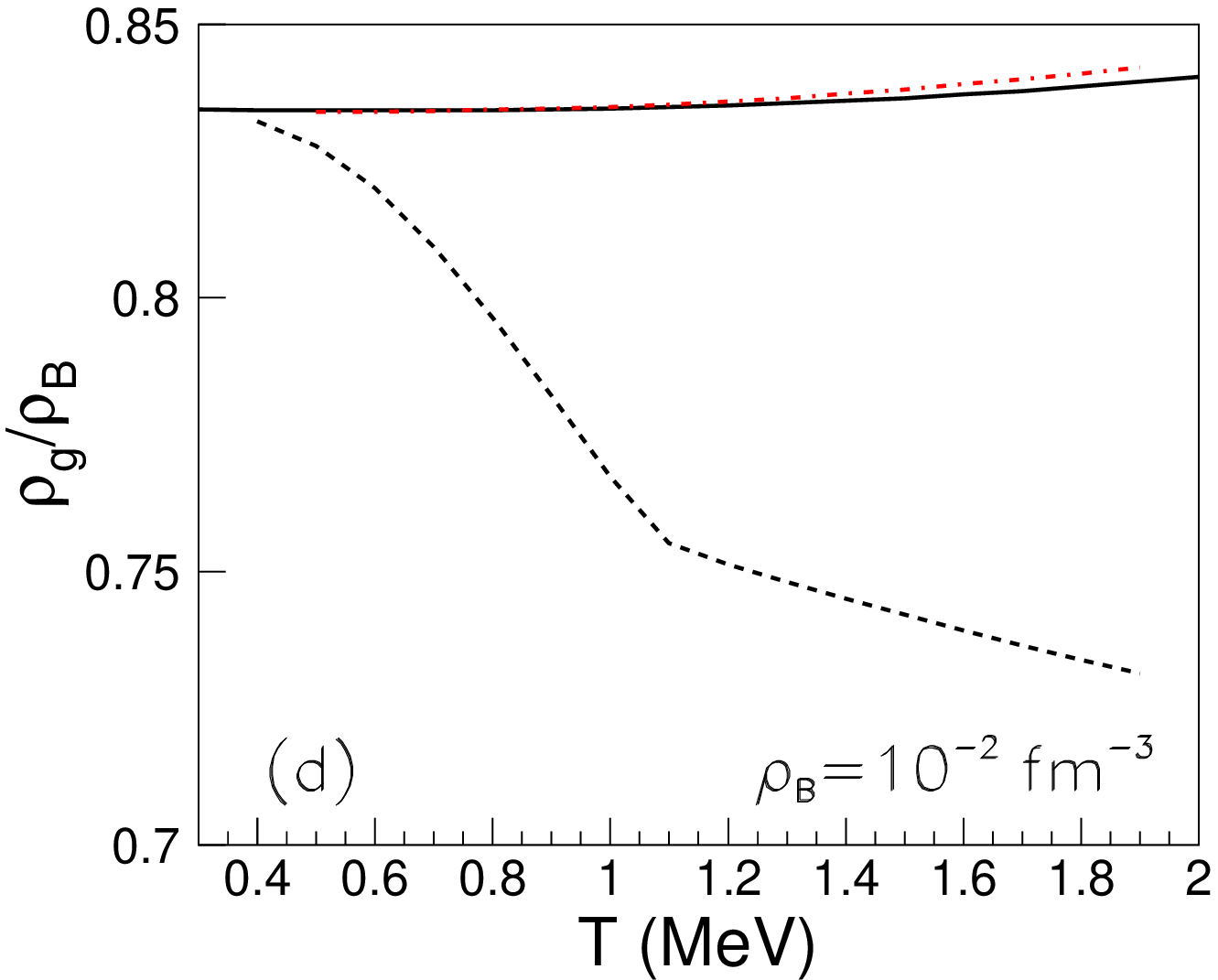}
\end{center}
\caption{(Color online). The same as in figure \ref{fig:NSEvsSNA_A_vs_T_betaeq} for the unbound nucleons
mass fraction.
}
\label{fig:NSEvsSNA_gas_vs_T_betaeq}
\end{figure}

A more detailed comparison between SNA and NSE is given by Figs. \ref{fig:NSEvsSNA_A_vs_T_betaeq}
and \ref{fig:NSEvsSNA_gas_vs_T_betaeq}
in terms of the unique/most probable cluster mass and, respectively, relative mass fraction of 
unbound nucleons.
For NSE, the most abundant cluster mass is plotted against the average mass of heavy clusters 
arbitrarily defined as clusters with $A\geq 20$.

As in the previous figure, Fig. \ref{fig:NSEvsSNA_A_vs_T_betaeq} shows that, whatever the density, 
increasing temperature leads
to an increased deviation between the average SNA 
composition and the most probable NSE cluster.
A huge part of this difference can be explained by the importance of accounting for (a variety of) light clusters 
which are entropically favored at increasing temperature. This is confirmed by red curves that correspond to
a "modified" NSE obtained by artificially switching to zero the statistical weight of all 
clusters lighter than $A=20$. We can see that neglecting light clusters considerably approaches SNA to NSE, 
even if residual differences still persist partially because of the shift in chemical potentials discussed above.
A complementary view is offered Fig. \ref{fig:NSEvsSNA_gas_vs_T_betaeq}.
At intermediate densities ($\rho_B=10^{-4}, 10^{-3}$ fm$^{-3}$) and $T>0.5$ MeV, SNA and NSE predictions agree 
in the percentage of unbound nucleons, thus indicating that the chemical potentials 
of the two models have close values (recall that at a given temperature the gas density only depends 
on the chemical potential).

At variance with this, the extreme densities show a strong reduction of the nucleon gas in NSE with increasing temperature. 
At the lowest density displayed $\rho_B=10^{-6}$ fm$^{-3}$ where
matter as a whole is close to isospin symmetry, this comes from the enhanced production of 
$^2$H and $^4$He at the cost of unbound nucleons. At the highest density $\rho_B=10^{-2}$ fm$^{-3}$ the opposite holds.
The extreme neutron enrichment of $\beta$-equilibrated matter favors copious production of isospin asymmetric
hydrogen and helium isotopes leaving thus less unbound nucleons.
The suppresion of clusters with $A<20$ (red curves) confirms the above reasoning by showing 
a perfect agreement between SNA and NSE
everywhere except $\rho_B=10^{-6}$ fm$^{-3}$ and $T \geq 1.3$ MeV where $\rho_{cl}/\rho_B \to 0$ .

\begin{figure}
\begin{center}
\includegraphics[angle=0, width=0.8\columnwidth]{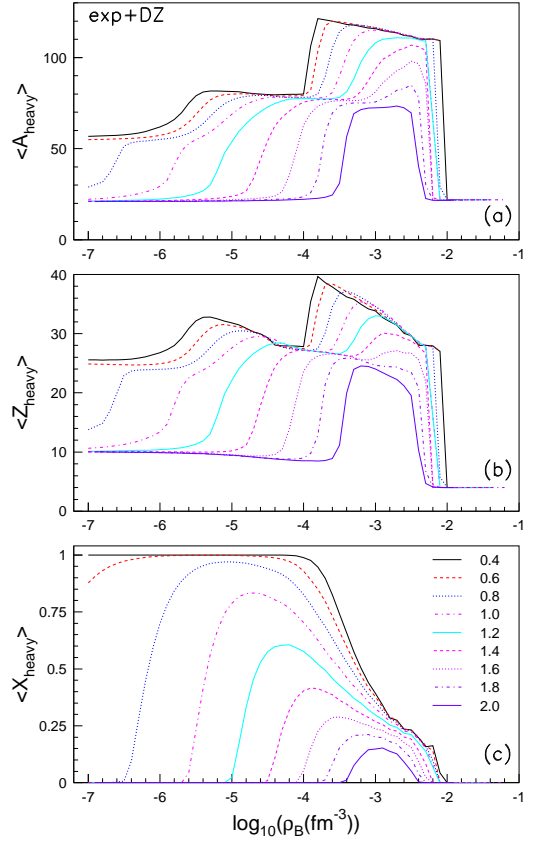}
\end{center}
\caption{(Color online). NSE results at $\beta-$equilibrium for different densities and temperatures
(expressed in MeV and listed in the key legend).
Upper and middle panels: average mass and atomic numbers of clusters heavier than $A=20$.
Lower panel: corresponding mass fraction.
Experimental and DZ10 \cite{DZ} nuclear masses have been considered for clusters.
The SLY4 effective interaction was used for the unbound nucleon component.}
\label{fig:NSE_compoheavy}
\end{figure}

\begin{figure}
\begin{center}
\includegraphics[angle=0, width=0.99\columnwidth]{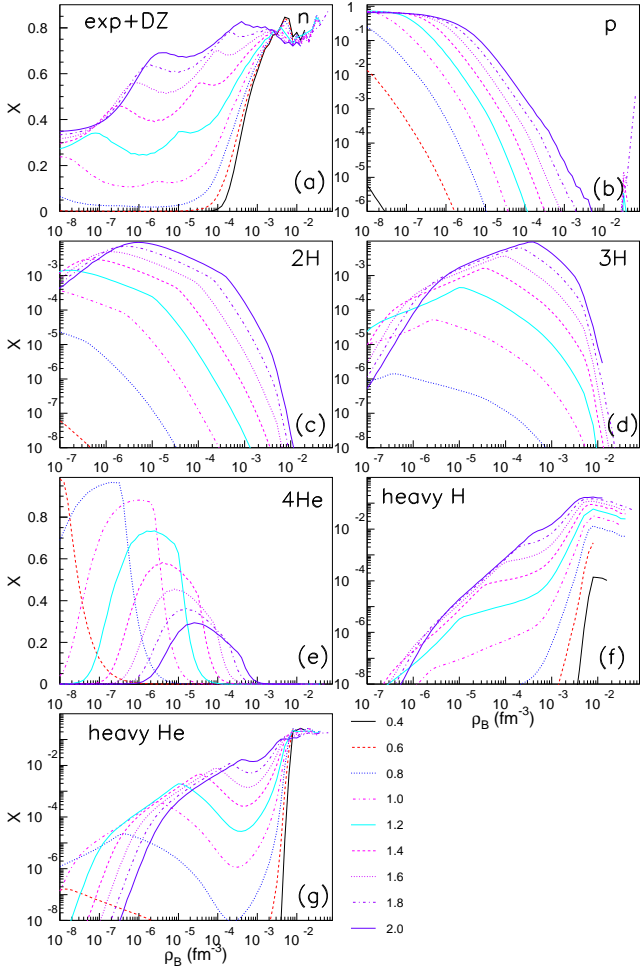}
\end{center}
\caption{(Color online).
NSE mass fractions of unbound nucleons and $^{2}$H, $^{3}$H, $^{3}$He, $^{4}$He, 
$^{A\geq 4}$H and $^{A \geq 5}$He 
at $\beta-$equilibrium for different densities and temperatures 
expressed in MeV and listed in the key legend.
Experimental \cite{nudat} and DZ10 \cite{DZ} data have been used for the binding energies. 
The SLY4 effective interaction was used for the unbound nucleon component.
Note that X-axis range is not the same in all panels.
}
\label{fig:NSE_compolight}
\end{figure}

The global behavior of the $\beta$-equilibrated matter composition in the NSE 
model is shown in Figs. \ref{fig:NSE_compoheavy}, \ref{fig:NSE_compolight}.
In Fig. \ref{fig:NSE_compoheavy} average mass, charge and mass fraction of heavy ($A \geq 20$) nuclei
are plotted as a function of density for temperatures ranging from 0.4 to 2 MeV.
Fig. \ref{fig:NSE_compolight} presents, for the same temperatures, the mass fractions
of unbound neutrons and protons together with the mass fraction of different light species
($^2$H, $^3$H, $^4$He, $^{A \geq 4}$H, $^{A \geq 5}$He) as a function of density.
As mentioned in the figure captions, in these cases experimental values \cite{nudat}
and DZ10 \cite{DZ} predictions have been used for nuclear masses.
The unbound nucleon component is treated according to SLY4.

We can again observe the nice convergence towards the zero temperature composition of the Wigner-Seitz cell, 
as well as the complex behavior as a function of density for all temperatures, 
leading to a melting of the clusters in the nuclear medium at a density of the order of $\rho_B=0.01$ fm$^{-3}$.
As it is well known in the literature, the exact value of the transition density  depends on the effective interaction. 
We do not try to make such a study here because the presence of deformation degrees of freedom in the 
form of pasta phases, here neglected, would most probably modify the value of the transition density. 
Inspection of Fig. \ref{fig:NSE_compolight} reveals the importance of accounting for all the different 
light nuclear species and not limiting to deuteron and $\alpha$-particles. 
This is true for any proton fraction, but particularly clear in the very neutron rich matter implied 
by $\beta$-equilibrium, where light unbound resonances completely dominate, together with unbound neutrons, 
the matter composition at high temperature. 
The consideration of light particles of all species, including heavy hydrogens and helions, is natural and easy in the context of a NSE model. However to our knowledge no SNA approach includes such particles in the description of the average Wigner-Seitz cell, even if a very promising step in this direction was recently undertaken in refs.\cite{pasta_light,typel_light}. 
This underlines again the importance of going beyond the SNA approximation in the finite temperature stellar problem.

\begin{figure}
\begin{center}
\includegraphics[angle=0, width=0.8\columnwidth]{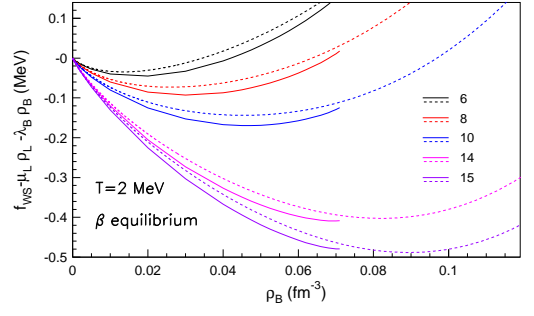}
\end{center}
\caption{(Color online). Behavior of the constrained free energy in $\beta$-equilibrium ($\mu_L=0$) 
for the clusterized phase (full line) and the homogeneous phase (dashed line) at different values
of the baryonic chemical potential $\lambda_B$ mentioned in the key legend (in MeV).
Experimental binding energies and predictions of the 10-parameter mass model of 
Duflo-Zuker are used for the nuclear clusters. 
For the unbound nucleon gas the SLY4 effective interaction has been used. 
}
\label{fig:barf_T=2_betaeq}
\end{figure}

\begin{figure}
\begin{center}
\includegraphics[angle=0, width=0.8\columnwidth]{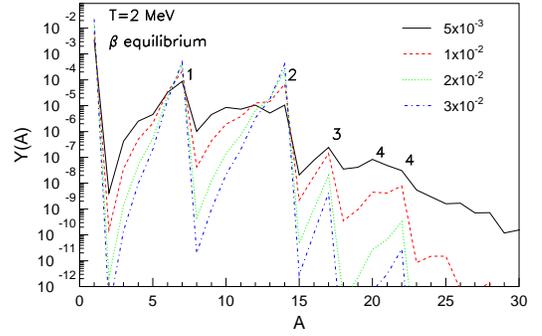}
\end{center}
\caption{(Color online). Fragment mass distributions corresponding to $\beta$-equilibrium, $T$=2 MeV 
and different baryonic densities (expressed in fm$^{-3}$) as listed in the key legend.
The numbers next to peaks specify the charge number of the most abundant nucleus.
}
\label{fig:YA_T=2_betaeq}
\end{figure}

\subsection{Thermodynamics and electrons} \label{sec:electrons}

The problem of the grandcanonical formulation which has been recently observed\cite{inequivalence,stone_2014}
 is that baryonic matter at sub-saturation densities presents a first order liquid-gas phase transition which is signaled 
by the fact that a huge part of the phase diagram is jumped over if one imposes constant chemical potentials\cite{nse_us,inequivalence}.

As we have discussed in section \ref{section:phase_transition}, this instability is not physical and only comes from the fact that 
the electron contribution is neglected in the instability analysis. If the electron free energy is accounted for, the dependence of the free energy density on the baryonic density reads 
\begin{equation}
f(\rho_B,\rho_p) =f_B(\rho_B,\rho_p) +f_{el}(\rho_p)  , 
\end{equation}
where $f_B$ denotes the baryonic part and the Coulomb interaction part between protons and electrons, and $f_{el}=-T\ln z_\beta^{el}$.
The relations (\ref{chemical}) between density and chemical potential are shifted because of the electron contribution
\begin{equation}
\mu_B \to \mu'_B=\mu_B +\frac{1}{2} \mu_{el} \;\; ;\;\; \mu_3\to \mu'_I=\mu_3-\frac{1}{2}  \mu_{el},
\end{equation}
and the curvature of the constrained free energy density is augmented of a positive term as:
\begin{equation}
\frac{\partial^2 f}{\partial \rho_B^2} = \frac{\partial \mu_{B}}{\partial \rho_B}+ \frac{1}{2}
\frac{\partial \mu_{e}}{\partial \rho_{el}}. \label{fcon}
\end{equation}

This quenching of the phase transition has as a practical consequence that a one-to-one correspondence 
between density and chemical potential exists in stellar matter, meaning that it is possible to describe 
all the possible density configurations in a grandcanonical treatment, provided the electron contribution 
is accounted for. In that case, the ensemble equivalence is recovered and the associated partitions 
are by construction identical to the ones obtained in a canonical model, as we have 
explicitly demonstrated in sections \ref{section:NSE_can},\ref{section:NSE}. 

It is however in principle perfectly possible that a residual convexity persists in the constrained 
free energy (\ref{fcon}). In that case, a first order phase transition reminiscent of liquid-gas 
would survive in stellar matter. 
Such a phenomenology was recently suggested in ref.\cite{stone_2014} 
and evidenced, for T=0, in Fig.6(b).

To answer to this question, we show in Fig.\ref{fig:barf_T=2_betaeq} the comparison between the 
constrained free energy energy density of the clusterized Wigner-Seitz cell, and the one corresponding to 
homogeneous matter, for different values of the baryonic chemical potential (mentioned in the key legend),
at a representative temperature of 2 MeV.
We can see that the clusterized phase systematically presents a lower free energy density than the 
homogeneous system, for all chemical potentials up to about $\mu_B=14$ MeV.
For the highest considered chemical potential, 15 MeV, the constrained energy minimum corresponds to homogeneous matter.
This means that the first order phase transition is restricted to a density domain between about 
$\rho_B=0.07$ and $\rho_B=0.09$ fm$^{-3}$. These values obviously depend on the temperature and 
on the effective interaction, but still the associated density discontinuity is too small to have any 
observable effects. Moreover, we have to stress again that we have disregarded deformation degrees of freedom in this model. 
The inclusion of highly deformed pasta clusters would lead to a lowering of the clusterized phase, 
and an extra shrinking of the possible transition domain. 
 
Figure  \ref{fig:YA_T=2_betaeq} shows the detailed matter composition in the high density region close to the 
transition to homogeneous matter. 
Dominance of exotic light nuclei as $^7$H, $^{14}$He, $^{17}$Li, $^{20}$Be, $^{22}$Be is worthwhile to note meaning that
it is very important to account for light clusters in that domain. 
It is therefore possible that smoother transitions would be observed between the different pasta phases, 
and between the pasta phase and homogeneous matter, if light clusters were accounted for.
We leave this point to future developments.

\section{Conclusions}\label{section:conclusions}

In this paper we have presented a unified treatment of the stellar matter composition 
and equation of state in the sub-saturation regime, which can be applied at any temperature, density and proton fraction.

The basic idea of the model is to consider stellar matter as a statistical mixing of independent Wigner-Seitz cells. 
The individual composition in terms of bound and unbound particles does not minimize the free energy density, but the combination 
of different cells does. 

The result is a set of NSE-like equations for the cluster abundancies, where both the bulk and the surface part of the cluster self-energies are modified by the presence of free nucleon scattering states, and a high energy cut naturally appears in the cluster internal state partition sum.
The model dependence of the finite temperature model is thus limited to the model dependence of the treatment of the Wigner-Seitz cell, 
which in turn is very well constrained by microscopic calculations, with a residual uncertainty limited to the density dependence of the symmetry energy in the underlying effective interaction, and the detailed treatment of the isospin dependent surface tension.  
In the present applications, the in-medium modifications are  treated in the local density approximation, but it will be extremely interesting 
to map them from a more sophisticated microscopic, and possibly beyond mean-field treatment in the next future.

We have analytically shown that, for a given set of chemical potentials, the most probable cell composition coincides with the one which is obtained by the standard variational procedure assuming one single representative cluster. This guarantees that the model has the correct zero temperature limit.

However, the simultaneous presence of many different clusters in each thermodynamic condition modifies the relation between density and chemical potential with respect to the single nucleus approximation. As a consequence, stellar matter predictions of this improved NSE model differ from the SNA approximation even at the level of the most probable composition, and even at temperatures lower than 1 MeV. 
 
We have specifically shown quantitative applications in $\beta$-equilibrium.
The dominant configuration is a mixture of  clusters of different 
mass and atomic numbers.
This effect is due at low density to the non-monotonic behavior of the cluster energies due 
to shell and sub-shell closures, and at high density to the flatness or multi-minima
of the free-energy landscape for very neutron rich matter. None of these features
can be accounted in a SNA approach.
In addition to the multi-peaked cluster distribution we have seen that very light clusters appear with a probability 
compared to the one of heavier clusters. This feature is accounted in the Lattimer-Swesty SNA model by including
$\alpha$-particles in the single representative Wigner-Seitz cell. 
We can see that this is physically correct at the lowest densities, which at $\beta$-equilibrium correspond to matter close 
to isospin symmetry. Conversely, in very asymmetric matter as it can be found at $\beta$-equilibrium at higher density,
the most probable light cluster is never a $\alpha$ particle, but rather the last bound isotope of H and He.
It is therefore clear that at finite temperature other light particles than $\alpha$ have to be included 
in the equilibrium.

\acknowledgements
Discussions with Stefan Typel are gratefully acknowledged. 
This work has been partially funded by the SN2NS project 
ANR-10-BLAN-0503 and it has been supported by
NewCompstar, COST Action MP1304.
Ad. R. R acknowledges partial support from the Romanian National
Authority for Scientific Research under grants 
PN-II-ID-PCE-2011-3-0092 and PN 09 37 01 05
and kind hospitality from LPC-Caen.

\end{document}